\documentclass[a4paper, 12pt]{article}

\usepackage{amsmath}
\usepackage{graphicx,psfrag,epsf}
\usepackage{enumerate}
\usepackage[round]{natbib}
\usepackage{url} % not crucial - just used below for the URL 
\usepackage{bm}
\usepackage[dvipsnames]{xcolor}
\usepackage[colorlinks = true]{hyperref}
\hypersetup{citecolor = blue,
			linkcolor = blue,
			 urlcolor = blue}
\usepackage{mathrsfs, mathtools}	
\usepackage{amsthm, amssymb, amsfonts}

\newtheorem*{lemma*}{Lemma}

\newtheorem*{cor*}{Corollary}
\newtheorem{thm}{Theorem}

\usepackage{sectsty}
\allsectionsfont{\normalsize\bfseries}
\usepackage{caption}
\usepackage[normalem]{ulem}
\newcommand{\GG}[1]{} % for bickel 2008a and 2008b

\newcommand{\blind}{0}

\begin{document}

	\def\spacingset#1{\renewcommand{\baselinestretch}%
		{#1}\small\normalsize} \spacingset{1}

	%%%%%%%%%%%%%%%%%%%%%%%%%%%%%%%%%%%%%%%%%%%%%%%%%%%%%%%%%%%%%%%%%%%%%%%%%%%%%%
	
	\if0\blind
	{
		\title{%\bf Covariance Estimation with Uniform Blocks 
			%\tred{Covariance Matrix Estimation for High-throughput Biomedical Data with Interconnected Community Structure}
			\bf{Covariance Matrix Estimation for High-Throughput Biomedical Data with Interconnected Communities}}
		\author{Yifan Yang \\
			Department of Mathematics, University of Maryland, College Park, MD \\
			and \\
			Chixiang Chen \\
			School of Medicine, University of Maryland, Baltimore, MD \\
			and \\
			Shuo Chen \thanks{
				Emails: yiorfun@umd.edu; chixiang.chen@som.umaryland.edu; shuochen@som.umaryland.edu; 
				and * is for the corresponding author}\hspace{.2cm} \\
			School of Medicine, University of Maryland, Baltimore, MD}
		\maketitle
	} \fi
	
	\if1\blind
	{
		\bigskip
		\bigskip
		\bigskip
		\begin{center}
			{\LARGE\bf Covariance Matrix Estimation for High-Throughput Biomedical Data with Interconnected Communities}
		\end{center}
		\medskip
	} \fi
	
	\bigskip
	
	\begin{abstract}
		Estimating a covariance matrix is central to high-dimensional data analysis.
		Empirical analyses of high-dimensional biomedical data, including genomics, proteomics, microbiome, and neuroimaging, among others, consistently reveal strong modularity in the dependence patterns. 
		In these analyses, intercorrelated high-dimensional biomedical features often form communities or modules that can be interconnected with others.
		While the interconnected community structure has been extensively studied in biomedical research (e.g., gene co-expression networks), its potential to assist in the estimation of covariance matrices remains largely unexplored. 
		To address this gap, we propose a procedure that leverages the commonly observed interconnected community structure in high-dimensional biomedical data to estimate large covariance and precision matrices.
		We derive the uniformly minimum variance unbiased estimators for covariance and precision matrices in closed forms and provide theoretical results on their asymptotic properties. 
		Our proposed method enhances the accuracy of covariance- and precision-matrix estimation and demonstrates superior performance compared to the competing methods in both simulations and real data analyses.
	\end{abstract}
	
	\noindent%
	\vfill
	
	\newpage
	\spacingset{1.45} % DON'T change the spacing!
	
	\section{Introduction}
	
	\label{Sec:introduction}
	
	Technological innovations in biomedicine have facilitated the generation of high-dimensional datasets with simultaneous measurements of up to millions of biological features \citep{FanLv2008}. 
	In the past few decades, numerous statistical methods have been developed to analyze these large-dimensional datasets. 
	Estimating a covariance matrix (or a precision matrix) is fundamental to these analyses \citep{FanLiaoLiu2016, CaiRenZhou2016, Wainwright2019} because a covariance matrix not only describes the complex interactive relations among variables but also leads to accurate inferential and predictive results for clinical outcomes \citep{HeKangHong2019, KeRenQi2022}. 
	Since the dimensionality of the variables is much larger than the sample size, we resort to advanced statistical methods rather than traditional covariance estimation strategies \citep{Johnstone2001, JohnstonePaul2018}. 
	The shrinkage and thresholding methods can provide a reliable and robust covariance estimator under the sparsity assumption \citep{LedoitWolf2004, BickelLevina2008b, RothmanLevinaZhu2009, CaiLiu2011}. 
	In addition, prior knowledge of a covariance structure can greatly improve the accuracy of estimation and statistical inference \citep{Fan2005, Bien2019}. 
	For example, recent methods can accommodate Toeplitz, banded, block-diagonal, or separable covariance structures \citep{CaiRenZhou2013, BickelLevina2008a, DevijverGallopin2018, KongAnZhang2020, ZhangShenKong2022}.

\begin{figure}[!htb]
		\begin{center}
			\includegraphics[width = 1.0\linewidth]{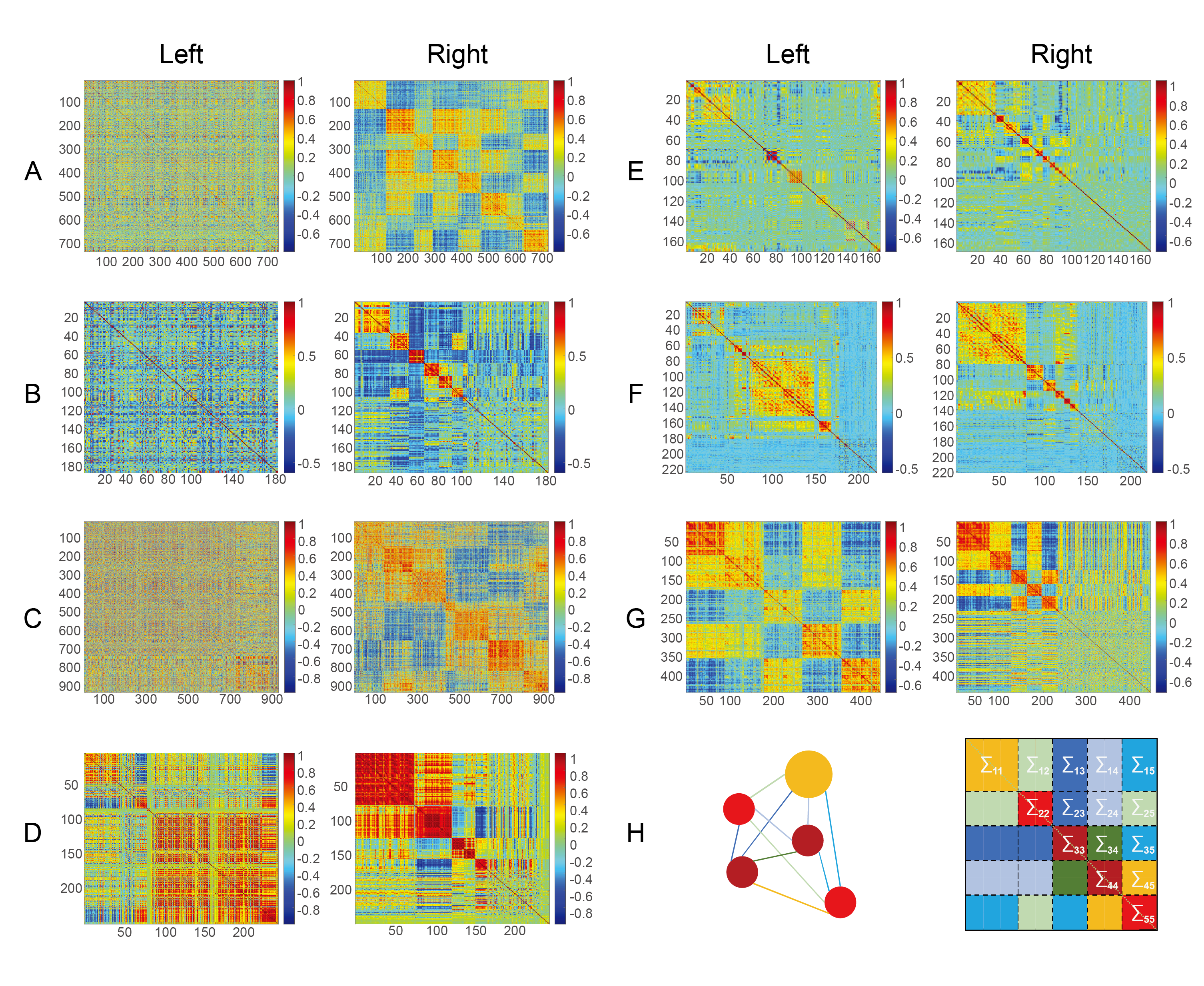} 
			\caption{ 
				We demonstrate that the interconnected community structure is widespread across various types of high-throughput biomedical data, including 
				genomics \citep[A;][]{SpellmanSherlockZhang1998}, 
				proteomics \citep[B;][]{YildizShyrRahman2007}, 
				multi-omics \citep[C;][]{PerrotLevyRajjou2022}, 
				nuclear magnetic resonance plasma metabolomics \citep[D;][]{RitchieSurendranKarthikeyan2023}, 
				exposome \citep[E;][]{ISG2021}, 
				liquid chromatography-tandem mass spectrometry serum metabolomics \citep[F;][]{ISG2021}, 
				brain imaging \citep[G;][]{ChiappelliRowlandWijtenburg2019}, 
				where the left plots are the heatmaps of correlation matrices of raw variables, 
				and the right plots are the heatmaps of correlation matrices of reordered variables (as the results of applying the community detection or clustering algorithms).
				The left plot in H illustrates a network model with $5$ interconnected communities,
				and the right plot in H displays the corresponding population matrix.}
				\label{Fig:data_instances}
		\end{center}
	\end{figure}

	In the present research, we consider a well-organized block structure named \emph{interconnected community structure} that holds considerable importance in biomedical applications due to its widespread prevalence across diverse fields and its challenges in accurate estimation. 
	The interconnected community structure is commonly found in a wide range of high-dimensional datasets with various data types, including genetics, proteomics, brain imaging, and RNA expression data, among others \citep[][please see the examples in Figure ~\ref{Fig:data_instances}]{SpellmanSherlockZhang1998, YildizShyrRahman2007, ChiappelliRowlandWijtenburg2019, ChenBowmanMayberg2016, HeKangHong2019, HeLiZhu2015, WuMaLiu2021}.
	Although latent, this structure can be accurately extracted using recently developed network pattern detection methods, as proposed by \citet{LeiRinaldo2015}, \citet{WuMaLiu2021}, and \citet{LiLeiBhattacharyya2022}. 
	However, the potential of utilizing this structure to enhance large covariance matrix estimation remains unexplored, making it a primary motivation for our current research.

	An interconnected community structure exhibits several properties of highly organized networks. 
	For example, it demonstrates high modality as some variables are clustered in the multiple and coherently correlated communities; it exhibits small-worldness as these communities are interconnected; and the network is scale-free as the remaining variables are isolated if singletons (see the right parts of B, D, E, F, and G in Figure~\ref{Fig:data_instances}) are detected \citep{Newman2006}. 
	Therefore, we can specify this interconnected community structure by assigning the high-dimensional variables to multiple interconnected communities, which are more informative, and a set of singletons. 
	The interconnected community structures might not be directly available from the high-dimensional biomedical data, but they can be estimated by several clustering algorithms and network detection methods \citep{LeiRinaldo2015, WuMaLiu2021}.
	Although there are potential benefits to leveraging the estimated interconnected community structure to enhance the estimation of large covariance matrices, existing statistical methods are restricted to establishing a connection between this structure and the covariance or precision parameters and providing precise estimates.
	To address this methodological gap, we propose a novel statistical procedure that enables closed-form estimators of large covariance and precision matrices and supports robust statistical inference.

	%%%%%%%%%%%%%%%%%%
	%%%% Figure 1 %%%%
	%%%%%%%%%%%%%%%%%%
	%%%%%%%%%%%%%%%%%%%%%
	%%%End of Figure 1 %%
	%%%%%%%%%%%%%%%%%%%%%
	
	We propose a parametric covariance model that subdivides the covariance matrix into blocks or submatrices and assigns each block to either a community or the interconnection between two communities based on the observed interconnected community structure. 
	By linking the covariance parameters and the underlying network topological structure, we can facilitate a closed-form estimator for each covariance parameter (focusing solely on the elements within the corresponding block) and establish the asymptotic properties for the proposed estimators.
	Specifically, we derive explicit estimators employing advanced matrix theories, i.e., the block Hadamard product representation of a covariance matrix with the above structure, and the covariance matrix regularization in the high-dimensional setting, where the number of diagonal blocks exceeds the sample size.

	Our method makes at least three novel contributions. 
	Firstly, we have developed a fast, closed-form, and accurate procedure for estimating a large covariance (and precision) matrix with a particular structure that is applicable to various high-throughput biomedical data, including genomics, metabolomics, proteomics, neuroimaging, and others.
	Our method quantitatively characterizes the interconnected community structure by estimating parameters in both diagonal submatrices (i.e., correlations among features within communities) and off-diagonal submatrices (i.e., interactions among features between communities). 
	Thus, it offers better insights into the interactive mechanisms of a complex biosystem.
	By utilizing this interconnected community structure, our large covariance matrix estimation procedure outperforms comparable methods in terms of accuracy and is numerically robust to model misspecification. 
	Consequently, our approach can lead to a more precise selection of biological features (e.g., cancer-related gene expressions) in the context of multiple testing,  which relies on accurate and reliable large covariance- and precision-matrix estimation \citep{FanHanGu2012, FanHan2017}. 
	Secondly, we have derived the exact variance estimators and established the asymptotic properties for the covariance parameter estimators, which enables us to evaluate covariance patterns and provide confidence intervals. 
	Lastly, we have extended our method to accommodate scenarios where the number of diagonal blocks exceeds the sample size, allowing for scalability to accommodate ultra-high-dimensional datasets.

	The rest of the paper is organized as follows.
	Section~\ref{Sec:methodology} introduces our proposed method. 
	We first mathematically define the interconnected community structure (with uniform blocks) and the uniform-block matrix in Section~\ref{Subsec:nmatrix}.
	In Section~\ref{Subsec:low}, we derive the uniformly minimum variance unbiased covariance- and precision-matrix estimators in the ``small number of communities” setting by taking advantage of the block Hadamard product representation.
	We then generalize the estimation procedure to the ``large number of communities” setting with a diverging number of diagonal blocks in Section~\ref{Subsec:high}.
	Section~\ref{Sec:numerical} contains thorough numerical evaluations of our method under various scenarios.
	Section~\ref{Sec:real} illustrates the use of our method in two real-world applications.
	We provide a discussion in Section~\ref{Sec:discussion}.
	The properties of uniform-block matrices, the exact covariance estimators for the proposed estimators, additional simulation studies, and all the technical proofs are provided in the \href{Supplementary Material.pdf}{Supplementary Material}.

	\section{Methodology}
	
	\label{Sec:methodology}
	
	\subsection{\textit{A Parametric Covariance Model with the Uniform-Block Structure}}
	
	\label{Subsec:nmatrix}
	
	Suppose $\textbf{X}_{n \times p}$ is an $n$ by $p$ observed data matrix containing $n$ independent and identically distributed $p$-variate normal vectors $\bm{X}_{1}, \ldots, \bm{X}_{n} \sim N(\bm{\mu}, \bm{\Sigma})$ with mean $\bm{\mu} \coloneqq \operatorname{E}(\bm{X}_1) \in \mathbb{R} ^ {p}$, positive definite covariance matrix $\bm{\Sigma} \coloneqq \operatorname{cov}(\bm{X}_1) \in \mathbb{R} ^ {p \times p}$, and precision matrix $\bm{\Omega} \coloneqq \bm{\Sigma} ^ {-1}$.
	Without loss of generality, let $\bm{\mu} = \textbf{0}_{p \times 1}$ be known, where $\textbf{0}_{r \times s}$ denotes an $r$ by $s$ zero matrix.  
	Furthermore, let $\textbf{S}_{p \times p} \coloneqq n ^ {-1} \textbf{X} ^ \top \textbf{X}$ denote the unbiased sample covariance matrix, where $\textbf{M} ^ \top$ denotes the transpose of a matrix (or a vector) $\textbf{M}$. 
	Moreover, we require that the covariance matrix has the \emph{interconnected community structure with uniform blocks} (abbreviated as the \emph{uniform-block structure}, distinguishing it from cases with singletons), as described below. 
	The parameterization of this covariance structure is illustrated in two steps.
	
	We specify the parametric covariance matrix based on our previous discussions of the uniform-block structure.
	First, we use a vector to characterize the community sizes.
	Specifically, given the dimension $p$ of the covariance matrix $\bm{\Sigma}$ and the number of communities $K$, let $p_1, \ldots, p_K$ be positive integers, satisfying $p_k > 1\ (k = 1, \ldots, K)$ and $p = p_1 + \cdots + p_K$, and let $\bm{p} \coloneqq (p_1, \ldots, p_K) ^ \top$ be the \emph{partition-size vector}, which is assumed to remain fixed throughout this paper. 
	Given $\bm{p}$, we can express $\bm{\Sigma}$ \emph{partitioned by $\bm{p}$ in block form}:
	\begin{equation}
		\label{Eq:partition}
		(\bm{\Sigma}_{kk'}) 
		\coloneqq 
		\begin{pmatrix}
			\bm{\Sigma}_{11} & \bm{\Sigma}_{12} & \ldots & \bm{\Sigma}_{1K} \\
			\bm{\Sigma}_{21} & \bm{\Sigma}_{22} & \ldots & \bm{\Sigma}_{2K} \\
			\vdots & \vdots & \ddots & \vdots \\
			\bm{\Sigma}_{K1} & \bm{\Sigma}_{K2} & \ldots & \bm{\Sigma}_{KK} \\
		\end{pmatrix},
	\end{equation}
	where $\bm{\Sigma}_{kk'} \in \mathbb{R} ^ {p_k \times p_{k'}}\ (k, k' = 1, \ldots, K)$.
	Second, following~\eqref{Eq:partition}, we specify the diagonal submatrix $\bm{\Sigma}_{kk} \coloneqq a_{kk} \textbf{I}_{p_k} + b_{kk} \textbf{J}_{p_k}$ for every $k$ and the off-diagonal submatrix $\bm{\Sigma}_{kk'} \coloneqq b_{kk'} \textbf{1}_{p_{k} \times p_{k'}}$ with $b_{kk'} = b_{k'k}$ for every $k \neq k'$, where $\textbf{I}_{s}$, $\textbf{J}_{s}$, and $\textbf{1}_{r \times s}$ are an $s$ by $s$ identity matrix, an $s$ by $s$ all-one matrix, and an $r$ by $s$ all-one matrix, respectively. 
	Using $K < p_1 + \cdots + p_K = p$, we can represent a large covariance matrix $\bm{\Sigma}$ by a smaller diagonal matrix $\textbf{A} \coloneqq \operatorname{diag}(a_{11}, \ldots, a_{KK})$, a smaller symmetric matrix $\textbf{B} \coloneqq (b_{kk'})$ with $b_{kk'} = b_{k'k}$ for every $k \neq k'$, and a known vector $\bm{p}$:
	\begin{equation}
		\label{Eq:nsigma}
		\bm{\Sigma}(\textbf{A}, \textbf{B}, \bm{p})
		\coloneqq
		(\bm{\Sigma}_{kk'})
		= \begin{pmatrix}
			a_{11} \textbf{I}_{p_1} + b_{11} \textbf{J}_{p_1} & b_{12} \textbf{1}_{p_1 \times p_2} & \ldots & b_{1K} \textbf{1}_{p_1 \times p_K} \\
			b_{21} \textbf{1}_{p_2 \times p_1} &  a_{22} \textbf{I}_{p_2} + b_{22} \textbf{J}_{p_2} & \ldots & b_{2K} \textbf{1}_{p_2 \times p_K} \\
			\vdots & \vdots & \ddots & \vdots \\
			b_{K1} \textbf{1}_{p_K \times p_1} & b_{K2} \textbf{1}_{p_K \times p_2} & \ldots & a_{KK} \textbf{I}_{p_K} + b_{KK} \textbf{J}_{p_K}
		\end{pmatrix}.
	\end{equation}
	
	In this paper, we say that the pattern in~\eqref{Eq:nsigma} is the \emph{uniform-block structure}. 
	If a matrix has the structure in~\eqref{Eq:nsigma}, it is called a \emph{uniform-block matrix}.
	This covariance parameterization strategy for $\bm{\Sigma}_{kk}$ and $\bm{\Sigma}_{kk'}$ has been widely employed in the field of statistics.
	For example, it is utilized in the generalized estimation equations where the working correlation structure has a compound symmetry, as well as in linear mixed-effects models with a random intercept both have this pattern.
	In practice, this parameterization strategy can characterize the covariance knowledge well using a parsimonious model (as shown in Figure~\ref{Fig:data_instances}).
	In Section~\ref{Sec:numerical}, we demonstrate that the performance of this parameterization remains robust under misspecification and matrix perturbation. 
	By building the parsimonious and effective covariance-matrix specification, we can develop reliable and accurate covariance-matrix estimators using a likelihood-based approach, and these estimators can achieve optimistic theoretical properties.

	Notice that the partition-size vector $\bm{p}$ is assumed to be known throughout this article. 
	In practice, it can be learned using a preliminary algorithm (e.g., a $K$-medoids clustering algorithm by \citet{LeiRinaldo2015} or a network detection algorithm by \citet{WuMaLiu2021}) before estimating the covariance matrix.
	Alternatively, given the heatmap, community sizes can be consistently estimated using a least-square approach \citep{BraultDelattreLebarbier2017}.
	Also, the above definition of a uniform-block matrix does not guarantee its positive definiteness in general.
	Thus, additional constraints should be imposed on the uniform-block matrix $\bm{\Sigma}(\textbf{A}, \textbf{B}, \bm{p})$ in~\eqref{Eq:nsigma} to ensure that it is a valid covariance matrix. 
	We defer the discussion of these constraints to the next section.
	
	\subsection{\textit{Matrix Estimation for the Uniform-Block Structure with a Small \texorpdfstring{$K$}{K}}}
	
	\label{Subsec:low}
	
	Given a partition-size vector $\bm{p}$ and the parametric covariance matrix $\bm{\Sigma}(\textbf{A}, \textbf{B}, \bm{p})$ with the uniform-block structure~\eqref{Eq:nsigma}, we define a $q$-dimensional parameter vector 
	\begin{equation*}
		\bm{\theta} \coloneqq (a_{11}, \ldots, a_{KK}, b_{11}, \ldots, b_{1K}, b_{22}, \ldots, b_{KK}) ^ \top
	\end{equation*}
	that includes all the covariance parameters in the blocks.
	That is, the parameters of interest are in the upper triangular part of $\bm{\Sigma}(\textbf{A}, \textbf{B}, \bm{p})$.
	Also, $q = K + K (K + 1) / 2$.
	Thus, the problem of estimating a $p$ by $p$ symmetric covariance matrix reduces to the estimation of the $q$-dimensional parameter vector $\bm{\theta}$.
	In practice, $q$ is considerably smaller than $p (p + 1) / 2$, thereby remarkably reducing the dimensionality of the parameters of interest.
	
	The small $K$ setting can be specified as follows: $q < n$, while $K$, $p$, and $q$ are fixed, and $p$ can be greater than $n$. 
	In other words, both the number of diagonal blocks $K$ and the number of parameters in the blocks $q = K + K(K + 1) / 2$ are smaller than the sample size $n$. 
	Moreover, we require that $K$ and the dimension of the covariance matrix $p$, which is proportional to $K$, are fixed, so that $q$ is also fixed.
	Also, $p$ can be large enough to exceed $n$ with a small $K$.
	In this section, we first introduce an explicit maximum likelihood estimator of $\bm{\theta}$ with asymptotic properties and then show it to be the uniformly minimum variance unbiased estimator (UMVUE), for which we also provide the exact variance estimator in closed form. 
	Finally, we present the uniformly minimum variance unbiased covariance- and precision-matrix estimators for the setting of a small $K$.
	
	We begin with maximum likelihood estimation.
	Specifically, let $(\textbf{S}_{kk'})$ be the block form of $\textbf{S}$ partitioned by $\bm{p}$.
	Since the data $\textbf{X}$ are normally distributed, the log-likelihood function of the data can be expressed by
	\begin{equation*}
		%\label{Eq:loglike_original}
		\ell_n (\bm{\theta}; \textbf{X})
		\propto \frac{n}{2} \log \left( \det \left[ \left \{\bm{\Sigma}(\textbf{A}, \textbf{B}, \bm{p}) \right \} ^ {-1} \right] \right) 
		- \frac{n}{2} \operatorname{tr}\left[{(\textbf{S}_{kk'})}  \left\{\bm{\Sigma}(\textbf{A}, \textbf{B}, \bm{p}) \right\} ^ {-1} \right].
		%+ \text{constant}.
	\end{equation*}
	A typical approach in the literature to estimate $\bm{\theta}$ is to derive the score function by taking the first-order partial derivative of the log-likelihood function with respect to $\bm{\theta}_j$:   
	\begin{equation}
		\label{Eq:score_original}
		\frac{\partial}{\partial \bm{\theta}_j} \ell_n (\bm{\theta}; \textbf{X})
		= \frac{n}{2} \operatorname{tr}\left[ \{\bm{\Sigma}(\textbf{A}, \textbf{B}, \bm{p}) - (\textbf{S}_{kk'})\}
		\frac{\partial \left\{ \bm{\Sigma}(\textbf{A}, \textbf{B}, \bm{p}) \right\} ^ {-1} }{\partial \bm{\theta}_j} \right]
		\quad (j = 1, \ldots, q),
	\end{equation}
	where $\bm{\theta}_j \in \{a_{11}, \ldots, a_{KK}, b_{11}, \ldots, b_{1K}, b_{22}, \ldots, b_{KK}$\} denotes the $j$th element of $\bm{\theta}$ and $\partial \{\bm{\Sigma}(\textbf{A}, \textbf{B}, \bm{p})\} ^ {-1} / \partial \bm{\theta}_j \in \mathbb{R} ^ {p \times p}$ is a $p$ by $p$ matrix whose entries are functions of these $\bm{\theta}_j$.
	
	However, solving the score equation derived from~\eqref{Eq:score_original} is challenging.
	Although the unknown entries $a_{kk}$ and $b_{kk '}$ are uniformly and elegantly arranged in $\bm{\Sigma}(\textbf{A}, \textbf{B}, \bm{p})$, they become entangled in a complex way in the precision matrix $\bm{\bm{\Omega}} = \{\bm{\Sigma}(\textbf{A}, \textbf{B}, \bm{p})\} ^ {-1}$.
	In other words, $\bm{\theta}_j$ is implicit in $\bm{\Omega}$, making the closed form of $\bm{\Omega}$ generally inaccessible.
	The complexity of the calculation increases as the dimension of the precision matrix grows.
	Alternatively, existing numerical algorithms for solving $\bm{\theta}$ (e.g., the method of averaging proposed by \citet{Szatrowski1980}) rely on iterative updating schemes, which demand a long computational time and may lead to unstable estimates.
	These facts motivate us to reconsider the possibility of deriving a closed-form estimator of $\bm{\theta}$.
	Therefore, we aim to find an explicit expression for $\bm{\Omega}$ in terms of $a_{kk}$ and $b_{kk '}$ by taking advantage of the special covariance structure in~\eqref{Eq:nsigma}.
	More precisely, we speculate that $\bm{\Omega}$ has an analogous form to~\eqref{Eq:nsigma}, which can indeed be fulfilled by realizing the following representation of the block Hadamard product.

	\begin{lemma*}
		
		\label{Lem:nrep}
			Given a partition-size vector $\bm{p} = \left(p_1, \ldots, p_K\right) ^ \top$ satisfying $p_k > 1$ for every $k$ and $p = p_1 + \cdots + p_K$, 
			suppose a $p$ by $p$ matrix $\textbf{N}$ partitioned by $\bm{p}$ exhibits the uniform-block structure in~\eqref{Eq:nsigma}, expressed by $\textbf{N}(\textbf{A}, \textbf{B}, \bm{p})$, 
			then the following representation is unique, 
			\begin{equation*}
				\textbf{N}(\textbf{A}, \textbf{B}, \bm{p}) = \textbf{A} \circ \textbf{I}(\bm{p}) + \textbf{B} \circ \textbf{J}(\bm{p}),
			\end{equation*}
			where $\textbf{I}(\bm{p}) \coloneqq \textbf{I}_p(\textbf{I}_K, \textbf{0}_{K \times K}, \bm{p}) = \operatorname{Bdiag}(\textbf{I}_{p_1}, \ldots, \textbf{I}_{p_K})$, 
			$\textbf{J}(\bm{p}) \coloneqq \textbf{J}_p (\textbf{0}_{K \times K}, \textbf{J}_{K}, \bm{p}) = (\textbf{1}_{p_k \times p_{k'}})$, 
			$\operatorname{Bdiag}(\cdot)$ denotes a block-diagonal matrix, 
			and $\circ$ denotes the block Hadamard product satisfying that  
			$\textbf{A} \circ \textbf{I}(\bm{p}) \coloneqq \operatorname{Bdiag} \left(a_{11} \textbf{I}_{p_1}, \ldots, a_{KK} \textbf{I}_{p_K}\right)$ and $\textbf{B} \circ \textbf{J}(\bm{p}) \coloneqq \left(b_{kk'} \textbf{1}_{p_k \times p_{k'}}\right)$.
	\end{lemma*}
	
	Based on the lemma, we derive several basic properties of a uniform-block matrix, which are summarized in the \href{Supplementary Material.pdf}{Supplementary Material}.
	These properties reveal how $\textbf{A}$, $\textbf{B}$, and $\bm{p}$ determine the algebraic operations for a uniform-block matrix $\textbf{N}(\textbf{A}, \textbf{B}, \bm{p})$ and how a collection of uniform-block matrices with the same $\bm{p}$ forms a quadratic subspace \citep{Seely1971}.
	If we view $\textbf{A}$, $\textbf{B}$, and $\bm{p}$ as the coordinates of a uniform-block matrix, then using the notation $\textbf{N}(\textbf{A}, \textbf{B}, \bm{p})$ can simplify the mathematical operations between $p$ by $p$ uniform-block matrices into those between their corresponding lower-dimensional $K$ by $K$ coordinates. 
	Following these properties of uniform-block matrices, we obtain a useful result for the precision matrix.
	
	\begin{cor*}
		
		\label{Cor:pd_inverse}
		
		Suppose $\bm{\Sigma}(\textbf{A}, \textbf{B}, \bm{p}) = \textbf{A} \circ \textbf{I}(\bm{p}) + \textbf{B} \circ \textbf{J}(\bm{p})$ is positive definite and $\bm{\Omega} = \{\bm{\Sigma}(\textbf{A}, \textbf{B}, \bm{p})\} ^ {-1}$ is the precision matrix,
		then, $\bm{\Omega}$ partitioned by $\bm{p}$ is a uniform-block matrix with expression $\bm{\Omega}(\textbf{A}_{\bm{\Omega}}, \textbf{B}_{\bm{\Omega}}, \bm{p}) = \textbf{A}_{\bm{\Omega}} \circ \textbf{I}(\bm{p}) + \textbf{B}_{\bm{\Omega}} \circ \textbf{J}(\bm{p})$, where
		$\textbf{A}_{\bm{\Omega}} = \textbf{A} ^ {-1}$, $\textbf{B}_{\bm{\Omega}} = - \bm{\Delta} ^ {-1}  \textbf{B} \textbf{A} ^ {-1}$, $\bm{\Delta} \coloneqq \textbf{A} + \textbf{B} \textbf{P}$, and $\textbf{P} \coloneqq \operatorname{diag}(p_1, \ldots, p_K)$.
	\end{cor*}
	The above corollary finalizes that $\bm{\Sigma}(\textbf{A}, \textbf{B}, \bm{p})$ is positive definite if and only if $\textbf{A}$ is positive definite (i.e., $a_{kk} > 0$ for all $k$) and $\bm{\Delta}$ has only positive eigenvalues.
	It also confirms that the precision matrix $\bm{\Omega} = \{\bm{\Sigma}(\textbf{A}, \textbf{B}, \bm{p})\} ^ {-1}$, partitioned by $\bm{p}$, is a uniform-block matrix, expressed by $\bm{\Omega}(\textbf{A}_{\bm{\Omega}}, \textbf{B}_{\bm{\Omega}}, \bm{p})$. 
	Furthermore, it provides the relations between $\textbf{A}_{\bm{\Omega}}$, $\textbf{B}_{\bm{\Omega}}$, and $\textbf{A}$, $\textbf{B}$.

	Therefore, by applying the representation of the precision matrix in the corollary, we can rewrite the partial derivative of the log-likelihood in~\eqref{Eq:score_original} as
	\begin{equation}
		\begin{split}
			\label{Eq:score}
			\frac{\partial}{\partial \bm{\theta}_j} \ell_n (\bm{\theta}; \textbf{X})
			= \frac{n}{2} \operatorname{tr}\left[\{\bm{\Sigma}(\textbf{A}, \textbf{B}, \bm{p}) - (\textbf{S}_{kk'})\}
			\left\{ \frac{\partial \textbf{A}_{\bm{\Omega}}}{\partial \bm{\theta}_j} \circ \textbf{I}(\bm{p})
			+ \frac{\partial \textbf{B}_{\bm{\Omega}}}{\partial \bm{\theta}_j} \circ \textbf{J}(\bm{p}) \right\} \right],  
		\end{split}
	\end{equation}
	for $j = 1, \ldots, q$, where $\bm{\theta}_j \in \{a_{11}, \ldots, a_{KK}, b_{11}, \ldots, b_{1K}, b_{22}, \ldots, b_{KK}\}$.
	In contrast to~\eqref{Eq:score_original}, the derivatives in~\eqref{Eq:score} can be calculated using explicit expressions of $\textbf{A}$, $\textbf{B}$, and $\bm{p}$ (details are provided in the \href{Supplementary Material.pdf}{Supplementary Material}).
	The explicit forms of the derivatives highlight the advantage of~\eqref{Eq:score} over~\eqref{Eq:score_original}.
	With~\eqref{Eq:score}, we can explicitly derive the analytic form of the maximum likelihood estimator for $\bm{\theta}$: 
	\begin{equation}
		\label{Eq:mle_ls}
		\begin{split}
			\widetilde{a}_{kk} \coloneqq  \frac{\operatorname{tr}(\textbf{S}_{kk})}{p_k} - \widetilde{b}_{kk}, \quad
			%- \frac{\operatorname{sum}({\color{red} \textbf{S}_{kk}})}{p_k (p_k - 1)}, \quad
			\widetilde{b}_{kk'} \coloneqq \begin{cases}
				\dfrac{\operatorname{sum}(\textbf{S}_{kk'})}{p_k p_{k'}}, & k \neq k' \\
				\dfrac{\operatorname{sum}(\textbf{S}_{kk'}) - \operatorname{tr}( \textbf{S}_{kk'})}{p_k (p_{k'} - 1)}, & k = k'
			\end{cases}, 
		\end{split}
	\end{equation}
	for $k, k' = 1, \ldots, K$, where $\operatorname{sum}(\textbf{M}) = \sum_{j=1}^{r} \sum_{j'=1}^{s} m_{jj'}$ denotes the sum of all entries in $\textbf{M} \coloneqq (m_{jj'}) \in \mathbb{R} ^ {r \times s}$.
	Technical details are referred to the \href{Supplementary Material.pdf}{Supplementary Material}.
	From~\eqref{Eq:mle_ls}, the maximum likelihood estimators are identical to the moment estimators: the estimator $\widetilde{b}_{kk'}$ is the average of all elements of $\textbf{S}_{kk'}$ for $k \neq k'$; the estimator $\widetilde{b}_{kk}$ is the average of all off-diagonal elements of $\textbf{S}_{kk}$, and the estimator $\widetilde{a}_{kk} + \widetilde{b}_{kk}$ is the average of all diagonal elements of $\textbf{S}_{kk}$ for $k' = k$.
	
	We denote the maximum likelihood estimator of $\bm{\theta}$ as 
	\begin{equation*}
		\widetilde{\bm{\theta}} \coloneqq (\widetilde{a}_{11}, \ldots, \widetilde{a}_{KK}, \widetilde{b}_{11}, \ldots, \widetilde{b}_{1K}, \widetilde{b}_{22}, \ldots, \widetilde{b}_{KK}) ^ \top.
	\end{equation*}
	The strong consistency, asymptotic efficiency, and asymptotic normality for $\widetilde{\bm{\theta}}$ can similarly be derived using standard procedures in the literature \citep{Ferguson1996, vanderVaartWellner1996, vanderVaart2000}. 
	Please refer to the \href{Supplementary Material.pdf}{Supplementary Material} for more details.	
	Moreover, the following theorem summarizes the optimal properties of the plug-in covariance- and precision-matrix estimators.
	
	\begin{thm}[Optimal properties of the proposed estimators]
		
		\label{Thm:ls}
		(1) $\widetilde{\bm{\theta}}$ is the UMVUE of $\bm{\theta}$. 
		
		(2) The plug-in covariance-matrix estimator
		\begin{equation}
			\label{Eq:ls_cov}
			\widetilde{\bm{\Sigma}} (\widetilde{\textbf{A}}, \widetilde{\textbf{B}}, \bm{p})
			= \widetilde{\textbf{A}} \circ \textbf{I}(\bm{p}) + \widetilde{\textbf{B}} \circ \textbf{J}(\bm{p})
		\end{equation}
		is the UMVUE of $\bm{\Sigma}(\textbf{A}, \textbf{B}, \bm{p})$, where $\widetilde{\textbf{A}} \coloneqq \operatorname{diag} (\widetilde{a}_{11}, \ldots, \widetilde{a}_{KK})$ and $\widetilde{\textbf{B}} \coloneqq (\widetilde{b}_{kk'})$ with $\widetilde{b}_{kk'} = \widetilde{b}_{k'k}$ for every $k \neq k'$ are the UMVUEs of $\textbf{A}$ and $\textbf{B}$, respectively.
		
		(3) If $\widetilde{\textbf{A}}$ is positive definite and $\widetilde{\bm{\Delta}} \coloneqq \widetilde{\textbf{A}} + \widetilde{\textbf{B}} \textbf{P}$ has only positive eigenvalues, then the plug-in precision-matrix estimator 
		\begin{equation}
			\label{Eq:ls_precision}
			\begin{split}
				\widetilde{\bm{\Omega}} (\widetilde{\textbf{A}}_{\bm{\Omega}}, \widetilde{\textbf{B}}_{\bm{\Omega}}, \bm{p})
				= \widetilde{\textbf{A}}_{\bm{\Omega}} \circ \textbf{I}(\bm{p}) + \widetilde{\textbf{B}}_{\bm{\Omega}} \circ \textbf{J}(\bm{p})
			\end{split}
		\end{equation}
		is the UMVUE of $\bm{\Omega}(\textbf{A}_{\bm{\Omega}}, \textbf{B}_{\bm{\Omega}}, \bm{p})$ with $\widetilde{\textbf{A}}_{\bm{\Omega}} = \widetilde{\textbf{A}} ^ {- 1}$ and $\widetilde{\textbf{B}}_{\bm{\Omega}} = - \widetilde{\bm{\Delta}} ^ {- 1} \widetilde{\textbf{B}} \widetilde{\textbf{A}} ^ {- 1}$.
	\end{thm}
	\citet{Geisser1963} and \citet{Morrison1972} derived the same estimator $\widetilde{\bm{\theta}}$ based on an analysis of the variance table, but they did not show the above optimal properties.
	
	The asymptotic covariance matrix of the proposed estimator $\widetilde{\bm{\theta}}$ can be calculated from the Fisher information matrix. 
	For a small $K$ (e.g., $K \leq 3$, empirically), the Fisher information matrix and its inverse may have explicit expressions.
	However, for a real application with a relatively large $K$ (e.g., $K > 3$), calculating the inverse of the Fisher information matrix may be burdensome and unstable.
	Alternatively, we provide exact variance (and covariance) estimators for the elements of $\widetilde{\bm{\theta}}$ in the \href{Supplementary Material.pdf}{Supplementary Material} under finite sample sizes.

	\subsection{\textit{Matrix Estimation for the Uniform-Block Structure with a Large \texorpdfstring{$K$}{K}}}
	
	\label{Subsec:high}
	
	In Section~\ref{Subsec:low}, we estimated the covariance matrix with the uniform-block structure and its precision matrix for a small $K$ setting. 
	However, there are applications where the covariance matrices exhibit uniform-block structures with more diagonal blocks than the sample size. %, that is, where $K$ is large.
	More specifically, a large $K$ setting occurs when both $K > n$ and $q > n$, and all $K$, $p$, and $q$ grow with $n$.
	In other words, the number of diagonal blocks $K$ is greater than the sample size $n$, and so is the number of covariance parameters in the blocks: $q = K + K(K + 1) / 2$.
	Moreover, we require that $K$, the dimension $p$ of the covariance matrix (which is proportional to $K$), and $q$ grow with $n$ and diverge as $n$ goes to infinity.
	In this section, we generalize the proposed small-$K$ estimation procedure and introduce a consistent covariance-matrix estimator by modifying the hard-thresholding method for a large $K$ setting. 
	Denote $\Vert \textbf{M} \Vert_{\text{F}} = (\sum_{j = 1} ^ {r} \sum_{j’ = 1} ^ {r} m_{jj’} ^ 2 ) ^ {1/2}$ and $\Vert \textbf{M} \Vert_{\text{S}} = {\max}_{\Vert \textbf{x} \Vert_2 = 1} \Vert \textbf{M} \textbf{x} \Vert_2$ as the Frobenius norm and spectral norm of $\textbf{M} \coloneqq (m_{jj’}) \in \mathbb{R} ^ {r \times r}$ respectively, where $\Vert \textbf{x} \Vert_2 \coloneqq (\sum_{j = 1} ^ {r} x_j ^ 2) ^ {1/2}$ for $\textbf{x} \coloneqq (x_1, \ldots, x_r)^\top \in \mathbb{R} ^ {r}$.
	
	For normal data $\textbf{X}_{n \times p}$ with the population mean $\bm{\mu} = \textbf{0}_{p \times 1}$, the unbiased sample covariance matrix $\textbf{S} = n ^ {-1} \textbf{X} ^ \top \textbf{X}$, and the population covariance matrix with the uniform-block structure $\bm{\Sigma}(\textbf{A}, \textbf{B}, \bm{p})$ for a large $K$, we propose a new thresholding approach based on the work by \citet{BickelLevina2008b}.
	Since covariance matrix $\bm{\Sigma}(\textbf{A}, \textbf{B}, \bm{p})$ is entirely determined by $\textbf{A}$, $\textbf{B}$, and $\bm{p}$ according to the lemma, we threshold the estimates of $\textbf{A}$ and $\textbf{B}$, rather than $\textbf{S}$, to yield a covariance-matrix estimate. 
	Specifically, given a thresholding level $\lambda = \lambda_{n} > 0$, let
	\begin{equation*}
		%\label{Eq:hard}
		\widehat{a}_{kk}(\lambda) \coloneqq \widetilde{a}_{kk} \times \mathbb{I}(\vert \widetilde{a}_{kk} \vert > \lambda), \quad 
		\widehat{b}_{kk'}(\lambda) \coloneqq \widetilde{b}_{kk'} \times \mathbb{I}(\vert \widetilde{b}_{kk'} \vert > \lambda)
	\end{equation*}
	be the hard-thresholding estimators of $a_{kk}$ and $b_{kk'}$, respectively, where $\widetilde{a}_{kk}$ and $\widetilde{b}_{kk'}$ are the UMVUEs defined in~\eqref{Eq:mle_ls} for every $k$ and $k'$ and $\mathbb{I}(\cdot)$ is the indicator function.
	We regard the new covariance-matrix estimator
	\begin{equation}
		\label{Eq:ls_hard}
		\widehat{\bm{\Sigma}}_{\lambda} (\widehat{\textbf{A}}_{\lambda} , \widehat{\textbf{B}}_{\lambda} , \bm{p})
		= \widehat{\textbf{A}}_{\lambda}  \circ \textbf{I}(\bm{p}) + \widehat{\textbf{B}}_{\lambda} \circ \textbf{J}(\bm{p})
	\end{equation}
	as the \emph{modified hard-thresholding estimator} of $\bm{\Sigma}(\textbf{A}, \textbf{B}, \bm{p})$, where $\widehat{\textbf{A}}_{\lambda} \coloneqq \operatorname{diag}\{\widehat{a}_{11}(\lambda), \ldots, \widehat{a}_{KK}(\lambda)\}$, $\widehat{\textbf{B}}_{\lambda} \coloneqq \{\widehat{b}_{kk'}(\lambda)\}$ with $\widehat{b}_{kk'}(\lambda) = \widehat{b}_{k'k}(\lambda)$ for every $k \neq k'$.
	The consistency of this modified hard-thresholding estimator is summarized below.
	
	\begin{thm}[Consistency of the modified hard-thresholding estimator]
		
		\label{Thm:hard}
		
		Consider a positive definite $\bm{\Sigma}(\textbf{A}, \textbf{B}, \bm{p})$ as defined in~\eqref{Eq:nsigma}. 
		If we choose $\lambda = C \{\log(K) / n\} ^ {1 / 2}$ for some positive constant $C$, and assume some mild regularity conditions are satisfied, 
		then $\widehat{\bm{\Sigma}}_{\lambda} (\widehat{\textbf{A}}_{\lambda}, \widehat{\textbf{B}}_{\lambda}, \bm{p})$ as defined in~\eqref{Eq:ls_hard} is (weakly) consistent in both Frobenius and spectral norms as $K > n$, $\log(K) / n \to 0$, $K, n \to \infty$.
	\end{thm}
	
	The performance of the modified hard-thresholding estimator~\eqref{Eq:ls_hard} relies on the choice of $\lambda$, which can be determined by applying the resampling rule outlined in \citet{BickelLevina2008a, BickelLevina2008b}.
	
	As demonstrated in the subfigures of B, D, E, F, and G of Figure~\ref{Fig:data_instances}, the high-dimensional features comprise a set of interconnected communities and a set of singletons.   
	Therefore, we represent the covariance matrix $\bm{\Sigma}_{\star} \coloneqq \{\bm{\Sigma}(\textbf{A}, \textbf{B}, \bm{p}), \textbf{D}_1; \textbf{D}_1 ^ \top, \textbf{D}_2\}$, where $\textbf{D}_1 \in \mathbb{R} ^ {p \times d}$, $\textbf{D}_2 \in \mathbb{R} ^ {d \times d}$, and $\bm{\Sigma}_{\star} \in \mathbb{R} ^ {(p + d) \times (p + d)}$, with $d$ denoting the number of singletons. 
	We estimate $\bm{\Sigma}_{\star}$ in two steps. First, we compute and partition the sample covariance matrix $\textbf{S}_{\star} \coloneqq \{ (\textbf{S}_{kk'}), \textbf{S}_{1}; \textbf{S}_1 ^\top, \textbf{S}_2\}$ based on the covariance structure. 
	Next, we calculate $\widetilde{\bm{\Sigma}}(\widetilde{\textbf{A}}, \widetilde{\textbf{B}}, \bm{p})$ using the consistent estimators in~\eqref{Eq:mle_ls} regarding $(\textbf{S}_{kk'})$ and~\eqref{Eq:ls_cov}, while performing conventional soft- or hard-thresholding on the singleton-related sample covariance matrix $\{\textbf{0}_{p \times p}, \textbf{S}_1; \textbf{S}_1 ^ \top, \textbf{S}_2\}$ \citep{BickelLevina2008b}. 
	Jointly, these two steps provide a consistent estimator of $\bm{\Sigma}_{\star}$.

	\section{Numerical Studies}
	
	\label{Sec:numerical}
	
	\subsection{\textit{Simulations}}
	
	\label{Subsec:study0}
	
	To evaluate the performance of the proposed method comprehensively, we simulate data and benchmark them against comparable estimation methods for large covariance (and precision) matrices in the following three scenarios.
	
	In Scenario 1 (Section~\ref{Subsec:study1}), we generate normal data using a covariance matrix with a uniform-block structure $\bm{\Sigma}_{0, 1}(\textbf{A}_0, \textbf{B}_0, \bm{p}_1)$ for $n$ subjects with a small $K$.
	%That is, the number of diagonal blocks $K$ is small so that the number of covariance parameters in the blocks $q$ is smaller than the sample size $n$, whereas the dimension of the covariance matrix $p$ can be greater than $n$. 
	We first focus on evaluating the finite-sample performance of the parameter vector estimator $\widetilde{\bm{\theta}}$ in~\eqref{Eq:mle_ls} by comparing the estimates with the ground truth.
	Additionally, we assess the accuracy of the exact covariance estimator for $\widetilde{\bm{\theta}}$, as presented in the \href{Supplementary Material.pdf}{Supplementary Material}.
	Subsequently, we present the performance of the covariance estimator $\widetilde{\bm{\Sigma}}_1(\widetilde{\textbf{A}}_1, \widetilde{\textbf{B}}_1, \bm{p}_1)$ in~\eqref{Eq:ls_cov} and the precision estimator $\widetilde{\bm{\Omega}}_1(\widetilde{\textbf{A}}_{\bm{\Omega}, 1}, \widetilde{\textbf{B}}_{\bm{\Omega}, 1}, \bm{p}_1)$ in~\eqref{Eq:ls_precision} by comparing their losses in the matrix norms with those of existing covariance- and precision-matrix estimators.
	
	In Scenario 2 (Section~\ref{Subsec:study2}), we simulate normal data using the covariance matrix $\bm{\Sigma}_{0, 2}(\textbf{A}_0, \textbf{B}_0, \bm{p}_2)$ with the structure of uniform blocks for a large $K$. 
	%That is, the number of diagonal blocks $K$ is greater than and grows with the sample size $n$, as does the number of parameters in the blocks $q$ and the dimension of the covariance matrix $p$.
	Next, we compare the modified hard-thresholding estimator $\widehat{\bm{\Sigma}}_2(\widehat{\textbf{A}}_2, \widehat{\textbf{B}}_2, \bm{p}_2)$ in ~\eqref{Eq:ls_hard} with several competing methods by computing the losses in the matrix norms. 
	
	In Scenario 3 (Section~\ref{Subsec:study3}), we perform a misspecification analysis for $\widetilde{\bm{\Sigma}}_3(\widetilde{\textbf{A}}_3, \widetilde{\textbf{B}}_3, \bm{p}_3)$ and $\widetilde{\bm{\Omega}}_3(\widetilde{\textbf{A}}_{\bm{\Omega}, 3}, \widetilde{\textbf{B}}_{\bm{\Omega}, 3}, \bm{p}_3)$ when the covariance matrix does not adhere to the uniform-block structure.
	
	\subsection{Scenario 1: Comparison for Covariance Matrices with a Small \texorpdfstring{$K$}{K}}
	
	\label{Subsec:study1}
	
	We first set the true covariance uniform-block matrix as $\bm{\Sigma}_{0, 1}(\textbf{A}_0, \textbf{B}_0, \bm{p}_1) = \textbf{A}_0 \circ \textbf{I}(\bm{p}_1) + \textbf{B}_0 \circ \textbf{J}(\bm{p}_1)$, where the number of diagonal blocks $K = 5$; the partition-size vector
	\begin{equation*}
		\bm{p}_1 \coloneqq (p_{\text{ind}}, p_{\text{ind}}, p_{\text{ind}}, p_{\text{ind}}, p_{\text{ind}}) ^ \top 
		= p_{\text{ind}} \textbf{1}_{K \times 1}
	\end{equation*}
	with individual component $p_{\text{ind}} = 30$, $45$, or $60$; the number of covariance parameters in the blocks $q = K + K(K + 1) / 2 = 20$; the dimension of the covariance matrix $p = K p_{\text{ind}} = 150$, $225$, or $300$; and $\textbf{A}_{0} \coloneqq \operatorname{diag}(a_{0, 11}, \ldots, a_{0, KK})$ and $\textbf{B}_{0} \coloneqq (b_{0, kk'})$ with $b_{0, kk'} = b_{0, k'k}$ for $k \neq k'$:
	\begin{equation*}
		\begin{split}
			\textbf{A}_{0} = \operatorname{diag}(0.016, 0.214, 0.749, 0.068, 0.100), \quad
			\textbf{B}_{0} = 
			\begin{pmatrix}
				6.731 & -1.690 & 0.696 & -2.936 & \phantom{-}1.913 \\
				& \phantom{-}5.215 & 3.815 & -1.010 & \phantom{-}0.703 \\
				& & 4.328 & -3.357 & -0.269 \\
				& & & \phantom{-}6.788 & \phantom{-}0.000 \\
				& & & & \phantom{-}3.954
			\end{pmatrix}.
		\end{split}
	\end{equation*}
	The true precision matrix $\bm{\Omega}_{0, 1}(\textbf{A}_{\bm{\Omega}, 0}, \textbf{B}_{\bm{\Omega}, 0}, \bm{p}_1) = \textbf{A}_{\bm{\Omega}, 0} \circ \textbf{I}(\bm{p}_1) + \textbf{B}_{\bm{\Omega}, 0} \circ \textbf{J}(\bm{p}_1)$ is given by $\textbf{A}_{\bm{\Omega}, 0} \coloneqq \textbf{A}_0 ^ {-1}$ and $\textbf{B}_{\bm{\Omega}, 0} \coloneqq - (\textbf{A}_0 + \textbf{B}_0 \textbf{P}_{0, 1}) ^ {-1} \textbf{B}_0 \textbf{A}_0 ^ {-1}$, where $\textbf{P}_{0, 1} \coloneqq \operatorname{diag}(p_{\text{ind}}, p_{\text{ind}}, p_{\text{ind}}, p_{\text{ind}}, p_{\text{ind}})$.
	Specifically, we choose $n = 50$, $100$, or $150$, while ensuring that $q = 20 < n$. 
	We generate the data matrix $\textbf{X}$ by drawing an independent and identically distributed sample from $N(\textbf{0}_{p \times 1}, \bm{\Sigma}_{0, 1}(\textbf{A}_0, \textbf{B}_0, \bm{p}_1))$ and repeat this procedure $1000$ times.
	
\begin{figure}[!htb]
		\centering
		\includegraphics[width = 1.0 \linewidth]{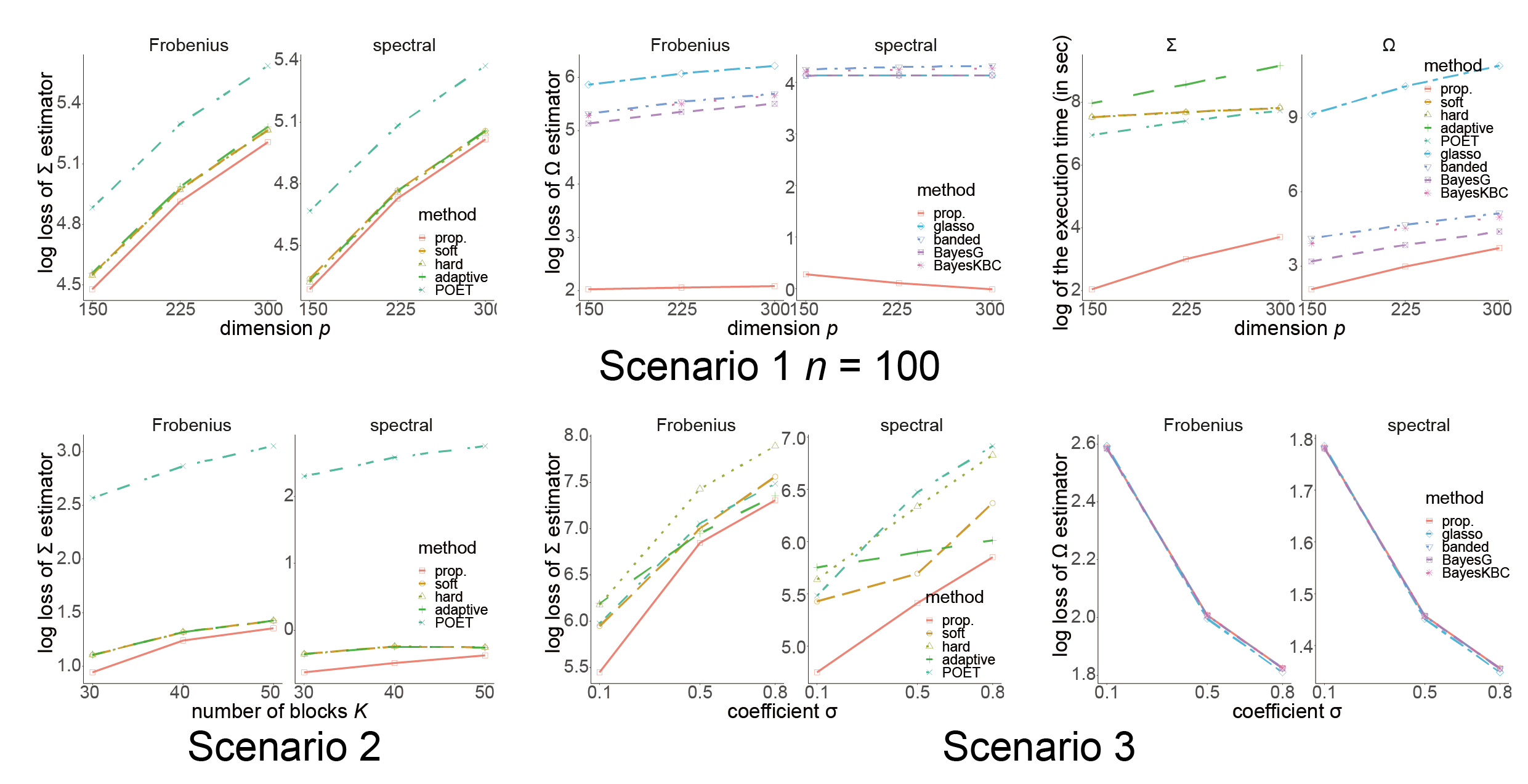} 
		\caption{
			We evaluate the estimated covariance and precision matrices using our approach and compare them with estimates from benchmark methods.
			In the first row, we present the results for Scenario 1 (the small $K$ setting) under $n = 100$ with various dimensions $p$ (the $x$-axis). 
			The left two and the middle two subfigures illustrate the logarithmic losses in both Frobenius and spectral norms for covariance matrix $\bm{\Sigma}$ and precision matrix $\bm{\Omega}$ estimations for all methods, respectively; 
			and the right two subfigures display the execution time for $\bm{\Sigma}$ and $\bm{\Omega}$ estimations. 
			In the second row, the left two subfigures demonstrate the logarithmic losses in both Frobenius and spectral norms for covariance matrix estimation in Scenario 2 (the large $K$ setting). 
			The middle two and the right two subfigures in the second row showcase the results under model misspecification (Scenario 3) by assessing the logarithmic losses in both Frobenius and spectral norms for covariance matrix $\bm{\Sigma}$ and precision matrix $\bm{\Omega}$ estimations.		
			In general, our method provides more accurate covariance- and precision-matrix estimates with reduced computational loads and demonstrates robustness to model specification.
		}
		\label{Fig:simulation_studies}
	\end{figure}

	For each replicate, we calculate $\widetilde{a}_{kk}$ and $\widetilde{b}_{kk'}$ using~\eqref{Eq:mle_ls} to obtain $\widetilde{\textbf{A}}_1 \coloneqq \operatorname{diag}(\widetilde{a}_{11}, \ldots, \widetilde{a}_{KK})$, $\widetilde{\textbf{B}}_1 \coloneqq (\widetilde{b}_{kk'})$, $\widetilde{\textbf{A}}_{\bm{\Omega}, 1} \coloneqq \widetilde{\textbf{A}}_1 ^ {-1}$, and $\widetilde{\textbf{B}}_{\bm{\Omega}, 1} \coloneqq - (\widetilde{\textbf{A}}_1 + \widetilde{\textbf{B}}_1 \textbf{P}_{0, 1}) ^ {- 1} \widetilde{\textbf{B}}_1 \widetilde{\textbf{A}}_1 ^ {- 1}$ and then calculate their standard errors by substituting the estimates $\widetilde{a}_{kk}$ and $\widetilde{b}_{kk'}$ for $a_{kk}$ and $b_{kk'}$, respectively, (please see the formulas in the \href{Supplementary Material.pdf}{Supplementary Material}).
	We also calculate the estimates of the covariance matrix and precision matrices using~\eqref{Eq:ls_cov} and~\eqref{Eq:ls_precision} and denote them by $\widetilde{\bm{\Sigma}}_{\text{prop.}} \coloneqq \widetilde{\bm{\Sigma}}_1(\widetilde{\textbf{A}}_1, \widetilde{\textbf{B}}_1, \bm{p}_1)$ and $\widetilde {\bm{\Omega}}_{\text{prop.}} \coloneqq \widetilde{\bm{\Omega}}_1(\widetilde{\textbf{A}}_{\bm{\Omega}, 1}, \widetilde{\textbf{B}}_{\bm{\Omega}, 1}, \bm{p}_1)$, respectively.
	In addition, we estimate the covariance matrix using conventional methods, which include the soft-thresholding method \citep[soft;][]{AntoniadisFan2001}, the hard-thresholding method (hard), the adaptive-thresholding method (adaptive), and the principal orthogonal complement thresholding method (POET), as shown in Figure~\ref{Fig:simulation_studies}.
	We also estimate the precision matrix using the graphical lasso method \citep[glasso;][]{FriedmanHastieTibshirani2008}, the bandwidth test \citep[banded;][]{AnGuoLiu2014}, Bayesian frameworks with G-Wishart prior \citep[BayesG;][]{BanerjeeGhosal2014} or with the $k$-banded Cholesky prior \citep[BayesKBC;][]{LeeLee2021}, as also illustrated in Figure~\ref{Fig:simulation_studies}.
	Finally, we evaluate the performance of all methods using the losses in the Frobenius and spectral norms respectively, denoted as $\Vert \widetilde{\bm{\Sigma}}_{*} - \bm{\Sigma}_{0, 1}(\textbf{A}_0, \textbf{B}_0, \bm{p}_1) \Vert_{\text{F}}$, $\Vert \widetilde{\bm{\Sigma}}_{*} - \bm{\Sigma}_{0, 1}(\textbf{A}_0, \textbf{B}_0, \bm{p}_1) \Vert_{\text{S}}$, $\Vert \widetilde{\bm{\Omega}}_{*} - \bm{\Omega}_{0, 1}(\textbf{A}_{\bm{\Omega}, 0}, \textbf{B}_{\bm{\Omega}, 0}, \bm{p}_1) \Vert_{\text{F}}$, and $\Vert \widetilde{\bm{\Omega}}_{*} - \bm{\Omega}_{0, 1}(\textbf{A}_{\bm{\Omega}, 0}, \textbf{B}_{\bm{\Omega}, 0}, \bm{p}_1) \Vert_{\text{S}}$ for the method $*$.
	
	For the $1000$ replicates, we assess the absolute relative (average) bias ($\%$), Monte Carlo standard deviation, average standard error, and empirical coverage probability based on $95\%$ Wald-type confidence intervals for each covariance parameter, as presented in Table~\ref{Tab:finite}.
	The results in Table~\ref{Tab:finite} demonstrate that the proposed estimators $\widetilde{a}_{kk}$ and $\widetilde{b}_{kk'}$ achieve satisfactory performance: absolute relative biases are generally small in contrast to Monte Carlo standard deviations; as the sample size $n$ increases, average standard errors decrease for all covariance parameters; average standard errors are approximately equal to Monte Carlo standard deviations with comparable corresponding $95\%$ empirical coverage probabilities. 
	We demonstrate the performance of covariance- and precision-matrix estimators in terms of the Frobenius and spectral norms, as well as computational times, for both the proposed procedure and competing methods in Figure~\ref{Fig:simulation_studies}.
	Our estimating procedure outperforms existing methods, as it is much faster and has smaller matrix norm losses. 
	Compared with conventional precision-matrix estimators, the proposed precision-matrix estimator has much fewer losses. This is probably because the true precision matrix contains many non-sparse blocks (e.g., the off-diagonal entries in $\textbf{B}_{\bm{\Omega}, 0}$ are very small but different from $0$). 
	As the dimension $p$ increases, the matrix norm losses between the proposed covariance estimate and $\bm{\Sigma}_{0, 1}(\textbf{A}_0, \textbf{B}_0, \bm{p}_1)$ increase, while those between the proposed precision estimate and $\bm{\Omega}_{0, 1}(\textbf{A}_{\bm{\Omega}, 0}, \textbf{B}_{\bm{\Omega}, 0}, \bm{p}_1)$ decrease slightly. 
	One possible explanation is that we fix $n$ and $q$, then $\widetilde{\bm{\Sigma}}_{\text{prop.}}$ is determined by $\widetilde{\textbf{A}}_1$, $\widetilde{\textbf{B}}_1$, and $\bm{p}_1$ while $\widetilde{\bm{\Omega}}_{\text{prop.}}$ is determined by $\widetilde{\textbf{A}}_1 ^ {-1}$, $\widetilde{\textbf{B}}_1$, $(\widetilde{\textbf{A}}_1 + \widetilde{\textbf{B}}_1 \textbf{P}_{0, 1}) ^ {-1}$, and $\bm{p}_1$.

	%%%%%%%%%%%%%%%%%
	%%%% Table 1 %%%%
	%%%%%%%%%%%%%%%%%
	\begin{table}[!htb]
		\scriptsize
		\begin{tabular}{crrrrrrrrrrrrrr}
			\hline
			& \multicolumn{4}{c}{$p = 150$} & \multicolumn{1}{c}{} & \multicolumn{4}{c}{$p = 225$} & \multicolumn{1}{c}{} & \multicolumn{4}{c}{$p = 300$} \\ \cline{2-5} \cline{7-10} \cline{12-15} 
			& \multicolumn{1}{c}{ARB} & \multicolumn{1}{c}{MCSD} & \multicolumn{1}{c}{ASE} & \multicolumn{1}{c}{$95\%$ CP} & \multicolumn{1}{c}{} & \multicolumn{1}{c}{ARB} & \multicolumn{1}{c}{MCSD} & \multicolumn{1}{c}{ASE} & \multicolumn{1}{c}{$95\%$ CP} & \multicolumn{1}{c}{} & \multicolumn{1}{c}{ARB} & \multicolumn{1}{c}{MCSD} & \multicolumn{1}{c}{ASE} & \multicolumn{1}{c}{$95\%$ CP} \\ \hline
			$a_{0,11}$ & 0.1 & 0.0  & 0.0  & 95.1 &  & 0.1 & 0.0  & 0.0  & 94.9 &  & 0.1 & 0.0 & 0.0 & 95.2 \\
			$a_{0,22}$ & 0.0 & 0.6  & 0.6  & 95.8 &  & 0.0 & 0.4  & 0.5  & 96.2 &  & 0.0 & 0.4 & 0.4 & 94.8 \\
			$a_{0,33}$ & 0.0 & 2.1  & 2.0  & 94.2 &  & 0.0 & 1.6  & 1.6  & 94.9 &  & 0.1 & 1.4 & 1.4 & 95.3 \\
			$a_{0,44}$ & 0.1 & 0.2  & 0.2  & 96.3 &  & 0.0 & 0.1  & 0.1  & 95.5 &  & 0.0 & 0.1 & 0.1 & 95.4 \\
			$a_{0,55}$ & 0.1 & 0.3  & 0.3  & 95.3 &  & 0.1 & 0.2  & 0.2  & 95.1 &  & 0.0 & 0.2 & 0.2 & 95.3 \\
			$b_{0,11}$ & 0.1 & 97.6 & 95.7 & 93.5 &  & 1.1 & 97.3 & 96.7 & 93.5 &  & 0.2 & 93.7 & 95.5 & 94.6 \\
			$b_{0,12}$ &-0.7 & 59.6 & 61.8 & 95.4 &  &-1.3 & 61.3 & 62.4 & 95.8 &  &-1.4 & 62.5 & 61.9 & 94.7 \\
			$b_{0,13}$ & 1.6 & 52.2 & 54.7 & 96.8 &  & 0.6 & 54.3 & 55.0 & 95.6 &  & 1.7 & 55.0 & 54.8 & 94.9 \\
			$b_{0,14}$ &-0.1 & 72.5 & 74.1 & 95.4 &  &-1.1 & 75.4 & 74.3 & 94.3 &  &-0.1 & 74.5 & 73.9 & 94.8 \\
			$b_{0,15}$ & 0.4 & 55.6 & 55.2 & 93.8 &  & 1.9 & 56.0 & 55.9 & 96.0 &  & 0.4 & 56.8 & 55.3 & 94.8 \\
			$b_{0,22}$ & 0.3 & 71.5 & 74.0 & 95.1 &  & 0.5 & 73.4 & 74.5 & 94.0 &  & 0.3 & 76.5 & 74.4 & 94.0 \\
			$b_{0,23}$ & 0.6 & 58.3 & 61.0 & 94.1 &  & 0.0 & 60.6 & 61.3 & 94.2 &  & 0.4 & 64.3 & 61.4 & 93.8 \\
			$b_{0,24}$ &-1.7 & 59.9 & 60.6 & 95.3 &  &-3.2 & 61.7 & 60.6 & 95.5 &  &-0.1 & 61.8 & 60.7 & 94.5 \\
			$b_{0,25}$ & 2.6 & 45.4 & 46.1 & 95.6 &  & 2.7 & 48.3 & 46.6 & 94.8 &  & 2.4 & 46.2 & 46.3 & 95.0 \\
			$b_{0,33}$ & 0.6 & 58.9 & 61.5 & 95.3 &  & 0.4 & 60.7 & 61.5 & 93.9 &  & 0.1 & 64.0 & 61.8 & 94.1 \\
			$b_{0,34}$ &-0.4 & 63.3 & 64.0 & 94.9 &  &-1.0 & 62.7 & 63.7 & 94.8 &  &-0.2 & 66.7 & 64.0 & 93.5 \\
			$b_{0,35}$ &-4.6 & 39.8 & 41.6 & 96.2 &  &-5.5 & 43.9 & 41.9 & 94.8 &  &-7.6 & 41.9 & 41.7 & 95.4 \\
			$b_{0,44}$ & 0.1 & 93.1 & 96.6 & 96.3 &  & 0.6 & 93.8 & 95.9 & 94.7 &  & 0.1 & 98.8 & 96.4 & 93.0 \\
			$b_{0,45}$ & NA  & 50.2 & 52.1 & 96.0 &  & NA  & 54.5 & 52.2 & 93.3 &  & NA  & 53.9 & 52.0 & 94.7 \\
			$b_{0,55}$ & 0.2 & 57.1 & 56.1 & 94.0 &  & 1.0 & 56.7 & 56.8 & 94.9 &  & 0.0 & 56.6 & 56.2 & 93.9 \\ \hline
		\end{tabular}
		\caption{
			We assess the estimated covariance parameters (of $a_{0, kk}$ and $b_{0,kk'}$) using our proposed method across $1000$ simulated datasets under $n = 100$:
			``ARB" represents the absolute relative bias (in $\%$), calculated as $\left \vert \widetilde{a}_{11} - a_{0, 11} \right \vert / a_{0, 11} \times 100\%$, where $\widetilde{a}_{11}$ denotes the averaged estimates over $1000$ datasets; 
			``MCSD" refers to the Monte Carlo standard deviation ($\times 100$); 
			``ASE" stands for the average standard error ($\times 100$) using the formulas presented in the \href{Supplementary Material.pdf}{Supplementary Material}; 
			``$95\%$ CP" represents the empirical coverage probability (in $\%$) based on a $95\%$ Wald-type confidence interval. 
			For the ARB of $b_{0, 45}$, denoted as ``NA" (not available) due to $b_{0, 45} = 0$, the absolute biases ($\times 100$) are $0.0$, $1.9$, and $1.2$ for $p = 150$, $225$, and $300$, respectively.
			In all settings, the estimation biases of the covariance parameters are small, generally less than $5\%$. 
			The standard errors, calculated using both MCSD and ASE, are close, indicating that our proposed variance estimators are accurate. 
			This is further supported by the results of the $95\%$ coverage probabilities.
			}
		\label{Tab:finite} 
	\end{table}
	%%%%%%%%%%%%%%%%%%%%%%%%
	%%%% End of Table 1 %%%%
	%%%%%%%%%%%%%%%%%%%%%%%%

	%%%%%%%%%%%%%%%%%%
	%%%% Figure 2 %%%%
	%%%%%%%%%%%%%%%%%%
	%%%%%%%%%%%%%%%%%%%%%
	%%%End of Figure 2 %%
	%%%%%%%%%%%%%%%%%%%%%
	
	\subsection{\textit{Scenario 2: Comparison for Covariance Matrices with a Large \texorpdfstring{$K$}{K}}}
	
	\label{Subsec:study2}
	
	The true covariance matrix is $\bm{\Sigma}_{0, 2}(\textbf{A}_{0, K}, \textbf{B}_{0, K}, \bm{p}_2) = \textbf{A}_{0, K} \circ \textbf{I}(\bm{p}_2) + \textbf{B}_{0, K} \circ \textbf{J}(\bm{p}_2)$, where $K = 30$, $40$, or $50$; $\bm{p}_2 \coloneqq (p_{\text{ind}}, \ldots, p_{\text{ind}}) ^ \top = p_{\text{ind}} \textbf{1}_{K \times 1}$ with the individual component $p_{\text{ind}} = 10$; $q = K + K(K + 1)/2 = 495$, $860$, or $1325$; $p = K p_{\text{ind}} = 300$, $400$, or $500$; and $\textbf{A}_{0, K}$ and $\textbf{B}_{0, K}$ are generated depending on the value of $K$.
	For each $K$, we first generate a $K$ by $K$ diagonal matrix $\textbf{A}_{0, K}$ and a symmetric matrix $\textbf{B}_{0, K}$ satisfying $\bm{\Sigma}_{0, 2} (\textbf{A}_{0, K}, \textbf{B}_{0, K}, \bm{p}_2)$ is positive definite, or equivalently, $\textbf{A}_{0, K}$ is positive definite and $\bm{\Delta}_{0, K} \coloneqq \textbf{A}_{0, K} + \textbf{B}_{0, K} \textbf{P}_{0, K}$ has only positive eigenvalues, where $\textbf{P}_{0, K} \coloneqq \operatorname{diag}(p_{\text{ind}}, \ldots, p_{\text{ind}}) \in \mathbb{R} ^ {K \times K}$.
	Then, we proceed to generate the data matrix $\textbf{X}$ based on an independently and identically distributed sample of size $n = 30$ from $N(\textbf{0}_{p \times 1}, \bm{\Sigma}_{0, 2}(\textbf{A}_{0, K}, \textbf{B}_{0, K}, \bm{p}_2))$, where all $K$, $p$, $q \geq n$.
	For each $K$, the above generation procedure is repeated $1000$ times.
	To estimate $\bm{\Sigma}_{0, 2} (\textbf{A}_{0, K}, \textbf{B}_{0, K}, \bm{p}_2)$, we adopt the modified hard-thresholding estimator in~\eqref{Eq:ls_hard}.
	The thresholding level $\lambda$ is chosen, following a similar procedure to that described in \citet{BickelLevina2008a, BickelLevina2008b}.
	We also compare our method with other large covariance matrix estimation techniques used in Scenario 1 (Section~\ref{Subsec:study1}). 
	For $1000$ replicates, we visualize the average losses in terms of matrix norms in Figure~\ref{Fig:simulation_studies}.
	The results in Figure~\ref{Fig:simulation_studies} reveal that the proposed modified estimator produces the smallest loss by taking advantage of the underlying structure.
	
	\subsection{\textit{Scenario 3: Simulation Analysis under Model Misspecification}}
	
	\label{Subsec:study3}
	
	In this scenario, we assess the performance of the proposed covariance- and precision-matrix estimators under model misspecification, where the true covariance matrix does not have a uniform-block structure.
	Specifically, we set the true positive definite covariance matrix $\bm{\Upsilon}_{0, \sigma} \coloneqq \bm{\Sigma}_{0, 3}(\textbf{A}_0, \textbf{B}_0, \bm{p}_3) + \textbf{M}_{\sigma}$, where $K = 5$; $\bm{p}_3 \coloneqq p_{\text{ind}} \textbf{1}_{K \times 1}$ with $p_{\text{ind}} = 30$; $q = 20$; $p = K  p_{\text{ind}} = 150$; and $\textbf{A}_0$ and $\textbf{B}_0$ are identical to those in Scenario 1 (Section~\ref{Subsec:study1}). 
	Matrix $\textbf{M}_{\sigma}$ follows a Wishart distribution with $p$ degrees of freedom and parameter $\sigma \textbf{I}_p$, where $\sigma = 0.1$, $0.5$, or $0.8$.   
	It is clear that if $\sigma = 0$, then $\bm{\Upsilon}_{0, \sigma} = \bm{\Sigma}_{0, 3} (\textbf{A}_0, \textbf{B}_0, \bm{p}_3)$ is a uniform-block matrix; if $\sigma > 0$, then $\bm{\Upsilon}_{0, \sigma}$ and the true precision matrix $\bm{\Upsilon}_{0, \sigma} ^ {-1}$ are not uniform-block matrices because uniformity does not hold.
	For each $\sigma$, we generate $\textbf{X}$ based on a random sample with a size of $n = 50$ (satisfying $q = 20 < n$) drawn from $N (\textbf{0}_{p \times 1}, \bm{\Upsilon}_{0, \sigma})$. 
	We fit the covariance-matrix estimator in~\eqref{Eq:ls_cov}, the precision-matrix estimator in~\eqref{Eq:ls_precision}, and the other estimators for large covariance and precision matrices described in Scenario 1 (Section~\ref{Subsec:study1}) to the data.
	We simulate $1000$ replicates for each $\sigma$.
	The average losses $\Vert \widetilde{\bm{\Sigma}}_{*} - \bm{\Upsilon}_{0, \sigma} \Vert_{\text{F}}$, $\Vert \widetilde{\bm{\Sigma}}_{*} - \bm{\Upsilon}_{0, \sigma} \Vert_{\text{S}}$, $\Vert \widetilde{\bm{\Omega}}_{*} - \bm{\Upsilon}_{0, \sigma} ^ {-1} \Vert_{\text{F}}$, and $\Vert \widetilde{\bm{\Omega}}_{*} - \bm{\Upsilon}_{0, \sigma} ^ {-1} \Vert_{\text{S}}$ are calculated among $1000$ replicates for the method $*$.
	The results are plotted in Figure~\ref{Fig:simulation_studies}.
	
	The results in Figure~\ref{Fig:simulation_studies} show that the proposed covariance-matrix estimator works well under the misspecified covariance structure.
	Compared with traditional covariance-matrix estimators, our method exhibits smaller losses in terms of matrix norms.
	The losses in terms of matrix norms for our precision-matrix estimator are comparable to those of the other methods.
	One possible reason is that the inverse of a non-uniform-block matrix is not a uniform-block matrix, so the proposed precision-matrix estimator cannot benefit from the underlying structure.
	
	In summary, our method can robustly and accurately estimate covariance matrices with the dependence structure of interconnected communities. 
	Since recent multiple testing correction methods, e.g., to control the false discovery proportion \citep[FDP;][]{FanHanGu2012, FanHan2017}, are based on the covariance-matrix estimate, we also evaluated the influence of covariance estimation on the accuracy of feature selection (please see the details in the additional simulation studies in the \href{Supplementary Material.pdf}{Supplementary Material}). 
	Our simulation results showed that our approach can largely improve the sensitivity while preserving the FDP compared to competing methods.

	\section{Data Examples}
	
	\label{Sec:real}
	
	\subsection{\textit{Proteomics Data Analysis}}
	
	\label{Subsec:real_1}
	
	We applied the proposed method to estimate the covariance matrix for high-throughput proteomics data used in cancer research \citep{YildizShyrRahman2007}.
	Specifically, this case--control study involved $288$ participants ($180$ male and $108$ female, aged $62.4 \pm 9.4$ years).
	Matrix-assisted laser desorption ionization mass spectrometry was employed to identify abundant peptides in human serum samples between case and control groups.
	
	Following preprocessing \citep{ChenLiHong2009} in the preliminary study, we selected $184$ features in the serum as candidate proteins and peptides.
	Out of these, a total of $107$ were identified as constituting $7$ interconnected communities, while the remaining $77$ were considered as singletons according to the network detection algorithm \citep{ChenKangXing2018}, as illustrated in the first two subfigures of A in Figure~\ref{Fig:real_data}.
	
	Our primary objective was to explore the interactive relationships within the $107$ features that constituted the interconnected communities through the $107$ by $107$ correlation matrix, denoted by $\bm{\Sigma}_0(\textbf{A}_0, \textbf{B}_0, \bm{p})$, as well as those within all $184$ features through the $184$ by $184$ correlation matrix, denoted as $\bm{\Sigma}_{\star} = \left(\bm{\Sigma}_0(\textbf{A}_0, \textbf{B}_0, \bm{p}), \textbf{D}_1; \textbf{D}_1 ^ \top, \textbf{D}_2\right)$. 
	We estimated both correlation matrices $\bm{\Sigma}_0(\textbf{A}_0, \textbf{B}_0, \bm{p})$ and $\bm{\Sigma}_{\star}$. 
	
	Firstly, we employed the proposed estimation procedure to calculate the $107$ by $107$ covariance matrix $\bm{\Sigma}_0\left(\textbf{A}_0, \textbf{B}_0, \bm{p}\right)$, where $n = 288$, $p = 107$, $K = 7$, $q = 35$, and $\bm{p} = (34, 18, 14, 14, 13, 10, 4) ^ \top$ as provided by the network detection algorithm \citep{ChenKangXing2018}.
	Let $\textbf{A}_{0} = \operatorname{diag}(a_{0, 11}, \ldots, a_{0, KK})$ and $\textbf{B}_{0} = (b_{0, kk'})$ with $b_{0, kk'} = b_{0, k'k}$ for $k' \neq k$ denote the $K$ by $K$ unknown diagonal matrix and symmetric matrix, respectively.
	Consequently, all diagonal entries were equal to $1$ (i.e., $a_{0, kk} + b_{0, kk} = 1$ for every $k$).
	As $q = 35 < n = 288$, we could obtain estimates and standard errors for $a_{0, kk}$ and $b_{0, kk'}$ using~\eqref{Eq:mle_ls} and their respective variance estimators.
	We summarized the results in the third and fourth subfigures of A in Figure~\ref{Fig:real_data}, noting that the sum of the estimates of $a_{0, kk}$ and $b_{0, kk}$ was $1$ for every $k$ because the diagonal entries in the sample correlation matrix were $1$.
	Moreover, the fourth subfigure of A in Figure~\ref{Fig:real_data} showed that the $95\%$ confidence intervals of the correlations between the $(1, 2)$, $(2, 4)$, and $(6, 7)$ blocks contained $0$.
	Additionally, the $(1, 3)$, $(1, 4)$, $(1, 5)$, $(1, 6)$, $(2, 3)$, $(2, 5)$, $(3, 4)$, $(3, 5)$, $(3, 6)$, $(3, 7)$, and $(5, 7)$ blocks exhibited negative correlations, while the remaining blocks displayed positive correlations.
	Lastly, we estimated the correlation matrix $\bm{\Sigma}_{\star}$ for all $184$ features by applying the soft-thresholding approach to the singleton-related sample covariance matrix and then merging it with the estimate of $\bm{\Sigma}_0\left(\textbf{A}_0, \textbf{B}_0, \bm{p}\right)$, as illustrated in the last subfigure of A in Figure~\ref{Fig:real_data}.

	\subsection{\textit{Brain Imaging Data Analysis}}
	
	\label{Subsec:real_2}
	
	The second example was from a brain imaging study based on echo-planar spectroscopic imaging, allowing simultaneous measurements of multiple neurometabolites across whole brain regions \citep{ChiappelliRowlandWijtenburg2019}.
	Data were collected from $78$ participants ($39$ male and $39$ female, aged $42.1 \pm 18.8$ years). 
	This study involved the participants' measurements of $5$ neurometabolites, which included choline, myo-inositol, creatine-containing compounds, $N$-acetylaspartate, and glutamate--glutamine, across $89$ brain regions for each participant.
	
\begin{figure}[!htb]
		\centering
		\includegraphics[width = 0.8\linewidth]{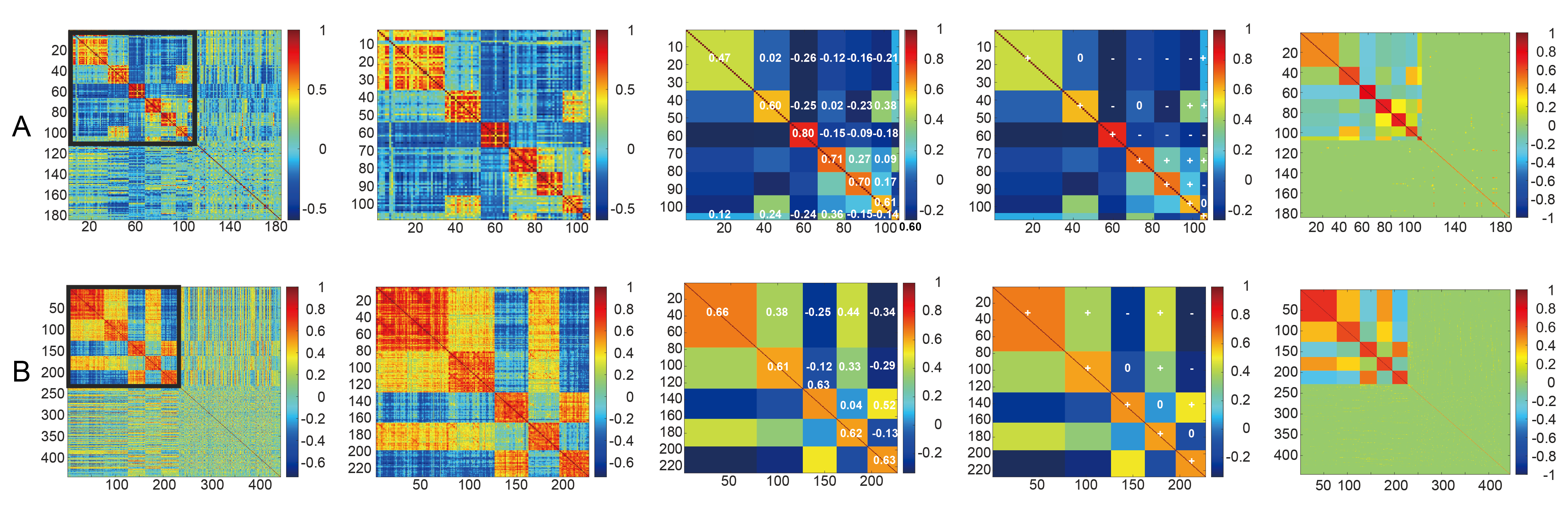} 
		\caption{
			We present the results for proteomics data analysis in the first row (A) and for brain imaging data analysis in the second row (B).
			A: 
			the first subfigure displays the heatmap of the sample correlation matrix for the proteomics dataset and the second one displays that of $107$ features within interconnected communities;
			the third subfigure exhibits $\widetilde{b}_{0, kk'}$ (with their standard errors ranging between $0.03$ and $0.07$, while for each $k$, $\widetilde{a}_{0, kk} = 1 - \widetilde{b}_{0, kk}$ with the standard error around $0.01$);
			the fourth subfigure provides confidence intervals for $b_{0, kk'}$;
			and the last subfigure exhibits the heatmap of the estimated correlation matrix for all $184$ features.
			B: the first subfigure exhibits the heatmap of the sample correlation matrix for the brain imaging dataset and the second one displays that of $227$ combinations within interconnected communities; 
			the third subfigure displays $\widetilde{b}_{0, kk'}$ (with their standard errors ranging between $0.07$ and $0.11$, while for each $k$, $\widetilde{a}_{0, kk} = 1 - \widetilde{b}_{0, kk}$ with the standard error around $0.01$);
			the fourth subfigure contains confidence intervals for $b_{0, kk'}$, where ``$+$" indicates the $95\%$ confidence interval to the right of $0$,
			``$-$" indicates the $95\%$ confidence interval to the left of $0$, 
			``$0$" indicates the $95\%$ confidence interval containing $0$;
			and the last subfigure exhibits the heatmap of the estimated correlation matrix for all $445$ combinations. 
		}
		\label{Fig:real_data}
	\end{figure}

	In the preliminary study, we calculated the sample correlation matrix for $445$ combinations of neurometabolites and brain regions (i.e., $445 = 5 \times 89$). 
	Employing the approach developed by \citet{ChenKangXing2018}, we extracted an interconnected community structure consisting of $227$ combinations from the initial $445$, while the remaining $218$ combinations were considered as singletons, as depicted in the first subfigure of B in Figure~\ref{Fig:real_data}.
	The interconnected community structure was characterized by $5$ diagonal blocks and $10$ off-diagonal blocks, as illustrated in the second subfigure of B in Figure~\ref{Fig:real_data}.

	We applied the proposed method to estimate two correlation matrices separately: the $227 \times 227$ block-form correlation matrix, denoted as $\bm{\Sigma}_{0}\left(\textbf{A}_0, \textbf{B}_0, \bm{p} \right)$, and the $445 \times 445$ correlation matrix, denoted as $\bm{\Sigma}_{\star} = \left(\bm{\Sigma}_{0}\left(\textbf{A}_0, \textbf{B}_0, \bm{p} \right), \textbf{D}_1; \textbf{D}_1 ^ \top, \textbf{D}_2 \right)$. 
	
	To estimate $\bm{\Sigma}_{0}\left(\textbf{A}_0, \textbf{B}_0, \bm{p} \right)$ in this application, we obtained $n = 78$ and $\bm{p} = (77, 49, 36, 33, 32) ^ \top$ from the preliminary network detection algorithm \citep{ChenKangXing2018}.
	Accordingly, we had $K = 5$, $p = 227$, and $q = 20$.
	We estimated $a_{0, kk}$ and $b_{0, kk'}$ using~\eqref{Eq:mle_ls} and their respective standard errors.
	The estimated correlations and their corresponding confidence intervals have been presented in the third and fourth subfigures of B in Figure~\ref{Fig:real_data}.
	Specifically, combinations within all diagonal blocks and those between the $(1,2)$, $(1, 4)$, $(2, 4)$, and $(3, 5)$ blocks were positively correlated, with $95\%$ confidence intervals excluding $0$, while combinations between the $(1, 3)$, $(1, 5)$, and $(2, 5)$ blocks were negatively correlated, with $95\%$ confidence intervals excluding $0$.
	Conversely, combinations within the $(2, 3)$, $(3, 4)$, and $(4, 5)$ blocks had $95\%$ confidence intervals for their correlations containing $0$.
	Furthermore, the $5$ diagonal blocks represented: (1) $74$ regions associated with choline (including three regions with other metabolites), (2) $49$ regions linked to myo-inositol, (3) $26$ regions associated with $N$-acetylaspartate (including $10$ regions with glutamate--glutamine), (4) $33$ regions related to creatine-containing compounds, and (5) an additional $32$ regions associated with $N$-acetylaspartate. 
	The off-diagonal blocks indicated positive relationships among choline, myo-inositol, and creatine-containing compounds, as well as between $N$-acetylaspartate and glutamate--glutamine across the brain. 
	Additionally, a global negative correlation existed between two sets of metabolites across the brain: (1) choline, myo-inositol, and creatine-containing compounds and (2) glutamate--glutamine and $N$-acetylaspartate. 
	The region-level correlations might assist in providing an understanding of the neurophysiological mechanisms relating to metabolites in the central nervous system.
	We employed a similar approach to estimate $\bm{\Sigma}_{\star}$ by combining the estimate of $\bm{\Sigma}_{0}\left(\textbf{A}_0, \textbf{B}_0, \bm{p} \right)$ with the soft-thresholding estimate of the singleton-related covariance matrix, as depicted in the last subfigure of B in Figure~\ref{Fig:real_data}.

	For comparison, we also applied existing methods for large covariance matrix estimation (e.g., the thresholding approach) to both proteomics and brain imaging datasets. 
	As outlined in an additional figure in the \href{Supplementary Material.pdf}{Supplementary Material}, these estimated covariance matrices, obtained using existing methods, appeared to miss a large proportion of correlations both within and between communities.

	%%%%%%%%%%%%%%%%%%
	%%%% Figure 3 %%%%
	%%%%%%%%%%%%%%%%%%
	%%%%%%%%%%%%%%%%%%%%%
	%%%End of Figure 3 %%
	%%%%%%%%%%%%%%%%%%%%%
	
	\section{Discussion}
	
	\label{Sec:discussion}
	
	We have developed a computationally efficient method for estimating large covariance and precision matrices with interconnected community structures. 
	In our empirical analyses of multiple types of high-throughput biomedical data, including gene expression, proteomics, neuroimaging, exposome, and many others, we have observed that most of these datasets exhibit a latent yet well-organized block pattern, as demonstrated in the examples in Figure~\ref{Fig:data_instances}. 
	By leveraging the interconnected community structure, we provided an accurate estimate of the parameter vector for the large covariance matrix with a drastically reduced number of parameters.
	We further derived the covariance- and precision-matrix estimators in closed forms, significantly reducing the computational burden and improving the accuracy of statistical inference.
	On the other hand, the proposed estimation procedure relies on the output of the chosen community detection or clustering algorithm. 
	Therefore, we aim to integrate community detection and covariance estimation into a unified framework to achieve more robust and reliable results in future studies.

	In a uniform-block structure, we assigned one parameter for the diagonal entries and one parameter for the off-diagonal entries in a diagonal block, and one parameter for all entries in an off-diagonal block. 
	This parameterization strategy is driven by the fact that the intra-block and the inter-block variances in the real applications are relatively small in the large sample correlation matrices (for additional results, refer to the \href{Supplementary Material.pdf}{Supplementary Material}). 
	Given the strong block patterns in the correlation matrices, as illustrated in the examples in Figure~\ref{Fig:data_instances}, this parameterization strategy appears valid. 
	Moreover, this strategy is analogous to the commonly used compound symmetry covariance structure in linear mixed-effect models and generalized estimating equation models. 
	Although our method is developed for the interconnected community structure, it is also applicable to other covariance structures (e.g., some hierarchical community structures in the \href{Supplementary Material.pdf}{Supplementary Material}).
	Additionally, we demonstrated that our method generally performs well when there are multiple parameters within each block, as in Scenario 3 (Section~\ref{Subsec:study3}). 
	Therefore, our approach is robust and computationally efficient for handling high-dimensional biomedical data, taking into account the frequently observed block-wise covariance structures.
	
	\bigskip
	\begin{center}
		{\large\bf Supplementary Material}
	\end{center}
	
	The supplementary material contains additional properties of uniform-block matrices, additional numerical results for both Numerical Studies and Data Examples sections, additional simulation studies, and technical proofs.
	R code is also provided.

	\bigskip
	\begin{center}
		{\large\bf Acknowledgments}
	\end{center}
	
	We appreciate the time and efforts that the editor, the associate editor, and the reviewers have dedicated to providing their valuable feedback on our manuscript.

	\bigskip
	\begin{center}
		{\large\bf Conflict of Interest Statement}
	\end{center}
	
	No potential competing interest was reported by the authors.

	\bigskip
	\begin{center}
		{\large\bf Funding}
	\end{center}
	
	This work was partially supported by the National Institute on Drug Abuse of the National Institutes of Health under Award Number 1DP1DA048968-01.
	
	\bibliographystyle{biom}	
	
	{\footnotesize \bibliography{paper_ref_UBCovEst}}

\end{document}

\maketitle
	
	\begin{abstract}
		The present supplementary material contains four sections. 
		Section~\ref{Sec:cor} includes the additional properties of the uniform-block matrices.
		Section~\ref{Sec:numerical} has the additional numerical results for both Numerical Studies and Data Examples sections. 
		Section~\ref{Sec:simulation} presents the additional simulation studies.
		Section~\ref{Sec:proof} provides the technical proofs.
		In addition to this document, the online material folder contains some additional simulation results presented in Excel files and the \proglang{R} code used for simulation studies in Section 3 and real data analyses in Section 4. 
		The code is also accessible at \url{https://github.com/yiorfun/UBCovEst}. 
	\end{abstract}
	
	\appendix
	
	\section{COROLLARIES}
	
	\label{Sec:cor}
	
	\begin{cors}[Algebraic properties of uniform-block matrices]
		
		\label{Cor:ops}
		
		Let $\textbf{N}(\textbf{A}, \textbf{B}, \bm{p})$, $\textbf{N}_1(\textbf{A}_1, \textbf{B}_1, \bm{p})$ and $\textbf{N}_2(\textbf{A}_2, \textbf{B}_2, \bm{p})$ be uniform-block matrices, where $\textbf{A} = \operatorname{diag}(a_{11}, \ldots, a_{KK})$, $\textbf{A}_1$, $\textbf{A}_2$ are diagonal matrices, $\textbf{B} = (b_{kk'})$ with $b_{kk'} = b_{k'k}$ for $k \neq k'$, $\textbf{B}_1$, $\textbf{B}_2$ are symmetric matrices. 
		Furthermore, $\bm{p} = (p_1, \ldots, p_K) ^ \top$ is the common pre-determined partition-size vector satisfying $p_k > 1$ for every $k$ and $p = p_1 + \cdots + p_K$.
		Let $\textbf{P} \coloneqq \operatorname{diag}(p_1, \ldots, p_K)$ and define $\bm{\Delta} \coloneqq \textbf{A} + \textbf{B} \textbf{P}$.  
		
		(1) Addition/Subtraction: suppose $\textbf{N} ^ * \coloneqq \textbf{N}_1 \pm \textbf{N}_2$, then $\textbf{N} ^ *$ partitioned by $\bm{p}$ is a uniform-block matrix, denoted by $\textbf{N} ^ *(\textbf{A} ^ *, \textbf{B} ^ *, \bm{p})$ with $\textbf{A} ^ * \coloneqq \textbf{A}_1 \pm \textbf{A}_2$ and $\textbf{B} ^ * \coloneqq \textbf{B}_1 \pm \textbf{B}_2$.
		
		(2) Square: suppose $\textbf{N} ^ * \coloneqq \textbf{N} ^ 2$, then $\textbf{N} ^ *$ partitioned by $\bm{p}$ is a uniform-block matrix, denoted by $\textbf{N} ^ *(\textbf{A} ^ *, \textbf{B} ^ *, \bm{p})$ with $\textbf{A} ^ * \coloneqq \textbf{A} ^ 2$ and $\textbf{B} ^ * \coloneqq \textbf{A} \textbf{B} + \textbf{B}\textbf{A} + \textbf{B} \textbf{P} \textbf{B}$.
		
		(3) Eigenvalues: $\textbf{N}(\textbf{A}, \textbf{B}, \bm{p})$ has a total of $p$ eigenvalues, 
		comprising $a_{kk}$ with multiplicity $(p_k - 1)$ for $k = 1, \ldots, K$ and the remaining $K$ eigenvalues are identical to those of $\bm{\Delta}$.
		
		(4) Positive Definiteness: $\textbf{N} (\textbf{A}, \textbf{B}, \bm{p})$ is positive definite if and only if $\textbf{A}$ is positive definite, and $\bm{\Delta}$ has positive eigenvalues only.
 		
		(5) Inverse: suppose $\textbf{N}$ is invertible, and $\textbf{N} ^ * \coloneqq \textbf{N} ^ {-1}$, then $\textbf{N} ^ *$ partitioned by $\bm{p}$ is a uniform-block matrix, denoted by $\textbf{N} ^ *(\textbf{A} ^ *, \textbf{B} ^ *, \bm{p})$ with $\textbf{A}^* = \textbf{A} ^ {- 1}$ and $\textbf{B}^* = - \bm{\Delta} ^ {- 1} \textbf{B} \textbf{A} ^ {- 1}$.
	\end{cors}
	
	\begin{cors}[Exact variance estimators]
		
		\label{Cor:vars}
		
		The exact variance estimators of $\widetilde{a}_{kk}$ and $\widetilde{b}_{kk'}$ are 
		\begin{equation*}
			\begin{split}
				\operatorname{var} (\widetilde{a}_{kk}) & = \frac{2 a_{kk} ^ 2}{(n - 1)(p_k - 1)},  \\
				\operatorname{var} (\widetilde{b}_{kk'})
				&  = 
				\begin{cases}
					\dfrac{2}{(n - 1)p_k(p_k - 1)} \left\{
					(a_{kk} + p_{k} b_{kk}) ^ 2 - (2 a_{kk} + p_k b_{kk}) b_{kk}
					\right\}, & k = k' \\
					\dfrac{1}{2 (n - 1) p_k p_{k'}} \left\{ p_k p_{k'} (b_{kk'} ^ 2 + b_{k'k} ^ 2) + 2 (a_{kk} + p_k b_{kk})(a_{k'k'} + p_{k'} b_{k'k'}) \right\}, & k \neq k'
				\end{cases}
			\end{split}
		\end{equation*}
		for every $k$ and $k'$.
			
		In a more general context, consider a partition-size vector $\bm{p} = (p_1, \ldots, p_K) ^ \top$ and a uniform-block matrix $\bm{\Sigma} = (\bm{\Sigma}_{kk'})$ partitioned by $\bm{p}$. 
		Let $\alpha_{kk} = \operatorname{tr}(\bm{\Sigma}_{kk})/p_k = a_{kk} + b_{kk}$ denote the average of the diagonal entries in $\bm{\Sigma}_{kk}$, 
		and let $\beta_{kk} = \operatorname{sum}(\bm{\Sigma}_{kk}) / (p_k ^ 2) = a_{kk}/p_k + b_{kk}$ denote the average of all entries in $\bm{\Sigma}_{kk}$ for every $k$. 
		Similarly, let $\beta_{kk'} = \operatorname{sum}(\bm{\Sigma}_{kk'}) / (p_k p_{k'}) = b_{kk'}$ denote the average of all entries in $\bm{\Sigma}_{kk'}$ for every $k \neq k'$.
		Furthermore, let $\widetilde{\alpha}_{kk} = \operatorname{tr}(\textbf{S}_{kk})/p_k$,
		$\widetilde{\beta}_{kk} = \operatorname{sum}(\textbf{S}_{kk}) / p_k ^ 2$ for every $k$, 
		and $\widetilde{\beta}_{kk'} = \operatorname{sum}(\textbf{S}_{kk'}) / (p_k p_{k'})$ for every $k \neq k'$.
		Then,	
			
		(1) $\widetilde{\alpha}_{kk}$, $\widetilde{\beta}_{kk'}$ are the UMVUEs of $\alpha_{kk}$, $\beta_{kk'}$ for every $k$ and $k'$; 
		the variance estimators of $\widetilde{\alpha}_{kk}$ and $\widetilde{\beta}_{kk'}$ are 
		\begin{equation*}
				\begin{split}
					\operatorname{var} \left(\widetilde{\alpha}_{kk} \right) & = \dfrac{2}{(n - 1) p_k} \left(a_{kk} ^ 2 + 2 a_{kk} b_{kk} + p_k b_{kk} ^ 2 \right),  \\
					\operatorname{var} \left(\widetilde{\beta}_{kk'} \right) & = 
					\begin{cases}
						\dfrac{2}{(n - 1) p_k ^ 2} \left(a_{kk} + p_{k} b_{kk}\right) ^ 2, & k = k' \\
						\dfrac{1}{2 (n - 1) p_k p_{k'}} \left\{ p_k p_{k'} \left(b_{kk'} ^ 2 + b_{k'k} ^ 2\right) + 2 \left(a_{kk} + p_k b_{kk}\right)\left(a_{k'k'} + p_{k'} b_{k'k'}\right) \right\}, & k \neq k'
					\end{cases},
				\end{split}
		\end{equation*}
		respectively, for every $k$ and $k'$;
		the covariance estimators between $\widetilde{\alpha}_{kk}$, $\widetilde{\alpha}_{k'k'}$, $\widetilde{\beta}_{kk}$, and $\widetilde{\beta}_{k'k'}$ are 
		\begin{equation*}
				%\scriptsize
				\begin{split}
					\operatorname{cov} \left(\widetilde{\alpha}_{kk}, \widetilde{\alpha}_{k'k'} \right) & = \frac{2}{n - 1} b_{kk'} b_{k'k} , \ \ \ \ 
					%%%%
					\operatorname{cov} \left(\widetilde{\beta}_{kk}, \widetilde{\beta}_{k'k'} \right) = \frac{2}{n - 1} b_{kk'} b_{k'k}, \ \ \ \ k \neq k' \\
					%%%%
					%%
					\operatorname{cov} \left(\widetilde{\alpha}_{kk}, \widetilde{\beta}_{k'k'} \right) & = 
					\begin{cases}
						\dfrac{2}{(n - 1) p_k ^ 2} \left(a_{kk} + p_k b_{kk}\right) ^ 2, & k = k' \\
						\dfrac{2}{n - 1} b_{kk'} b_{k'k}, & k \neq k'
					\end{cases},
					%\operatorname{cov}\left(\widehat{\beta}_{kk}, \widehat{\alpha}_{k'k'} \right) & = \frac{2}{n - 1} b_{kk'} b_{k'k}, \ \ \ \ k \neq k' \\
					%%
				\end{split}
		\end{equation*}
		respectively, for every $k$ and $k'$;
		the covariance estimators between $\widetilde{\beta}_{kk'}$ and the other estimators are 
		\begin{equation*}
				%\scriptsize
				\begin{split}
					\operatorname{cov} \left(\widetilde{\alpha}_{kk}, \widetilde{\beta}_{k'k''} \right)
					& = \frac{1}{(n - 1) p_{k'} p_{k''}} 
					\begin{cases}
						p_{k'} p_{k''} \left(b_{k'k}b_{kk''} + b_{k''k} b_{kk'}\right), & k \neq k', k \neq k'' \\
						p_{k''}\left(b_{k' k''} + b_{k'' k'}\right)\left(a_{kk} + p_k b_{kk}\right) , & k = k' \\
						p_{k'}\left(b_{k' k''} + b_{k'' k'}\right)\left(a_{kk} + p_k b_{kk}\right), & k = k'' 
					\end{cases}, \\
					%%%% 
					\operatorname{cov} \left(\widetilde{\beta}_{kk}, \widetilde{\beta}_{k'k''} \right)
					& = \frac{1}{(n - 1) p_{k}}
					\begin{cases}
						p_k \left(b_{k' k} b_{k k''} + b_{k'' k} b_{k k'}\right), & k \neq k', k \neq k'' \\
						\left(a_{kk} + p_k b_{kk} \right) \left(b_{k' k''} + b_{k'' k'} \right), & k = k' \\
						\left(a_{kk} + p_k b_{kk} \right) \left(b_{k' k''} + b_{k'' k'} \right), & k = k'' 
					\end{cases},
				\end{split}
		\end{equation*}
		respectively, for every $k$, $k' \neq k''$; and
		\begin{equation*}
				%\scriptsize
				\begin{split}
					& \operatorname{cov} \left( \widetilde{\beta}_{k_1 k_2}, \widetilde{\beta}_{l_1, l_2} \right) = \frac{1}{2 (n - 1)} \\
					& \begin{cases}
						b_{l_1 k_1} b_{k_2 l_2} + b_{l_2 k_1} b_{k_2 l_1} + b_{l_1 k_2} b_{k_1 l_2} + b_{l_2 k_2} b_{k_1 l_1}, 
						\\ & 
						\text{(1-1) } k_1 \neq l_1, l_2; k_2 \neq l_1, l_2 \\
						%%%
						\frac{1}{p_{l_1}}
						\left(a_{k_2 k_2} b_{l_2 k_1} + a_{k_2 k_2} b_{k_1 l_2} \right)
						+ 
						b_{l_1 k_1} b_{k_2 l_2} + b_{l_2 k_1} b_{k_2 l_1} + b_{l_1 k_2} b_{k_1 l_2} + b_{l_2 k_2} b_{k_1 l_1}, 
						\\ & 
						\text{(1-2) } k_1 \neq l_1, l_2; k_2 = l_1 \\
						%%%
						\frac{1}{p_{l_2}} 
						\left(a_{k_2 k_2} b_{l_1 k_1} + a_{k_2 k_2} b_{k_1 l_1} \right)
						+  
						b_{l_1 k_1} b_{k_2 l_2} + b_{l_2 k_1} b_{k_2 l_1} + b_{l_1 k_2} b_{k_1 l_2} + b_{l_2 k_2} b_{k_1 l_1}, 
						\\ & 
						\text{(1-3) } k_1 \neq l_1, l_2; k_2 = l_2 \\
						%%%
						\frac{1}{p_{l_1}}
						\left(a_{k_1 k_1} b_{l_2 k_2} + a_{k_1 k_1} b_{k_2 l_2} \right)
						+ 
						b_{l_1 k_2} b_{k_1 l_2} + b_{l_2 k_2} b_{k_1 l_1} + b_{l_1 k_1} b_{k_2 l_2} + b_{l_2 k_1} b_{k_2 l_1},
						\\ & \text{(2-1) } k_1 = l_1; k_2 \neq l_1, l_2;
						\\ & \text{switch } k_1, k_2 \text{ in (1-2)} \\
						%%%
						(b_{l_1 l_2} ^ 2 + b_{l_2 l_1} ^ 2) 
						+ \frac{2}{p_{l_1} p_{l_2}} \left(a_{l_1 l_1} + p_{l_1} b_{l_1 l_1} \right) \left(a_{l_2 l_2} + p_{l_2} b_{l_2 l_2} \right),
						\\ & \text{(2-2) } k_1 = l_1; k_2 = l_2;
						\\ & i.e., \operatorname{var} \left(\widetilde{\beta}_{l_1, l_2} \right) \\
						%%%
						\frac{1}{p_{l_2}} 
						\left(a_{k_1 k_1} b_{l_1 k_2} + a_{k_1 k_1} b_{k_2 l_1}\right)
						+
						b_{l_1 k_2} b_{k_1 l_2} + b_{l_2 k_2} b_{k_1 l_1} + b_{l_1 k_1} b_{k_2 l_2} + b_{l_2 k_1} b_{k_2 l_1},
						\\ & \text{(3-1) } k_1 = l_2; k_2 \neq l_1, l_2; 
						\\ & \text{switch } k_1, k_2 \text{ in (1-3)} \\
						%%%%
						\left(b_{l_2 l_1} ^ 2 + b_{l_1 l_2} ^ 2 \right) 
						+ \frac{2}{p_{l_2} p_{l_1}} \left(a_{l_2 l_2} + p_{l_2} b_{l_2 l_2} \right) \left(a_{l_1 l_1} + p_{l_1} b_{l_1 l_1} \right),
						\\ & \text{(3-2) } k_1 = l_2; k_2 = l_1;
						\\ & i.e., \operatorname{var} \left(\widetilde{\beta}_{l_2, l_1} \right).
					\end{cases}
				\end{split}
		\end{equation*} 
		for every $k_1 \neq k_2$ and $l_1 \neq l_2$.
			
		(2) Furthermore, the $q$ by $q$ covariance matrix of $\widetilde{\bm{\theta}}$, i.e.,
		\begin{equation*}
			\operatorname{var}(\widetilde{\bm{\theta}}) = \bm{\Phi}_{\bm{p}} \operatorname{var}\{(\widetilde{\alpha}_{11}, \ldots, \widetilde{\alpha}_{KK}, \widetilde{\beta}_{11}, \ldots, \widetilde{\beta}_{1K}, \widetilde{\beta}_{22}, \widetilde{\beta}_{KK})^\top\} \bm{\Phi}_{\bm{p}} ^ \top
		\end{equation*}
		can be obtained using the below linear transformations 
		\begin{equation*}
				\begin{split}
					\begin{pmatrix}
						\widetilde{a}_{kk} \\ \widetilde{b}_{kk}
					\end{pmatrix}
					= \frac{1}{p_k - 1}
					\begin{pmatrix}
						p_k & - p_k \\ - 1 & p_k
					\end{pmatrix}
					\begin{pmatrix}
						\widetilde{\alpha}_{kk} \\ \widetilde{\beta}_{kk}
					\end{pmatrix} \ \ \ \ \text{for every $k$,} \ \ \ \ 
					\widetilde{b}_{kk'} = \widetilde{\beta}_{kk'} \ \ \ \ \text{for every $k \neq k'$,}
				\end{split}
		\end{equation*}
		where $\bm{\Phi}_{\bm{p}} \in \mathbb{R} ^ {q \times q}$ is a matrix containing the elements of $\bm{p}$ only.
		In particular, by rearranging the order of the elements of $\widetilde{\bm{\theta}}$, we can obtain the results at the beginning. 	
	\end{cors}
	
	\section{ADDITIONAL NUMERICAL RESULTS}
	
	\label{Sec:numerical}
	
	\subsection{Finite-sample performance of the proposed covariance estimators}
	
	We include the additional results from simulation studies for Scenario 1 under various sample sizes, which are presented in Table~\ref{Tab:finite_n_50}, Table~\ref{Tab:finite_n_100}, and Table~\ref{Tab:finite_n_150}.
	
	\begin{table}[!h]
		\footnotesize		
		\begin{tabular}{crrrrrrrrrrrrrr}
			\hline
			& \multicolumn{4}{c}{$p = 150$} & \multicolumn{1}{c}{} & \multicolumn{4}{c}{$p = 225$} & \multicolumn{1}{c}{} & \multicolumn{4}{c}{$p = 300$} \\ \cline{2-5} \cline{7-10} \cline{12-15} 
			& \multicolumn{1}{c}{bias} & \multicolumn{1}{c}{MCSD} & \multicolumn{1}{c}{ASE} & \multicolumn{1}{c}{$95\%$ CP} & \multicolumn{1}{c}{} & \multicolumn{1}{c}{bias} & \multicolumn{1}{c}{MCSD} & \multicolumn{1}{c}{ASE} & \multicolumn{1}{c}{$95\%$ CP} & \multicolumn{1}{c}{} & \multicolumn{1}{c}{bias} & \multicolumn{1}{c}{MCSD} & \multicolumn{1}{c}{ASE} & \multicolumn{1}{c}{$95\%$ CP} \\ \hline
			$a_{0,11}$ & 0.0 & 0.1 & 0.1 & 95.5 &  & 0.0 & 0.0 & 0.0 & 95.2 &  & 0.0 & 0.0 & 0.0 & 95.7 \\
			$a_{0,22}$ & 0.0 & 0.8 & 0.8 & 93.9 &  & 0.0 & 0.7 & 0.7 & 95.0 &  & 0.0 & 0.6 & 0.6 & 94.7 \\
			$a_{0,33}$ & 0.1 & 2.8 & 2.8 & 95.7 &  & 0.1 & 2.3 & 2.3 & 95.3 &  & 0.0 & 1.9 & 2.0 & 96.3 \\
			$a_{0,44}$ & 0.0 & 0.3 & 0.3 & 94.8 &  & 0.0 & 0.2 & 0.2 & 95.3 &  & 0.0 & 0.2 & 0.2 & 95.4 \\
			$a_{0,55}$ & 0.0 & 0.4 & 0.4 & 94.6 &  & 0.0 & 0.3 & 0.3 & 94.7 &  & 0.0 & 0.3 & 0.3 & 94.3 \\
			$b_{0,11}$ &-1.7 & 139.4 & 135.7 & 91.4 &  & 0.8 & 135.3 & 136.2 & 93.1 &  & -0.4 & 134.1 & 135.9 & 93.2 \\
			$b_{0,12}$ & 6.0 & 86.9 & 87.5 & 94.8 &  & 4.3 & 85.7 & 87.2 & 95.1 &  & 1.0 & 86.2 & 88.2 & 95.0 \\
			$b_{0,13}$ & 5.6 & 79.4 & 78.0 & 95.4 &  & 3.5 & 77.7 & 77.9 & 95.9 &  & 2.3 & 75.9 & 78.0 & 96.9 \\
			$b_{0,14}$ &-2.6 &110.3 &105.1 & 91.8 &  & -0.8 & 104.0 & 105.3 & 93.7 &  & -0.1 & 103.6 & 104.8 & 95.0 \\
			$b_{0,15}$ &-2.0 & 80.7 & 78.2 & 93.5 &  & 1.9 & 78.1 & 78.4 & 94.0 &  & -3.3 & 77.6 & 78.4 & 93.7 \\
			$b_{0,22}$ &-0.9 &105.7 &105.3 & 93.2 &  & -7.4 & 105.7 & 104.0 & 91.2 &  & 5.7 & 106.3 & 106.6 & 93.9 \\
			$b_{0,23}$ &-0.1 & 88.6 & 87.1 & 92.7 &  & -3.1 & 86.6 & 86.3 & 93.3 &  & 4.1 & 89.5 & 87.8 & 93.7 \\
			$b_{0,24}$ & 1.3 & 85.6 & 86.1 & 95.4 &  & -2.7 & 86.2 & 85.8 & 95.3 &  & -0.4 & 88.7 & 86.4 & 95.2 \\
			$b_{0,25}$ & 3.2 & 66.1 & 65.5 & 95.7 &  & -3.3 & 63.4 & 64.9 & 95.4 &  & 2.9 & 67.9 & 66.1 & 95.7 \\
			$b_{0,33}$ & 0.7 & 88.1 & 88.1 & 93.6 &  & -0.5 & 86.6 & 87.7 & 93.3 &  & 3.3 & 90.1 & 88.3 & 93.5 \\
			$b_{0,34}$ & 0.4 & 91.6 & 91.1 & 94.4 &  & -2.6 & 91.3 & 91.3 & 93.0 &  & 1.2 & 91.4 & 90.9 & 93.5 \\
			$b_{0,35}$ & 2.9 & 61.4 & 59.3 & 95.6 &  & -0.9 & 58.1 & 59.1 & 95.8 &  & 1.9 & 60.2 & 59.5 & 96.3 \\
			$b_{0,44}$ & 0.1 & 140.5 & 137.2 & 92.1 &  & 2.6 & 137.0 & 137.7 & 94.2 &  & -4.9 & 132.2 & 136.2 & 93.1 \\.0
			$b_{0,45}$ & -0.5 & 78.6 & 74.0 & 94.2 &  & -1.5 & 73.5 & 73.9 & 96.2 &  & -2.1 & 73.8 & 73.8 & 96.0 \\
			$b_{0,55}$ & -1.6 & 80.7 & 79.6 & 92.7 &  & -2.2 & 81.4 & 79.5 & 92.7 &  & 0.4 & 80.4 & 80.0 & 93.0 \\ \hline
		\end{tabular}
		\caption{We assess the estimated covariance parameters ($a_{0, kk}$ and $b_{0,kk'}$) using our proposed method across $1000$ simulated datasets under $n = 50$: 
			``bias" represents the relative bias ($\times 100$); 
			``MCSD" refers to the Monte Carlo standard deviation ($\times 100$); 
			``ASE" stands for the average standard error ($\times 100$); 
			``$95\%$ CP" represents the empirical coverage probability (in $\%$) based on a $95\%$ Wald-type confidence interval.}
		\label{Tab:finite_n_50} 
	\end{table}

	\begin{table}[!h]
		\footnotesize
		\begin{tabular}{crrrrrrrrrrrrrr}
			\hline
			& \multicolumn{4}{c}{$p = 150$} & \multicolumn{1}{c}{} & \multicolumn{4}{c}{$p = 225$} & \multicolumn{1}{c}{} & \multicolumn{4}{c}{$p = 300$} \\ \cline{2-5} \cline{7-10} \cline{12-15} 
			& \multicolumn{1}{c}{bias} & \multicolumn{1}{c}{MCSD} & \multicolumn{1}{c}{ASE} & \multicolumn{1}{c}{$95\%$ CP} & \multicolumn{1}{c}{} & \multicolumn{1}{c}{bias} & \multicolumn{1}{c}{MCSD} & \multicolumn{1}{c}{ASE} & \multicolumn{1}{c}{$95\%$ CP} & \multicolumn{1}{c}{} & \multicolumn{1}{c}{bias} & \multicolumn{1}{c}{MCSD} & \multicolumn{1}{c}{ASE} & \multicolumn{1}{c}{$95\%$ CP} \\ \hline
			$a_{0,11}$ & 0.0 & 0.0 & 0.0 & 95.1 &  & 0.0 & 0.0 & 0.0 & 94.9 &  & 0.0 & 0.0 & 0.0 & 95.2 \\
			$a_{0,22}$ & 0.0 & 0.6 & 0.6 & 95.8 &  & 0.0 & 0.4 & 0.5 & 96.2 &  & 0.0 & 0.4 & 0.4 & 94.8 \\
			$a_{0,33}$ & 0.0 & 2.1 & 2.0 & 94.2 &  & 0.0 & 1.6 & 1.6 & 94.9 &  & 0.1 & 1.4 & 1.4 & 95.3 \\
			$a_{0,44}$ & 0.0 & 0.2 & 0.2 & 96.3 &  & 0.0 & 0.1 & 0.1 & 95.5 &  & 0.0 & 0.1 & 0.1 & 95.4 \\
			$a_{0,55}$ & 0.0 & 0.3 & 0.3 & 95.3 &  & 0.0 & 0.2 & 0.2 & 95.1 &  & 0.0 & 0.2 & 0.2 & 95.3 \\
			$b_{0,11}$ & 0.3 & 97.6 & 95.7 & 93.5 &  & 7.1 & 97.3 & 96.7 & 93.5 &  & -1.1 & 93.7 & 95.5 & 94.6 \\
			$b_{0,12}$ & 1.2 & 59.6 & 61.8 & 95.4 &  & -2.1 & 61.3 & 62.4 & 95.8 &  & 2.3 & 62.5 & 61.9 & 94.7 \\
			$b_{0,13}$ & 1.1 & 52.2 & 54.7 & 96.8 &  & -0.4 & 54.3 & 55.0 & 95.6 &  & 1.2 & 55.0 & 54.8 & 94.9 \\
			$b_{0,14}$ &-0.4 & 72.5 & 74.1 & 95.4 &  & -3.1 & 75.4 & 74.3 & 94.3 &  & -0.2 & 74.5 & 73.9 & 94.8 \\
			$b_{0,15}$ & -0.8 & 55.6 & 55.2 & 93.8 &  & 3.7 & 56.0 & 55.9 & 96.0 &  & 0.7 & 56.8 & 55.3 & 94.8 \\
			$b_{0,22}$ & -1.6 & 71.5 & 74.0 & 95.1 &  & 2.3 & 73.4 & 74.5 & 94.0 &  & 1.8 & 76.5 & 74.4 & 94.0 \\
			$b_{0,23}$ & -2.3 & 58.3 & 61.0 & 94.1 &  & -0.1 & 60.6 & 61.3 & 94.2 &  & 1.5 & 64.3 & 61.4 & 93.8 \\
			$b_{0,24}$ & 1.7 & 59.9 & 60.6 & 95.3 &  & 3.3 & 61.7 & 60.6 & 95.5 &  & -0.1 & 61.8 & 60.7 & 94.5 \\
			$b_{0,25}$ & 1.8 & 45.4 & 46.1 & 95.6 &  & 1.9 & 48.3 & 46.6 & 94.8 &  & 1.7 & 46.2 & 46.3 & 95.0 \\
			$b_{0,33}$ & -2.6 & 58.9 & 61.5 & 95.3 &  & -1.6 & 60.7 & 61.5 & 93.9 &  & 0.5 & 64.0 & 61.8 & 94.1 \\
			$b_{0,34}$ & 1.4 & 63.3 & 64.0 & 94.9 &  & 3.5 & 62.7 & 63.7 & 94.8 &  & 0.6 & 66.7 & 64.0 & 93.5 \\
			$b_{0,35}$ & 1.2 & 39.8 & 41.6 & 96.2 &  & 1.5 & 43.9 & 41.9 & 94.8 &  & 2.0 & 41.9 & 41.7 & 95.4 \\
			$b_{0,44}$ & 0.8 & 93.1 & 96.6 & 96.3 &  & -3.9 & 93.8 & 95.9 & 94.7 &  & -0.8 & 98.8 & 96.4 & 93.0 \\
			$b_{0,45}$ & 0.0 & 50.2 & 52.1 & 96.0 &  & -1.9 & 54.5 & 52.2 & 93.3 &  & -1.2 & 53.9 & 52.0 & 94.7 \\
			$b_{0,55}$ & -0.8 & 57.1 & 56.1 & 94.0 &  & 4.1 & 56.7 & 56.8 & 94.9 &  & 0.0 & 56.6 & 56.2 & 93.9 \\ \hline
		\end{tabular}
		\caption{We assess the estimated covariance parameters ($a_{0,kk}$ and $b_{0,kk'}$) using our proposed method across
			$1000$ simulated datasets under $n = 100$: 
			``bias" represents the relative bias ($\times 100$); 
			``MCSD" refers to the Monte Carlo standard deviation ($\times 100$); 
			``ASE" stands for the average standard error ($\times 100$); 
			``$95\%$ CP" represents the empirical coverage probability (in $\%$) based on a $95\%$ Wald-type confidence interval.}
		\label{Tab:finite_n_100} 
	\end{table}
	
	\begin{table}[!h]
		\footnotesize
		\begin{tabular}{crrrrrrrrrrrrrr}
			\hline
			& \multicolumn{4}{c}{$p = 150$} & \multicolumn{1}{c}{} & \multicolumn{4}{c}{$p = 225$} & \multicolumn{1}{c}{} & \multicolumn{4}{c}{$p = 300$} \\ \cline{2-5} \cline{7-10} \cline{12-15} 
			& \multicolumn{1}{c}{bias} & \multicolumn{1}{c}{MCSD} & \multicolumn{1}{c}{ASE} & \multicolumn{1}{c}{$95\%$ CP} & \multicolumn{1}{c}{} & \multicolumn{1}{c}{bias} & \multicolumn{1}{c}{MCSD} & \multicolumn{1}{c}{ASE} & \multicolumn{1}{c}{$95\%$ CP} & \multicolumn{1}{c}{} & \multicolumn{1}{c}{bias} & \multicolumn{1}{c}{MCSD} & \multicolumn{1}{c}{ASE} & \multicolumn{1}{c}{$95\%$ CP} \\ \hline
			$a_{0,11}$ & 0.0 & 0.0 & 0.0 & 95.0 &  & 0.0 & 0.0 & 0.0 & 96.0 &  & 0.0 & 0.0 & 0.0 & 94.6 \\
			$a_{0,22}$ & 0.0 & 0.5 & 0.5 & 93.7 &  & 0.0 & 0.4 & 0.4 & 95.5 &  & 0.0 & 0.3 & 0.3 & 94.1 \\
			$a_{0,33}$ & 0.0 & 1.6 & 1.6 & 95.1 &  & 0.0 & 1.3 & 1.3 & 93.6 &  & 0.0 & 1.1 & 1.1 & 96.3 \\
			$a_{0,44}$ & 0.0 & 0.1 & 0.1 & 94.7 &  & 0.0 & 0.1 & 0.1 & 95.1 &  & 0.0 & 0.1 & 0.1 & 96.0 \\
			$a_{0,55}$ & 0.0 & 0.2 & 0.2 & 95.3 &  & 0.0 & 0.2 & 0.2 & 95.2 &  & 0.0 & 0.2 & 0.2 & 94.4 \\
			$b_{0,11}$ & 2.6 & 75.1 & 78.3 & 95.6 &  & -2.8 & 77.3 & 77.7 & 94.3 &  & -5.8 & 76.5 & 77.3 & 94.3 \\
			$b_{0,12}$ & 1.1 & 49.7 & 50.6 & 95.1 &  & 1.3 & 49.4 & 50.2 & 95.1 &  & 1.1 & 51.6 & 50.3 & 93.7 \\
			$b_{0,13}$ & 2.3 & 44.1 & 44.9 & 96.1 &  & 1.0 & 44.4 & 44.7 & 96.2 &  & -1.5 & 45.9 & 44.4 & 94.0 \\
			$b_{0,14}$ &-3.6 & 58.3 & 60.6 & 95.9 &  & -0.5 & 61.9 & 60.4 & 94.2 &  & 4.6 & 60.3 & 59.8 & 93.7 \\
			$b_{0,15}$ & 0.6 & 46.1 & 45.1 & 94.5 &  & -1.6 & 45.9 & 44.9 & 94.9 &  & -1.4 & 44.9 & 44.9 & 93.7 \\
			$b_{0,22}$ & 1.1 & 61.7 & 60.6 & 95.0 &  & -2.7 & 60.5 & 60.2 & 93.6 &  & 1.2 & 61.7 & 60.6 & 93.9 \\
			$b_{0,23}$ & 1.1 & 51.4 & 50.1 & 94.6 &  & -0.8 & 49.5 & 49.9 & 94.2 &  & 0.9 & 50.0 & 50.0 & 94.2 \\
			$b_{0,24}$ &-0.5 & 48.6 & 49.6 & 95.8 &  & -2.0 & 48.3 & 49.5 & 95.3 &  & -1.2 & 49.5 & 49.4 & 94.4 \\
			$b_{0,25}$ & 1.4 & 36.7 & 37.7 & 95.2 &  & -1.1 & 37.2 & 37.5 & 94.5 &  & 2.1 & 38.7 & 37.8 & 95.6 \\
			$b_{0,33}$ & 1.1 & 52.1 & 50.6 & 94.4 &  & 1.3 & 49.5 & 50.5 & 94.7 &  & -0.2 & 49.3 & 50.3 & 94.4 \\
			$b_{0,34}$ &-1.0 & 52.4 & 52.4 & 94.9 &  & -3.6 & 51.5 & 52.6 & 95.1 &  & 1.0 & 51.4 & 52.1 & 94.4 \\
			$b_{0,35}$ & 2.3 & 33.2 & 34.1 & 95.6 &  & -1.8 & 33.7 & 34.1 & 95.3 &  & 1.0 & 35.5 & 34.0 & 94.1 \\
			$b_{0,44}$ & 1.5 & 77.2 & 78.8 & 95.0 &  & 3.2 & 81.2 & 79.0 & 93.9 &  & -3.9 & 79.7 & 78.2 & 93.2 \\
			$b_{0,45}$ &-2.6 & 42.3 & 42.5 & 95.5 &  & 2.7 & 43.1 & 42.5 & 94.5 &  & 0.4 & 42.9 & 42.4 & 95.2 \\
			$b_{0,55}$ &-0.6 & 46.9 & 45.8 & 93.4 &  & -0.7 & 46.8 & 45.8 & 94.2 &  & 0.7 & 46.4 & 45.9 & 94.5 \\ \hline
		\end{tabular}
		\caption{We assess the estimated covariance parameters ($a_{0,kk}$ and $b_{0,kk'}$) using our proposed method across
			$1000$ simulated datasets under $n = 150$: 
			``bias" represents the relative bias ($\times 100$); 
			``MCSD" refers to the Monte Carlo standard deviation ($\times 100$); 
			``ASE" stands for the average standard error ($\times 100$); 
			``$95\%$ CP" represents the empirical coverage probability (in $\%$) based on a $95\%$ Wald-type confidence interval.}
		\label{Tab:finite_n_150} 
	\end{table}
	
	\subsection{Comparison of the proposed method with existing methods on the simulated datasets}
	
\begin{figure}[!h]
		\centering
		\includegraphics[width = 1.0\linewidth]{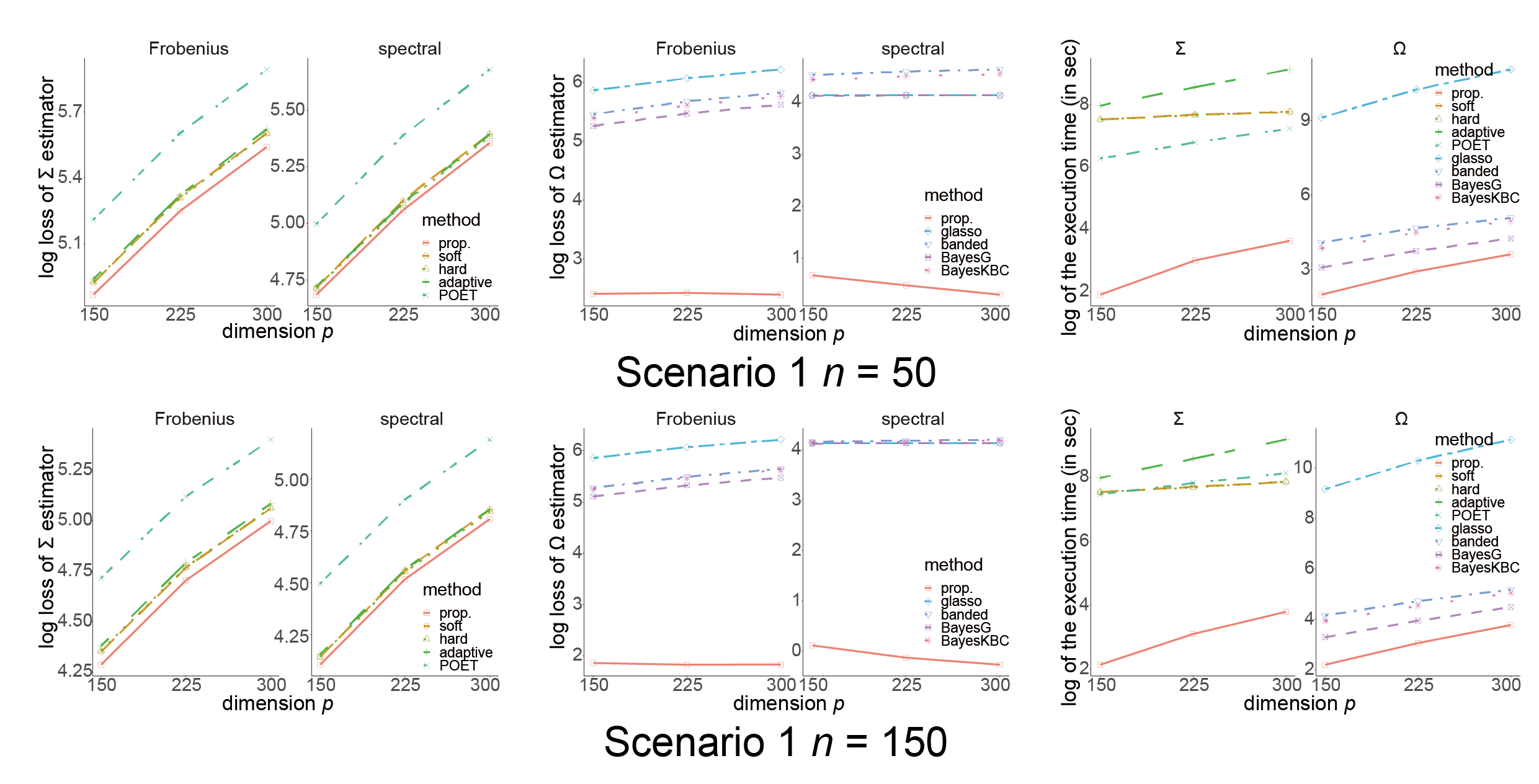} 
		\caption{We evaluate the estimated covariance and precision matrices using our approach and compare them with estimates from benchmark methods. 
		In the first and second rows, we present the results for Scenario 1 (the small $K$ setting) under $n = 50$ and $n = 150$, respectively. 
		The left two and the middle two subfigures illustrate the logarithmic losses in both Frobenius and spectral norms for covariance matrix $\bm{\Sigma}$ and precision matrix $\bm{\Omega}$ estimations for all methods, respectively;
		and the right two subfigures display the execution time for $\bm{\Sigma}$ and $\bm{\Omega}$ estimations.
		}
		\label{Fig:simulation_studies_n_50_150}
	\end{figure}

	We include the additional results from simulation studies for Scenario 1 under various sample sizes, 
	which are presented in Figure~\ref{Fig:simulation_studies_n_50_150}.
	
	\subsection{Comparison of the proposed method with existing methods on the real datasets}

\begin{figure}[!h]
		\centering
		\includegraphics[width = 1.0 \linewidth]{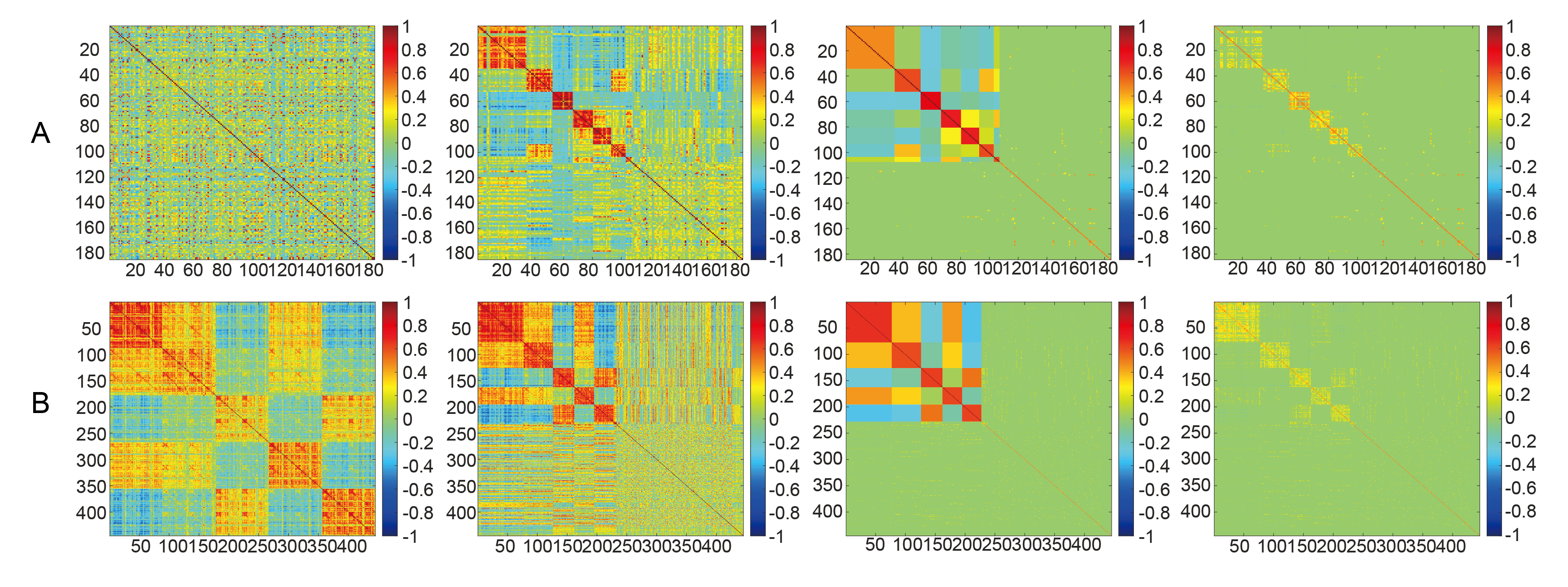} 
		\caption{
			We compare the proposed covariance estimates for all features in the proteomics and brain imaging studies with the conventional covariance estimates (i.e., the soft-thresholding estimates).
			The two rows correspond to the two data examples: A, the proteomics dataset; B, the brain imaging dataset.			
			From the left heatmaps to the right: (1) the raw correlation matrix, (2) the correlation matrix produced by a community detection algorithm, (3) the proposed covariance matrix estimate, and (4) the conventional soft-thresholding covariance matrix estimate.}
		\label{Fig:real_comparison}
	\end{figure}

	We include the additional results for real data examples in Figure~\ref{Fig:real_comparison}.
	
	\section{ADDITIONAL SIMULATION STUDIES}
	
	\label{Sec:simulation}
	
	\subsection{Examination of the accuracy of the covariance estimator}
	
	%\emph{Examination of the accuracy of the covariance estimator.}
	%
	Following the setup in Scenario 1,  
	we aim to access the accuracy of covariance estimators outlined in Corollary~\ref{Cor:vars}.
	This evaluation involve examining each pair of $\alpha_{0, 11}, \ldots, \alpha_{0, KK}$ and $\beta_{0, 11}, \ldots, \beta_{0, KK}$ for the sample sizes of $n = 50$, $100$, or $150$, with $1000$ Monte Carlo replicates.
	In each replicate, we compute the covariance estimate for every pair of $\alpha_0$'s and $\beta_0$'s by substituting the respective estimates $\widetilde{\alpha}_{kk}$ and $\widetilde{\beta}_{kk'}$ in Corollary~\ref{Cor:vars}.
	Averaging the estimated covariances over the $1000$ replicates provided the average estimated covariance (denoted as AC) for each pair of $\alpha_0$'s and $\beta_0$'s.
	Additionally, we calculate the covariances using the actual values in the covariance formulas (denoted as RC) for every pair of $\alpha_0$'s and $\beta_0$'s. 
	To establish a baseline, we compute the Monte Carlo covariance (denoted as MCC) for every pair of $\alpha_0$'s and $\beta_0$'s.
	The results for pairs involving $\alpha_0$'s, $\beta_0$'s, and $\alpha_0$'s-$\beta_0$'s are tabulated in an Excel file, available at \url{https://github.com/yiorfun/UBCovEst}. 
	
	The findings in the tables demonstrate that the covariances evaluated at the true values are closely aligned with the empirical covariances computed using the Monte Carlo method.
	Moreover, for relatively small sample sizes, say $n = 50$, the proposed method may slightly overestimate the covariances for certain pairs of $\alpha_0$'s and $\beta_0$'s.
	This occurrence may be attributed to the number of unknown parameters (i.e., $q = 20$) being comparable to $n = 50$.
	As the sample size increases beyond $50$, the proposed method becomes more effective in providing reasonable covariance estimation.
	
	\subsection{Evaluation of the estimated covariance matrix estimator on multiple testing}
	We conduct an additional simulation study to investigate the advantage of a covariance matrix estimate with less bias.
	Specifically, we examine how a covariance matrix estimate with a larger bias may potentially lead to an inappropriate type $1$ error or a loss in statistical power.
	For this study, we set $p_{\text{ind}} = 100$ (resulting in $p = 500$) and generate a sample with $n = 50$ from $N \left(\bm{\bm{\mu}}_0, \bm{\bm{\Sigma}}_{0, 1} \left(\textbf{A}_0, \textbf{B}_0, \bm{p}_1 \right)\right)$, where $\bm{\bm{\mu}}_0 = \left(\mu_{01}, \ldots, \mu_{0p} \right) ^ \top$, with $\mu_{0j} = 0.3$ for $j = 1, \ldots, 10$ and $\mu_{0j} = 0$ otherwise. 
	The other settings are kept consistent with those in Scenario 1.
	To simultaneously perform hypothesis tests 
	\begin{equation*}
		H_{0j}: \mu_j = 0, \quad \text{ against } \quad H_{1j}: \mu_j \neq 0, \quad j = 1, \ldots, p,
	\end{equation*} 
	we utilize the principal factor approximation (PFA) method, as described by \citet{FanHanGu2012} and \citet{FanHan2017}, to estimate the false discovery proportion (FDP) in the presence of dependence among test statistics.
	Across $100$ Monte Carlo replicates, for a sequence of threshold values $t$, 
	we calculate the medians and standard errors (s.d.) of the total number of rejections $R(t)$, 
	the number of correct rejections $S(t)$, 
	and the false discovery proportion $\operatorname{FDP}(t)$ using the true covariance matrix $\bm{\bm{\Sigma}}_{0, 1} \left(\textbf{A}_0, \textbf{B}_0, \bm{p}_1 \right)$, its POET estimator $\widetilde{\bm{\bm{\Sigma}}}_{\text{POET}}$, and the proposed estimator $\widetilde{\bm{\bm{\Sigma}}}_{\text{prop.}} = \widetilde{\bm{\bm{\Sigma}}}_1 \left(\widetilde{\textbf{A}}_1, \widetilde{\textbf{B}}_1, \bm{p}_1 \right)$, respectively.
	
	The results presented in Table~\ref{Tab:FDP} demonstrate that the proposed covariance matrix estimator behaves more similarly to the true covariance matrix than the POET estimator, primarily due to having lower bias.
	For example, at a threshold value $t = 0.0527$, the approximate $\operatorname{FDP}(t)$ are $0.00$, $0.79$, and $0.13$ for the true covariance matrix, the POET estimator, and the proposed estimator, respectively. 
	This suggests that the bias in the POET estimator may result in an inappropriate FDP.
	Moreover, the number of correct rejections $S(t)$ is estimated as $10$, $4$, and $10$ for the true covariance matrix, the POET estimator, and the proposed estimator, respectively.
	This indicates a loss in statistical power for the POET estimator.
	
	\begin{landscape}
		\begin{table}[!h]
			\tiny
			\begin{tabular}{ccccccccccccccccccc}
				\hline
				\multicolumn{1}{l}{} & \multicolumn{6}{c}{true} & \multicolumn{6}{c}{POET} & \multicolumn{6}{c}{proposed method} \\ \cline{2-19} 
				\begin{tabular}[c]{@{}c@{}}threshold\\ $t$\end{tabular} & \begin{tabular}[c]{@{}c@{}}median\\ $R(t)$\end{tabular} & \begin{tabular}[c]{@{}c@{}}s.d.\\ $R(t)$\end{tabular} & \begin{tabular}[c]{@{}c@{}}median\\ $S(t)$\end{tabular} & \begin{tabular}[c]{@{}c@{}}s.d.\\ $S(t)$\end{tabular} & \begin{tabular}[c]{@{}c@{}}median\\ $\operatorname{FDP}(t)$\end{tabular} & \begin{tabular}[c]{@{}c@{}}s.d.\\ $\operatorname{FDP}(t)$\end{tabular} & \begin{tabular}[c]{@{}c@{}}median\\ $R(t)$\end{tabular} & \begin{tabular}[c]{@{}c@{}}s.d.\\ $R(t)$\end{tabular} & \begin{tabular}[c]{@{}c@{}}median\\ $S(t)$\end{tabular} & \begin{tabular}[c]{@{}c@{}}s.d.\\ $S(t)$\end{tabular} & \begin{tabular}[c]{@{}c@{}}median\\ $\operatorname{FDP}(t)$\end{tabular} & \begin{tabular}[c]{@{}c@{}}s.d.\\ $\operatorname{FDP}(t)$\end{tabular} & \begin{tabular}[c]{@{}c@{}}median\\ $R(t)$\end{tabular} & \begin{tabular}[c]{@{}c@{}}s.d.\\ $R(t)$\end{tabular} & \begin{tabular}[c]{@{}c@{}}median\\ $S(t)$\end{tabular} & \begin{tabular}[c]{@{}c@{}}s.d.\\ $S(t)$\end{tabular} & \begin{tabular}[c]{@{}c@{}}median\\ $\operatorname{FDP}(t)$\end{tabular} & \begin{tabular}[c]{@{}c@{}}s.d.\\ $\operatorname{FDP}(t)$\end{tabular} \\ \hline
				0.0001 & 0.00 & 0.00 & 10.00 & 0.00 & 0.00 & 0.00 & 0.00 & 0.00 & 0.00 & 1.90 & 0.00 & 0.00 & 0.00 & 0.00 & 10.00 & 4.23 & 0.00 & 0.00 \\
				0.0054 & 0.00 & 3.47 & 10.00 & 0.00 & 0.00 & 0.22 & 0.00 & 11.40 & 0.00 & 3.21 & 0.00 & 0.31 & 0.00 & 10.97 & 10.00 & 3.18 & 0.00 & 0.23 \\
				0.0106 & 0.00 & 11.96 & 10.00 & 0.00 & 0.00 & 0.31 & 0.00 & 15.95 & 0.00 & 3.42 & 0.00 & 0.38 & 0.00 & 14.68 & 10.00 & 2.81 & 0.00 & 0.34 \\
				0.0159 & 0.00 & 20.02 & 10.00 & 0.00 & 0.00 & 0.37 & 0.00 & 24.75 & 0.00 & 3.70 & 0.00 & 0.41 & 0.00 & 22.95 & 10.00 & 2.76 & 0.00 & 0.39 \\
				0.0211 & 0.00 & 25.19 & 10.00 & 0.00 & 0.00 & 0.40 & 0.00 & 30.70 & 0.00 & 3.78 & 0.00 & 0.45 & 0.00 & 30.01 & 10.00 & 2.70 & 0.00 & 0.40 \\
				0.0264 & 0.00 & 30.84 & 10.00 & 0.00 & 0.00 & 0.43 & 0.00 & 35.79 & 0.00 & 3.96 & 0.00 & 0.47 & 0.00 & 34.22 & 10.00 & 2.46 & 0.00 & 0.42 \\
				0.0316 & 0.00 & 36.12 & 10.00 & 0.00 & 0.00 & 0.44 & 0.00 & 39.71 & 0.00 & 3.99 & 0.00 & 0.47 & 0.00 & 38.52 & 10.00 & 2.41 & 0.00 & 0.43 \\
				0.0369 & 0.00 & 40.75 & 10.00 & 0.00 & 0.00 & 0.45 & 0.00 & 44.15 & 0.00 & 4.19 & 0.00 & 0.47 & 0.00 & 42.58 & 10.00 & 2.36 & 0.00 & 0.44 \\
				0.0422 & 0.00 & 45.65 & 10.00 & 0.00 & 0.00 & 0.45 & 0.00 & 49.61 & 1.50 & 4.29 & 0.00 & 0.48 & 0.00 & 47.61 & 10.00 & 2.18 & 0.00 & 0.45 \\
				0.0474 & 0.00 & 49.77 & 10.00 & 0.00 & 0.00 & 0.46 & 0.50 & 55.04 & 2.00 & 4.43 & 0.03 & 0.48 & 0.00 & 53.13 & 10.00 & 2.09 & 0.00 & 0.45 \\
				0.0527 & 1.00 & 52.49 & 10.00 & 0.00 & 0.00 & 0.46 & 1.50 & 59.48 & 4.00 & 4.52 & 0.79 & 0.48 & 1.00 & 57.79 & 10.00 & 2.04 & 0.13 & 0.46 \\
				0.0579 & 1.00 & 55.14 & 10.00 & 0.00 & 0.00 & 0.46 & 2.00 & 63.10 & 5.00 & 4.61 & 0.85 & 0.48 & 2.00 & 61.26 & 10.00 & 1.97 & 0.42 & 0.46 \\
				0.0632 & 4.00 & 57.63 & 10.00 & 0.00 & 0.59 & 0.46 & 4.00 & 65.92 & 6.50 & 4.65 & 0.87 & 0.47 & 3.50 & 64.55 & 10.00 & 1.96 & 0.44 & 0.46 \\
				0.0685 & 8.00 & 59.61 & 10.00 & 0.00 & 0.64 & 0.46 & 8.00 & 67.90 & 7.00 & 4.63 & 0.97 & 0.46 & 5.50 & 66.84 & 10.00 & 1.96 & 0.60 & 0.45 \\
				0.0737 & 10.00 & 62.86 & 10.00 & 0.00 & 0.74 & 0.46 & 10.00 & 69.55 & 7.50 & 4.61 & 1.00 & 0.45 & 8.00 & 68.44 & 10.00 & 1.96 & 0.73 & 0.46 \\
				0.0790 & 10.00 & 65.62 & 10.00 & 0.00 & 0.74 & 0.45 & 12.00 & 71.26 & 8.00 & 4.45 & 1.00 & 0.44 & 10.00 & 70.06 & 10.00 & 1.96 & 0.84 & 0.45 \\
				0.0842 & 10.50 & 68.02 & 10.00 & 0.00 & 0.80 & 0.45 & 13.00 & 72.40 & 9.00 & 4.37 & 1.00 & 0.43 & 12.00 & 71.64 & 10.00 & 1.94 & 0.88 & 0.44 \\
				0.0895 & 10.50 & 70.18 & 10.00 & 0.00 & 0.87 & 0.45 & 16.00 & 73.83 & 10.00 & 4.26 & 1.00 & 0.43 & 13.00 & 72.85 & 10.00 & 1.94 & 0.82 & 0.44 \\
				0.0947 & 14.00 & 71.77 & 10.00 & 0.00 & 0.88 & 0.44 & 21.50 & 75.19 & 10.00 & 4.16 & 1.00 & 0.42 & 18.00 & 74.40 & 10.00 & 1.94 & 0.79 & 0.43 \\
				0.1000 & 20.00 & 72.97 & 10.00 & 0.00 & 0.88 & 0.44 & 27.00 & 77.10 & 10.00 & 3.95 & 1.00 & 0.40 & 22.00 & 75.90 & 10.00 & 1.93 & 0.83 & 0.43 \\ \hline
			\end{tabular}
			\caption{We present the total number of rejections $R(t)$, the number of correct rejections $S(t)$, and the false discovery proportion (FDP) using the true covariance matrix, the POET estimator, and the proposed estimator}
			\label{Tab:FDP}
		\end{table}
	\end{landscape}
		
	\subsection{Justification of the proposed parameterization strategy for the population covariance matrix}
\begin{figure}[!h]
		\centering
		\includegraphics[width = 1.0 \linewidth]{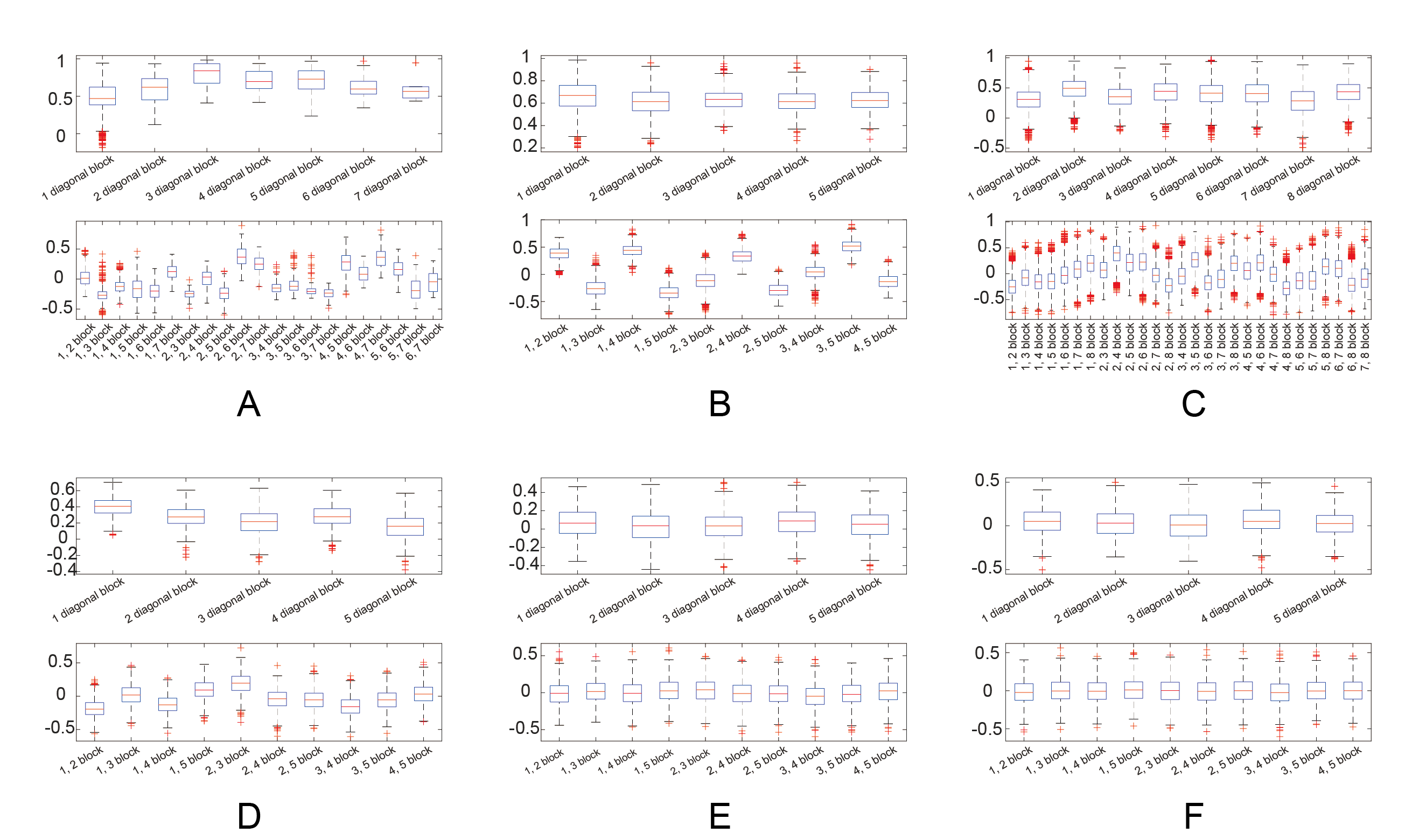} 
		\caption{
			We display box plots of the off-diagonal entries within the diagonal blocks and those of all entries in the off-diagonal blocks.
			A: Box plots for the proteomics dataset; 
			B: Box plots for the brain imaging dataset;
			C: Box plots for the Spellman dataset;
			D: Box plots for the simulated dataset with $\sigma = 0.1$;
			E: Box plots for the simulated dataset with $\sigma = 0.5$;
			F: Box plots for the simulated dataset with $\sigma = 0.8$.}
		\label{Fig:simulation_boxplots}
	\end{figure}

	To validate the proposed parameterization for the uniform-block structure, 
	we compared box plots of the off-diagonal entries in the diagonal blocks and those of all entries in the off-diagonal blocks, using both real and simulated datasets from Scenario 3 (Section 3.4). 
	The resulting box plots, displayed in Figure~\ref{Fig:simulation_boxplots}, illustrated that although the true covariance matrix $\bm{\Upsilon}_{0, \sigma}$ might not precisely have a uniform-block structure, 
	the proposed estimation procedure yielded competitive covariance-matrix estimates and comparable precision-matrix estimates to conventional methods.
	This observation held particularly true when variations within blocks were small. 
	These findings suggested that considering the structure of uniform blocks for real datasets was reasonable, especially given that the variations within the blocks closely resembled those observed in the simulated datasets with $\sigma = 0.8$ for both the proteomics study and the brain imaging study.

	\subsection{Justification of the proposed approach among alternative covariance structures}
	
	We assess the performance of our proposed approach across three distinct community structures: unbalanced interconnected community structure, 
	independent community structure, and hierarchical community structure, denoted as ``unbalanced", ``independent", and ``hierarchical", respectively. 
	
	To generate the simulated datasets, initially, we define the true covariance matrices:
	
	(1) for unbalanced interconnected community (or uniform-block) structure $\bm{\Sigma}_{0, \text{U}} = \textbf{A}_{0, \text{U}} \circ \textbf{I}(\bm{p}_4) + \textbf{B}_{0, \text{U}} \circ \textbf{J}(\bm{p}_4)$; 
	
	(2) for independent community structure $\bm{\Sigma}_{0, \text{I}} = \textbf{A}_{0, \text{I}} \circ \textbf{I}(\bm{p}_4) + \textbf{B}_{0, \text{I}} \circ \textbf{J}(\bm{p}_4)$;
	
	(3) for hierarchical community structure $\bm{\Sigma}_{0, \text{H}} = \textbf{A}_{0, \text{H}} \circ \textbf{I}(\bm{p}_4) + \textbf{B}_{0, \text{H}} \circ \textbf{J}(\bm{p}_4)$. 
	
	Here, the partition-size vector is defined as $\bm{p}_4 = \left(10, 15, 15, 20, 25 \right) ^ \top$, indicating that $K = 5$, $p = 85$, and community sizes are unbalanced.
	Moreover,  
	$\textbf{A}_{0, \text{U}} = \textbf{A}_{0, \text{I}} = \textbf{A}_{0, \text{H}} = \textbf{A}_0$ that is defined in Scenario 1, 
	$\textbf{B}_{0, \text{U}} = \textbf{B}_0$ that is defined in Scenario 1,
	$\textbf{B}_{0, \text{I}} = \operatorname{diag}(\textbf{B}_0)$,
	and 
	\begin{equation*}
		\begin{split}
			\textbf{B}_{0, \text{H}} = 
			\begin{pmatrix}
				6.731 & -1.690 & 0.530 & 0.530 & 0.530 \\
				& \phantom{-}5.215 & 0.530 & 0.530 & 0.530 \\
				& & 4.328 & 0.790 & 0.790 \\
				& & & 6.788 & 0.790 \\
				& & & & 3.954
			\end{pmatrix}.
		\end{split}
	\end{equation*}
	
\begin{figure}[!h]
		\centering
		\includegraphics[width = 0.6 \linewidth]{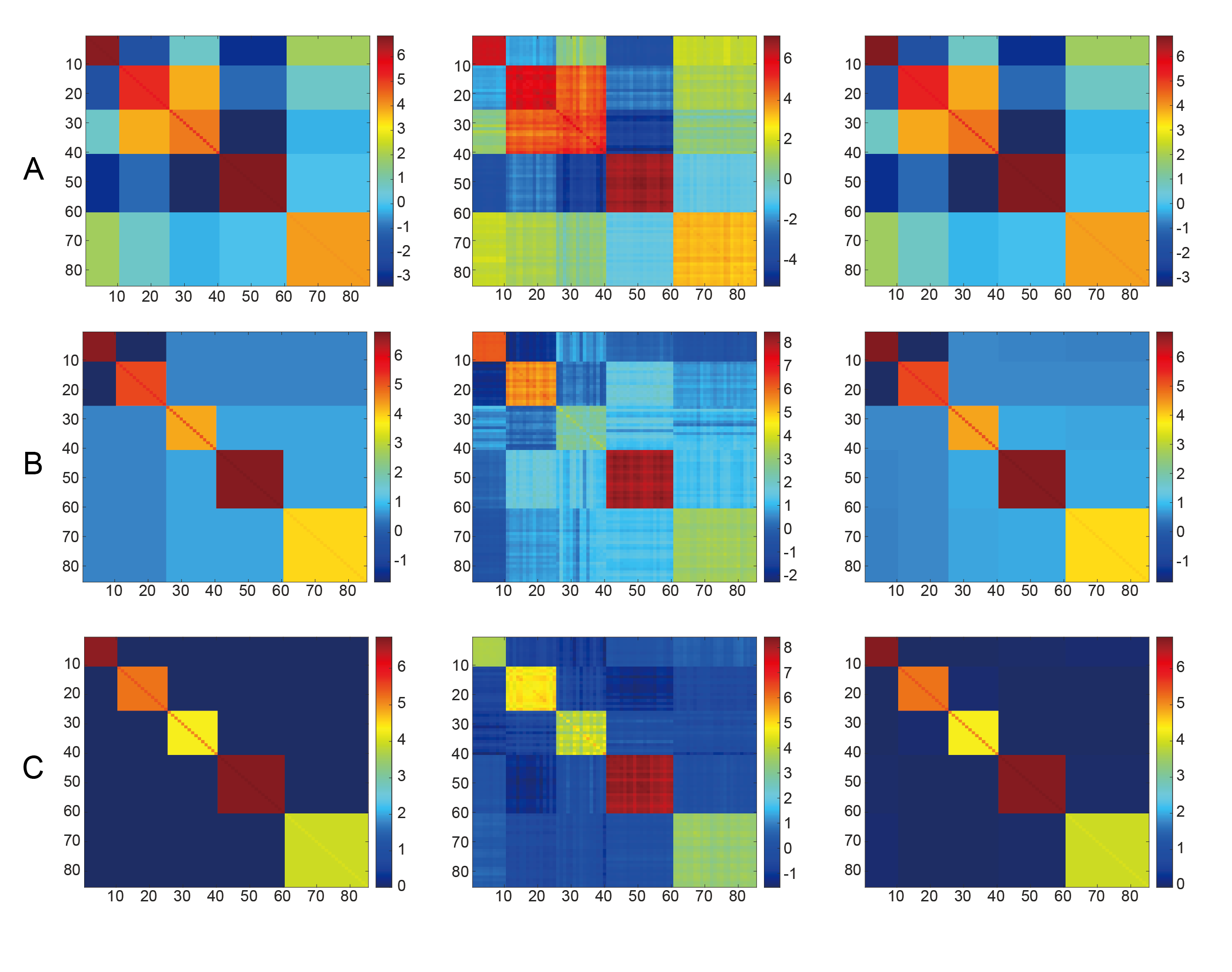} 
		\caption{
			We compare the estimated covariance matrix using our proposed method across
			$1000$ simulated datasets under $n = 50$ for the unbalanced interconnected community structure, the hierarchical community structure, and the independent community structure.
			A: the unbalanced interconnected community structure, from left to right: the population, the sample, and the estimated covariance matrices; 
			B: the hierarchical community structure, from left to right: the population, the sample, and the estimated covariance matrices;
			C: the independent structure, from left to right: the population, the sample, and the estimated covariance matrices.}
		\label{Fig:alternative}
	\end{figure}

	We then proceed to generate datasets of size $n = 50$ with a mean vector $\bm{\mu} = \textbf{0}_{p \times 1}$. 
	Subsequently, multivariate normal distributed datasets of size $n$ are generated with mean $\bm{\mu}$ and covariance matrices $\bm{\Sigma}_{0, \text{U}}$, $\bm{\Sigma}_{0, \text{I}}$, and $\bm{\Sigma}_{0, \text{H}}$, respectively.
	We apply the proposed approach to estimate the covariance parameters for each structure.
	Across $1000$ replicates, we evaluate the average relative bias, Monte Carlo standard deviation, average standard error, and empirical covariance probability based on $95\%$ Wald-type confidence intervals for each parameter. 
	The results are demonstrated in Table~\ref{Tab:alternative}.  
	Furthermore, we visually represented the population, sample, and estimated covariance matrices for three structures in Figure~\ref{Fig:alternative}.
	
	\begin{table}[!h]
		\footnotesize
		\begin{tabular}{crrrrrrrrrrrrrr}
			\hline
			& \multicolumn{4}{c}{Unbalanced} & \multicolumn{1}{c}{} & \multicolumn{4}{c}{Hierarchical} & \multicolumn{1}{c}{} & \multicolumn{4}{c}{Independent} \\ \cline{2-5} \cline{7-10} \cline{12-15} 
			& \multicolumn{1}{c}{bias} & \multicolumn{1}{c}{MCSD} & \multicolumn{1}{c}{ASE} & \multicolumn{1}{c}{$95\%$ CP} & \multicolumn{1}{c}{} & \multicolumn{1}{c}{bias} & \multicolumn{1}{c}{MCSD} & \multicolumn{1}{c}{ASE} & \multicolumn{1}{c}{$95\%$ CP} & \multicolumn{1}{c}{} & \multicolumn{1}{c}{bias} & \multicolumn{1}{c}{MCSD} & \multicolumn{1}{c}{ASE} & \multicolumn{1}{c}{$95\%$ CP} \\ \hline
			$a_{0,11}$ & 0.0 & 0.1 & 0.1 & 93.9 &  & 0.0 & 0.1 & 0.1 & 93.9 &  & 0.0 & 0.1 & 0.1 & 93.9 \\
			$a_{0,22}$ & 0.0 & 1.2 & 1.2 & 94.8 &  & 0.0 & 1.2 & 1.2 & 94.8 &  & 0.0 & 1.2 & 1.2 & 94.8 \\
			$a_{0,33}$ &-0.2 & 4.0 & 4.0 & 93.9 &  & -0.3 & 4.0 & 4.0 & 95.0 &  & -0.3 & 4.0 & 4.0 & 95.0 \\
			$a_{0,44}$ & 0.0 & 0.3 & 0.3 & 94.3 &  & 0.0 & 0.3 & 0.3 & 94.3 &  & 0.0 & 0.3 & 0.3 & 94.3 \\
			$a_{0,55}$ & 0.0 & 0.4 & 0.4 & 95.1 &  & 0.0 & 0.4 & 0.4 & 95.1 &  & 0.0 & 0.4 & 0.4 & 95.1 \\
			$b_{0,11}$ & 3.9 &137.3 &136.8 & 94.0 &  & 7.7 & 138.9 & 137.6 & 93.7 &  & 6.3 & 132.9 & 137.3 & 94.2 \\
			$b_{0,12}$ &-3.1 & 87.6 & 88.7 & 94.9 &  & -4.1 & 91.8 & 88.5 & 93.9 &  & -4.7 & 85.1 & 85.0 & 96.2 \\
			$b_{0,13}$ & 0.7 & 78.1 & 78.5 & 95.9 &  & 2.0 & 80.3 & 78.6 & 95.5 &  & -2.1 & 77.7 & 77.9 & 96.2 \\
			$b_{0,14}$ & 0.2 &103.7 &105.2 & 94.3 &  & -2.0 & 95.4 & 97.4 & 96.5 &  & -0.1 & 101.0 & 97.2 & 94.9 \\
			$b_{0,15}$ &-3.4 & 78.5 & 78.2 & 93.3 &  & -6.6 & 75.0 & 74.2 & 96.0 &  & 3.2 & 73.7 & 73.9 & 96.3 \\
			$b_{0,22}$ & 4.2 &108.3 &106.5 & 94.2 &  & -2.6 & 105.4 & 105.1 & 92.3 &  & -1.5 & 105.0 & 105.3 & 92.2 \\
			$b_{0,23}$ & 2.1 & 88.1 & 87.7 & 94.2 &  & -0.6 & 66.6 & 68.7 & 96.8 &  & 2.6 & 68.9 & 68.1 & 95.7 \\
			$b_{0,24}$ &-2.4 & 86.7 & 86.6 & 95.0 &  & 2.1 & 84.9 & 85.3 & 95.6 &  & -3.1 & 84.7 & 85.1 & 96.8 \\
			$b_{0,25}$ & 1.5 & 66.1 & 65.7 & 95.8 &  & 1.0 & 63.4 & 65.0 & 96.6 &  & -4.0 & 66.2 & 64.8 & 95.5 \\
			$b_{0,33}$ & 1.2 & 88.3 & 88.7 & 92.7 &  & 2.1 & 91.1 & 88.9 & 93.3 &  & -0.8 & 88.2 & 88.3 & 92.7 \\
			$b_{0,34}$ &-1.0 & 92.1 & 91.3 & 93.8 &  & 1.4 & 81.0 & 78.9 & 95.8 &  & -2.4 & 79.3 & 77.9 & 95.4 \\
			$b_{0,35}$ & 2.2 & 59.5 & 59.3 & 95.4 &  & -1.6 & 59.6 & 60.3 & 95.3 &  & -3.8 & 59.2 & 59.3 & 96.3 \\
			$b_{0,44}$ &-1.5 &136.1 &136.9 & 93.5 &  & 0.7 & 138.1 & 137.3 & 93.3 &  & 1.5 & 139.4 & 137.5 & 92.4 \\
			$b_{0,45}$ &-2.4 & 71.0 & 73.6 & 96.2 &  & 2.0 & 74.6 & 74.7 & 96.2 &  & 0.5 & 69.4 & 73.9 & 97.0 \\
			$b_{0,55}$ &-4.0 & 79.4 & 79.2 & 93.3 &  & -2.7 & 77.6 & 79.4 & 92.6 &  & -1.7 & 79.6 & 79.6 & 92.8 \\ \hline
		\end{tabular}
		\caption{We assess the estimated covariance parameters ($a_{0,kk}$ and $b_{0,kk'}$) using our proposed method across
			$1000$ simulated datasets under $n = 50$ for the unbalanced interconnected community structure, the hierarchical community structure, and the independent community structure: 
			``bias" represents the relative bias ($\times 100$); 
			``MCSD" refers to the Monte Carlo standard deviation ($\times 100$); 
			``ASE" stands for the average standard error ($\times 100$); 
			``$95\%$ CP" represents the empirical coverage probability (in $\%$) based on a $95\%$ Wald-type confidence interval.}
		\label{Tab:alternative} 
	\end{table}

	To conduct the simulation studies in Section 3 and Section~\ref{Sec:simulation}, we utilized the \proglang{R} packages \pkg{pfa} \citep{FanHanGu2012} (with some modifications),  \pkg{CovTools} \citep{RLeeLinYou2021}, \pkg{POET} \citep{RFanLiaoMincheva2016}, \pkg{CVTuningCov} \citep{RWang2015} (with some modifications).
	The estimation procedures were implemented using the \proglang{R} software \citep{R}.
	The \proglang{R} code used for simulation studies in Section 3 and real data analyses in Section 4 can be accessed at \url{https://github.com/yiorfun/UBCovEst}.

	\newpage
	
	\section{PROOFS}
	
	\label{Sec:proof}
	
	\subsection{Proofs of Lemma, Corollary, Corollary~\ref{Cor:ops}}
	
	Please refer to the arguments and proofs in the work of \citet{YangChenLee2024} for further details.
	
	\subsection{Derivations of maximum likelihood estimators and their properties}
	
	We start with equation (4), 
	\begin{equation*}
		\begin{split}
			%\label{Eq:score}
			\frac{\partial}{\partial \bm{\theta}_j} \ell_n (\bm{\theta}; \textbf{X})
			= \frac{n}{2} \operatorname{tr}\left[\{\bm{\Sigma}(\textbf{A}, \textbf{B}, \bm{p}) - (\textbf{S}_{kk'})\}
			\left\{ \frac{\partial \textbf{A}_{\bm{\Omega}}}{\partial \bm{\theta}_j} \circ \textbf{I}(\bm{p}) 
			+ \frac{\partial \textbf{B}_{\bm{\Omega}}}{\partial \bm{\theta}_j} \circ \textbf{J}(\bm{p}) \right\} \right].
		\end{split}
	\end{equation*}
	Let $(\textbf{M}_{kk'})$ denote $\textbf{M} = \bm{\Sigma}(\textbf{A}, \textbf{B}, \bm{p}) - (\textbf{S}_{kk'})$ partitioned by $\bm{p}$.
	Then, the system of score equations is
	\begin{equation}
		\label{Eq:score}
		\textbf{S}_n \left(\bm{\theta}; \textbf{X}\right) = \frac{n}{2} 
		\begin{pmatrix}
			\operatorname{tr}\left \{ (\textbf{M}_{kk'}) \frac{\partial \bm{\Omega}(\textbf{A}_{\bm{\Omega}}, \textbf{B}_{\bm{\Omega}}, \bm{p})}{\partial a_{11}} \right \} \\
			\operatorname{tr}\left \{ (\textbf{M}_{kk'}) \frac{\partial \bm{\Omega}(\textbf{A}_{\bm{\Omega}}, \textbf{B}_{\bm{\Omega}}, \bm{p})}{\partial a_{22}} \right \} \\
			\vdots \\
			\operatorname{tr}\left \{ (\textbf{M}_{kk'}) \frac{\partial \bm{\Omega}(\textbf{A}_{\bm{\Omega}}, \textbf{B}_{\bm{\Omega}}, \bm{p})}{\partial a_{KK}} \right \} \\
			\operatorname{tr}\left \{ (\textbf{M}_{kk'}) \frac{\partial \bm{\Omega}(\textbf{A}_{\bm{\Omega}}, \textbf{B}_{\bm{\Omega}}, \bm{p})}{\partial b_{11}} \right \} \\
			\vdots \\
			\operatorname{tr}\left \{ (\textbf{M}_{kk'}) \frac{\partial \bm{\Omega}(\textbf{A}_{\bm{\Omega}}, \textbf{B}_{\bm{\Omega}}, \bm{p})}{\partial b_{KK}} \right \}
		\end{pmatrix} = \textbf{0}_{q \times 1}.
	\end{equation}
	Recall the notations $\textbf{P} = \operatorname{diag}(p_1, \ldots, p_K)$ and $\bm{\Delta} = \textbf{A} + \textbf{B} \textbf{P}$. 
	Using a fact that $(\textbf{A} \textbf{P}) ^ {-1} - \bm{\Delta} \textbf{B} \textbf{A} ^{-1} = (\textbf{P} \bm{\Delta}) ^ {-1}$ and doing some algebra,
	we obtain the following derivatives,
	\begin{equation*}
		\begin{split}
			\frac{\partial \textbf{A}_{\bm{\Omega}}}{\partial a_{kk}} & = - a_{kk} ^ {-2} \textbf{E}_{kk}, \quad
			\frac{\partial \textbf{B}_{\bm{\Omega}}}{\partial a_{kk}} = \bm{\Delta} ^ {-1} \textbf{E}_{kk} \bm{\Delta} ^ {-1} \textbf{B} \textbf{A} ^{-1} + a_{kk} ^ {-2} \bm{\Delta} ^ {-1} \textbf{B} \textbf{E}_{kk}, \quad \text{for every $k$,} \\
			\frac{\partial \textbf{A}_{\bm{\Omega}}}{\partial b_{kk'}} & = \textbf{0}_{K \times K}, \quad
			\frac{\partial \textbf{B}_{\bm{\Omega}}}{\partial b_{kk'}} = \begin{cases}
				- \bm{\Delta} ^ {-1} \textbf{E}_{kk} \textbf{P} \bm{\Delta} ^ {-1} \textbf{P} ^ {- 1}, & k = k' \\
				- \bm{\Delta} ^ {-1} (\textbf{E}_{kk'} + \textbf{E}_{k'k}) \textbf{P} \bm{\Delta} ^ {-1} \textbf{P} ^ {-1}, & k \neq k'
			\end{cases}, \quad \text{for every $k$ and $k'$.}
		\end{split}
	\end{equation*}
	Thus, the individual equations in~\eqref{Eq:score} can be simplified as 
	\begin{equation*}
		\begin{split}
			\operatorname{tr}\left\{(\textbf{M}_{kk'}) \frac{\partial \bm{\Omega}(\textbf{A}_{\bm{\Omega}}, \textbf{B}_{\bm{\Omega}}, \bm{p})}{\partial a_{kk}} \right\}
			& = \left(- a_{kk}^{-2}\right)\operatorname{tr}\left(\textbf{M}_{kk}\right)
			+ \sum_{\ell = 1}^{K}\sum_{\ell' = 1} ^ {K} \operatorname{sum}( \textbf{M}_{\ell \ell'}) \left( \frac{\partial \textbf{B}_{\bm{\Omega}}}{\partial a_{kk}}\right)_{\ell', \ell} = 0, \\ %\ \ \ \ k = 1, 2, \ldots, K; \\
			%%%%
			\operatorname{tr}\left\{(\textbf{M}_{kk'}) \frac{\partial \bm{\Omega}(\textbf{A}_{\bm{\Omega}}, \textbf{B}_{\bm{\Omega}}, \bm{p})}{\partial b_{kk'}} \right\}
			& = \sum_{\ell = 1}^{K}\sum_{\ell' = 1} ^ {K} \operatorname{sum}( \textbf{M}_{\ell \ell'} ) \left( \frac{\partial \textbf{B}_{\bm{\Omega}}}{\partial b_{kk'}}\right)_{\ell', \ell} = 0,  %\ \ \ \ k, k' = 1, 2, \ldots, K.
		\end{split}
	\end{equation*}
	for every $k$ and $k'$, where $(\partial \textbf{B}_{\bm{\Omega}}/\partial a_{kk}), (\partial \textbf{B}_{\bm{\Omega}} / \partial b_{kk'}) \in \mathbb{R} ^ {K \times K}$ and the subscript $(\ell', \ell)$ denotes the $(\ell', \ell)$ element of $(\partial \textbf{B}_{\bm{\Omega}}/\partial a_{kk})$ or $(\partial \textbf{B}_{\bm{\Omega}} / \partial b_{kk'})$. 
	
	Now, we claim the system of score equations in~\eqref{Eq:score} has a unique solution.
	By definition, $\textbf{A} = \operatorname{diag}\left(a_{11}, \ldots, a_{KK}\right)$, then $- \textbf{A} ^{-2} = \operatorname{diag}\left(- a_{11} ^ {-2}, \ldots, - a_{KK} ^ {-2} \right)$, denoted by $\textbf{A} ^\star$. 
	Let $\bm{\beta} ^ \star \in \mathbb{R} ^ {q}$ denote a vector as below,
	\begin{equation*}
		\begin{split}
			\bm{\beta} ^ \star = 
			\left\{
			\operatorname{tr}(\textbf{M}_{11}), 
			\ldots,
			\operatorname{tr}(\textbf{M}_{KK}), 
			\operatorname{sum}(\textbf{M}_{11}),
			\ldots,
			\operatorname{sum}(\textbf{M}_{1K}),
			\operatorname{sum}(\textbf{M}_{22}),
			\ldots,
			\operatorname{sum}(\textbf{M}_{KK})
			\right \} ^\top.
		\end{split}	 
	\end{equation*}
	Let $\textbf{B} ^{(1), \star} \in \mathbb{R} ^ {K \times (q - K)}$ denote a matrix with $k$th row 
	\begin{equation*}
		\begin{split}
			\bigg\{
			& \left( \frac{\partial \textbf{B}_{\bm{\Omega}}}{\partial a_{kk}}\right)_{1, 1}, 
			\left( \frac{\partial \textbf{B}_{\bm{\Omega}}}{\partial a_{kk}}\right)_{1, 2} + \left( \frac{\partial \textbf{B}_{\bm{\Omega}}}{\partial a_{kk}}\right)_{2, 1}, 
			\ldots, 
			\left( \frac{\partial \textbf{B}_{\bm{\Omega}}}{\partial a_{kk}}\right)_{1, K} + \left( \frac{\partial \textbf{B}_{\bm{\Omega}}}{\partial a_{kk}}\right)_{K, 1}, 
			\left( \frac{\partial \textbf{B}_{\bm{\Omega}}}{\partial a_{kk}}\right)_{2, 2}, 
			\ldots, \\
			& \left( \frac{\partial \textbf{B}_{\bm{\Omega}}}{\partial a_{kk}}\right)_{K - 1, K} + \left( \frac{\partial \textbf{B}_{\bm{\Omega}}}{\partial a_{kk}}\right)_{K, K - 1}, 
			\left( \frac{\partial \textbf{B}_{\bm{\Omega}}}{\partial a_{kk}}\right)_{K, K}
			\bigg \}, 
		\end{split}
	\end{equation*}
	for $k = 1, \ldots, K$, and let $\textbf{B} ^{(2), \star} \in \mathbb{R} ^ {(q - K) \times (q - K)}$ denote a matrix with rows 
	\begin{equation*}
		\begin{split}
			\bigg\{
			& \left( \frac{\partial \textbf{B}_{\bm{\Omega}}}{\partial b_{kk'}}\right)_{1, 1}, 
			\left( \frac{\partial \textbf{B}_{\bm{\Omega}}}{\partial b_{kk'}}\right)_{1, 2} + \left( \frac{\partial \textbf{B}_{\bm{\Omega}}}{\partial b_{kk'}}\right)_{2, 1}, 
			\ldots, 
			\left( \frac{\partial \textbf{B}_{\bm{\Omega}}}{\partial b_{kk'}}\right)_{1, K} + \left( \frac{\partial \textbf{B}_{\bm{\Omega}}}{\partial b_{kk'}}\right)_{K, 1}, 
			\left( \frac{\partial \textbf{B}_{\bm{\Omega}}}{\partial b_{kk'}}\right)_{2, 2}, 
			\ldots, \\
			& \left( \frac{\partial \textbf{B}_{\bm{\Omega}}}{\partial b_{kk'}}\right)_{K - 1, K} + \left( \frac{\partial \textbf{B}_{\bm{\Omega}}}{\partial b_{kk'}}\right)_{K, K - 1}, 
			\left( \frac{\partial \textbf{B}_{\bm{\Omega}}}{\partial b_{kk'}}\right)_{K, K}
			\bigg\}
		\end{split}
	\end{equation*}
	for every $k \leq k'$. 
	Thus, the system of score equations in~\eqref{Eq:score} can be rewritten as 
	\begin{equation*}
		\textbf{S}_n\left( \bm{\theta}; \textbf{X} \right) = \begin{pmatrix}
			\textbf{A} ^\star & \textbf{B} ^{(1), \star} \\ 
			\textbf{0}_{(q - K) \times K} & \textbf{B} ^{(2), \star}
		\end{pmatrix}
		\bm{\beta} ^ \star = \textbf{0}_{q \times 1}, 
	\end{equation*}
	and the $q$ by $q$ coefficient matrix has $\operatorname{rank}\begin{pmatrix}
		\textbf{A} ^\star & \textbf{B} ^{(1), \star} \\ 
		\textbf{0}_{(q - K) \times K} & \textbf{B} ^{(2), \star}
	\end{pmatrix} = \operatorname{rank}(\textbf{A} ^\star) + \operatorname{rank}(\textbf{B} ^{(2), \star})$ due to $\textbf{I}_K - \textbf{A} ^{\star} (\textbf{A} ^{\star}) ^ {-1} = \textbf{0}_{K \times K}$, which satisfies the condition provided in \citet[Theorem 3.10, p.334]{BuaphimOnsaardSongoen2018}.
	
	On the one hand, $\operatorname{rank}(\textbf{A} ^\star) = K$ by the positive definiteness that $a_{kk} > 0$ for every $k$.
	On the other hand, to compute $\operatorname{rank}(\textbf{B} ^{(2), \star})$, 
	use the fact $\operatorname{vech}(A X B) = (B ^\top \otimes A) \times \operatorname{vech}(X)$, where the matrices $A, B$ and $X$ with suitable sizes, and $\otimes$ denotes the Kronecker product.
	Since 
	\begin{equation*}
		\frac{\partial \textbf{B}_{\bm{\Omega}}}{\partial b_{kk'}} = \begin{cases}
			- \bm{\Delta} ^ {-1} \textbf{E}_{kk} \textbf{P} \bm{\Delta} ^ {-1} \textbf{P} ^ {- 1}, & k = k' \\
			- \bm{\Delta} ^ {-1} (\textbf{E}_{kk'} + \textbf{E}_{k'k}) \textbf{P} \bm{\Delta} ^ {-1} \textbf{P} ^ {-1}, & k \neq k'
		\end{cases},
	\end{equation*}
	for every $k$ and $k'$,
	where $\textbf{E}_{kk}$ and $\textbf{E}_{kk'} + \textbf{E}_{k'k}$ span the entire matrix space of $K$ by $K$ symmetric matrices. 
	Thus, $\operatorname{rank}(\textbf{B} ^{(2), \star}) = (K + 1) K / 2$ due to the equivalence of the matrix space and the vector space spanned by the matrices.
	Therefore, the coefficient matrix of the given homogeneous system of linear equations has full rank and has a unique solution $\bm{\beta} ^ \star = \textbf{0}_{q \times 1}$ with the definition $q = K + (K + 1) K / 2$.
	Finally, the solution to the system in~\eqref{Eq:score}, denoted by $\widetilde{\bm{\theta}}$, must be the maximum likelihood estimator because the system of score equations has a unique solution.
	
	Following~\eqref{Eq:score}, the first-order partial derivative of the score function with respect to $\bm{\theta}$ is 
	\begin{equation}
		\label{Eq:hess}
		\frac{\partial \textbf{S}_n\left(\bm{\theta}; \textbf{X}\right)}{\partial \bm{\theta}}
		= \begin{pmatrix}
			\dfrac{\partial ^ 2 \ell_n(\bm{\theta}; \textbf{X})}{\partial \bm{\theta} ^ {\text{(A)}} \partial \bm{\theta} ^ {\text{(A)},\top}} &
			\dfrac{\partial ^ 2 \ell_n(\bm{\theta}; \textbf{X})}{\partial \bm{\theta} ^ {\text{(A)}} \partial \bm{\theta} ^ {\text{(B)},\top}} \\
			\dfrac{\partial ^ 2 \ell_n(\bm{\theta}; \textbf{X})}{\partial \bm{\theta} ^ {\text{(B)}} \partial \bm{\theta} ^ {\text{(A)},\top}} & 
			\dfrac{\partial ^ 2 \ell_n(\bm{\theta}; \textbf{X})}{\partial \bm{\theta} ^ {\text{(B)}} \partial \bm{\theta} ^ {\text{(B)},\top}}
		\end{pmatrix} 
		= \frac{n}{2} \begin{pmatrix}
			\textbf{H}_1 & \textbf{H}_2 \\ 
			\textbf{H}_2 ^\top & \textbf{H}_3
		\end{pmatrix},
	\end{equation}
	where $\bm{\theta} ^ {\text{(A)}} = (a_{11}, \ldots, a_{KK}) ^\top$, $\bm{\theta} ^ {\text{(B)}} = (b_{11}, \ldots, b_{1K}, b_{22}, \ldots, b_{KK}) ^\top$, and therefore $\bm{\theta} = (\bm{\theta} ^ {\text{(A)},\top}, \bm{\theta} ^ {\text{(B)},\top}) ^\top$, 
	the blocks $\textbf{H}_1 \in \mathbb{R} ^ {K \times K}$, $\textbf{H}_2 \in \mathbb{R} ^ {K \times (q - K)}$, and $\textbf{H}_3 \in \mathbb{R} ^ {(q - K) \times (q - K)}$.
	In particular, given $k$ and $k'$, 
	\begin{equation*}
		\begin{split}
			\textbf{H}_1 = \begin{cases}
				2 a_{kk} ^ {- 3} \operatorname{tr}(\textbf{M}_{kk}) 
				- a_{kk} ^ {-2}  \left\{ \dfrac{\partial }{\partial a_{mm}} \operatorname{tr}(\textbf{M}_{kk}) \right\} \\
				%\quad \quad
				+ \sum_{\ell = 1}^{K} \sum_{\ell' = 1}^{K} 
				\left[ 
				\left\{ \dfrac{\partial }{\partial a_{mm}} \operatorname{sum}(\textbf{M}_{\ell \ell'})  \right\}
				\left(\dfrac{\partial \textbf{B}_{\bm{\Omega}}}{\partial a_{kk}} \right)_{\ell', \ell}
				+ \operatorname{sum}(\textbf{M}_{\ell \ell'})
				\left\{\dfrac{\partial }{\partial a_{mm}} \left(\dfrac{\partial \textbf{B}_{\bm{\Omega}}}{\partial a_{kk}} \right) \right\}_{\ell', \ell}
				\right], & m = k \\
				\sum_{\ell = 1}^{K} \sum_{\ell' = 1}^{K} 
				\left[ 
				\left\{ \dfrac{\partial }{\partial a_{mm}} \operatorname{sum}(\textbf{M}_{\ell \ell'})  \right\}
				\left(\dfrac{\partial \textbf{B}_{\bm{\Omega}}}{\partial a_{kk}} \right)_{\ell', \ell}
				+ \operatorname{sum}(\textbf{M}_{\ell \ell'})
				\left\{\dfrac{\partial }{\partial a_{mm}} \left(\dfrac{\partial \textbf{B}_{\bm{\Omega}}}{\partial a_{kk}} \right) \right\}_{\ell', \ell}
				\right], & m \neq k
			\end{cases}
		\end{split}
	\end{equation*}
	for $m = 1, 2, \ldots, K$; and 
	\begin{equation*}
		\begin{split}
			\textbf{H}_2
			& = 
			\sum_{\ell = 1}^{K} \sum_{\ell' = 1}^{K} 
			\left[
			\left\{ \dfrac{\partial }{\partial a_{mm}} \operatorname{sum}(\textbf{M}_{\ell \ell'})  \right\}
			\left(\dfrac{\partial \textbf{B}_{\bm{\Omega}}}{\partial b_{kk'}} \right)_{\ell', \ell}
			+ \operatorname{sum}(\textbf{M}_{\ell \ell'})
			\left\{\dfrac{\partial }{\partial a_{mm}} \left(\dfrac{\partial \textbf{B}_{\bm{\Omega}}}{\partial b_{kk'}} \right) \right\}_{\ell', \ell}
			\right], \\
			%%%
			\textbf{H}_3
			& = \sum_{\ell = 1}^{K} \sum_{\ell' = 1}^{K} 
			\left[ 
			\left\{\dfrac{\partial }{\partial b_{mm'}} \operatorname{sum}(\textbf{M}_{\ell \ell'})  \right\}
			\left(\dfrac{\partial \textbf{B}_{\bm{\Omega}}}{\partial b_{kk'}} \right)_{\ell', \ell}
			+ \operatorname{sum}(\textbf{M}_{\ell \ell'})
			\left\{\dfrac{\partial }{\partial b_{mm'}} \left(\dfrac{\partial \textbf{B}_{\bm{\Omega}}}{\partial b_{kk'}} \right) \right\}_{\ell', \ell}
			\right],
		\end{split}
	\end{equation*}
	for $m, m' = 1, 2, \ldots, K$.
	The first-order partial derivatives $\partial \operatorname{sum}(\textbf{M}_{\ell \ell'}) / \partial a_{mm}$ and $\partial \operatorname{sum}(\textbf{M}_{\ell \ell'}) / \partial b_{mm'}$ are easily computed, where $\textbf{M}_{\ell \ell'} = \bm{\Sigma}_{\ell \ell'} - \textbf{S}_{\ell \ell'}$ for $m, m', \ell, \ell' = 1, 2, \ldots, K$.
	The second-order partial derivatives $\partial (\partial \textbf{B}_{\bm{\Omega}} / \partial a_{kk}) / \partial a_{mm}$, 
	$\partial (\partial \textbf{B}_{\bm{\Omega}} / \partial b_{kk'}) / \partial a_{mm}$, 
	and $\partial (\partial \textbf{B}_{\bm{\Omega}} / \partial b_{kk'}) / \partial b_{mm'}$ are matrices for $m, m', k, k' = 1, 2, \ldots, K$, shown respectively as below. 
	\begin{equation*}
		\begin{split}
			\frac{\partial }{\partial a_{mm}} \left( \frac{\partial \textbf{B}_{\bm{\Omega}}}{\partial a_{kk}} \right) 
			= \begin{cases}
				- \bm{\Delta} ^ {- 1} \textbf{E}_{kk} \bm{\Delta} ^ {- 1} \left( 2 \textbf{E}_{kk} \bm{\Delta} ^ {-1} \textbf{B} \textbf{A} ^{-1} + \textbf{B} \textbf{A} ^{- 1} \textbf{E}_{kk} \textbf{A} ^{- 1} \right) \\
				\quad \quad - \left( 2 a_{kk} ^ {- 3} \textbf{I}_K + a_{kk} ^ {- 2}  \bm{\Delta} ^ {- 1} \textbf{E}_{kk} \right) \bm{\Delta} ^ {- 1} \textbf{B} \textbf{E}_{kk}
				, & k = m \\
				- \bm{\Delta} ^ {- 1} \left( \textbf{E}_{mm} \bm{\Delta} ^ {- 1} \textbf{E}_{kk} + \textbf{E}_{kk} \bm{\Delta} ^ {- 1} \textbf{E}_{mm} \right)\bm{\Delta} ^ {-1} \textbf{B} \textbf{A} ^{-1} \\
				\quad \quad  - \bm{\Delta} ^ {- 1} \textbf{E}_{kk} \bm{\Delta} ^ {-1} \textbf{B} \textbf{A} ^{- 1} \textbf{E}_{mm} \textbf{A} ^{- 1} 
				- a_{kk} ^ {- 2} \bm{\Delta} ^ {- 1} \textbf{E}_{mm} \bm{\Delta} ^ {- 1} \textbf{B} \textbf{E}_{kk}
				, & k \neq m
			\end{cases},
		\end{split}
	\end{equation*}
	and, 
	\begin{equation*}
		\begin{split}
			%%%
			\frac{\partial }{\partial a_{mm}} \left( \frac{\partial \textbf{B}_{\bm{\Omega}}}{\partial b_{kk'}} \right) 
			= \begin{cases}
				\bm{\Delta} ^ {- 1} \textbf{E}_{mm}  \bm{\Delta} ^ {- 1} \textbf{E}_{kk} \textbf{P} \bm{\Delta} ^ {-1} \textbf{P} \\
				\quad \quad 
				- \bm{\Delta} ^ {- 1} \textbf{E}_{kk} \textbf{P} \bm{\Delta} ^ {- 1} ( \textbf{E}_{mm} \bm{\Delta} ^ {- 1} \textbf{B} - \textbf{P} \textbf{E}_{mm} ) \textbf{A} ^{-1}, & k = k' \\
				\bm{\Delta} ^ {- 1} \textbf{E}_{mm} \bm{\Delta} ^ {- 1} \left( \textbf{E}_{kk'} + \textbf{E}_{k'k} \right)
				\textbf{P} \bm{\Delta} ^ {-1} \textbf{P} \\
				\quad \quad
				- \bm{\Delta} ^ {- 1} \left( \textbf{E}_{kk'} + \textbf{E}_{k'k} \right) \textbf{P} \bm{\Delta} ^ {- 1} (\textbf{E}_{mm} \bm{\Delta} ^ {- 1} \textbf{B} - \textbf{P} \textbf{E}_{mm} ) \textbf{A} ^{-1}, & k \neq k'
			\end{cases} 
		\end{split}
	\end{equation*}
	and, 
	\begin{equation*}
		\begin{split}
			& \frac{\partial }{\partial b_{mm'}} \left( \frac{\partial \textbf{B}_{\bm{\Omega}}}{\partial b_{kk'}} \right) \\
			& = \begin{cases}
				\begin{cases}
					\bm{\Delta} ^ {- 1}
					(\textbf{E}_{mm} + \textbf{E}_{kk}) \textbf{P} \bm{\Delta} ^ {- 1} (\textbf{E}_{kk} + \textbf{E}_{mm})
					\textbf{P} \bm{\Delta} ^ {-1} \textbf{P}, & m = m' \\
					\bm{\Delta} ^ {- 1}
					(\textbf{E}_{mm'} + \textbf{E}_{m'm} + \textbf{E}_{kk}) \textbf{P} \bm{\Delta} ^ {- 1} (\textbf{E}_{kk} + \textbf{E}_{mm'} + \textbf{E}_{m'm})
					\textbf{P} \bm{\Delta} ^ {-1} \textbf{P}, & m \neq m'
				\end{cases}, & k = k' \\
				\begin{cases}
					\bm{\Delta} ^ {- 1}
					(\textbf{E}_{mm} + \textbf{E}_{kk'} + \textbf{E}_{k'k}) \textbf{P} \bm{\Delta} ^ {- 1} (\textbf{E}_{kk'} + \textbf{E}_{k'k} + \textbf{E}_{mm})
					\textbf{P} \bm{\Delta} ^ {-1} \textbf{P}, & m = m' \\	
					\bm{\Delta} ^ {- 1}
					(\textbf{E}_{mm'} + \textbf{E}_{m'm} + \textbf{E}_{kk'} + \textbf{E}_{k'k}) \textbf{P} \bm{\Delta} ^ {- 1} (\textbf{E}_{kk'} + \textbf{E}_{k'k} + \textbf{E}_{mm'} + \textbf{E}_{m'm})
					\textbf{P} \bm{\Delta} ^ {-1} \textbf{P}, & m \neq m'
				\end{cases}, & k \neq k'
			\end{cases}.
		\end{split}
	\end{equation*}
	
	By the unbiasedness from Theorem 1, the expectations of $\textbf{H}_1$, $\textbf{H}_2$, and $\textbf{H}_3$ depend only on the terms $\operatorname{E}(\operatorname{sum}(\textbf{M}_{\ell \ell})) = (1 - c_n) p_{\ell}(a_{\ell \ell} + b_{\ell \ell} p_{\ell})$, 
	$\operatorname{E}(\operatorname{sum}(\textbf{M}_{\ell \ell'})) = (1 - c_n)p_{\ell} p_{\ell'} b_{\ell \ell'}$ and $\operatorname{E}(\operatorname{tr}(\textbf{M}_{\ell \ell})) = (1 - c_n ) p_{\ell} (a_{\ell} + b_{\ell})$, 
	where $c_n = (n - 1) / n$.
	Then Fisher's information matrices for $n$ observations, denoted as $\mathbb{I}_n(\bm{\theta})$, and one observation, $\mathbb{I}_1(\bm{\theta})$, can be calculated by $\mathbb{I}_n(\bm{\theta}) = - \operatorname{E}(\partial \textbf{S}_n(\bm{\theta}; \textbf{X}) / \partial \bm{\theta})$ and $\mathbb{I}_1(\bm{\theta}) = \mathbb{I}_n(\bm{\theta}) / n$, respectively.
	Furthermore, since both the determinant operator and the trace operator are continuous, 
	the log-likelihood function $\ell_n(\bm{\theta}; \textbf{X})$ is continuous with respect to $\bm{\theta}$.
	In addition, there is a unique solution to the score equation for every $n$.
	Thus, $\widetilde{\bm{\theta}}$ is strongly consistent, asymptotically efficient, and asymptotically normal, 
	following the classical arguments in \citet{Ferguson1996, vanderVaartWellner1996, StuartOrdArnold1999, BickelDoksum20151, BickelDoksum20152}, 
	i.e., $\widetilde{\bm{\theta}} \to \bm{\theta}$ almost surely as $n \to \infty$ and $\sqrt{n}(\widetilde{\bm{\theta}} - \bm{\theta}) \to N(\textbf{0}_{q \times 1}, \mathbb{I}_1^{-1}(\bm{\theta}))$ in distribution as $n \to \infty$.

	\subsection{Proof of Theorem 1}
	
	\begin{proof}[Proof of Theorem 1]
		
		Consider the i.i.d. copy $\bm{X}_{1}, \ldots, \bm{X}_n$ of $p$-dimensional $\bm{X} \sim N (\bm{\mu}, \bm{\Sigma}(\textbf{A}, \textbf{B}, \bm{p}))$, 
		where $\bm{\Sigma}(\textbf{A}, \textbf{B}, \bm{p})$ is a uniform-block matrix with a $K$ by $K$ diagonal matrix $\textbf{A}$, a $K$ by $K$ symmetric matrix $\textbf{B}$, 
		and a predetermined $K$-dimensional $\bm{p} = (p_1, \ldots, p_K) ^\top$ partition-size vector satisfying that $p_k > 1$ for every $k$ and $p = p_1 + \cdots + p_K$.
		Let $\bar{\bm{X}} = n ^ {-1} (\bm{X}_1 + \cdots + \bm{X}_n)$ denote the sample mean.
		Let $\bm{X}_{i, p_k}, \bar{\bm{X}}_{p_k}, \bm{\mu}_{p_k} \in \mathbb{R} ^ {p_k}$ satisfying $\bm{X}_{i} = (\bm{X}_{i, p_1} ^\top, \ldots, \bm{X}_{i, p_K} ^\top) ^\top$, $\bar{\bm{X}} = (\bar{\bm{X}}_{p_1} ^\top, \ldots, \bar{\bm{X}}_{p_K} ^\top) ^\top$, and $\bm{\mu} = (\bm{\mu}_{p_1} ^\top, \ldots, \bm{\mu}_{p_K} ^\top) ^\top$, respectively.
		
		\emph{Unbiasedness.}
		We firstly prove that $\widetilde{\bm{\theta}}$ is unbiased.
		By $\bm{\Sigma}(\textbf{A}, \textbf{B}, \bm{p}) = (\bm{\Sigma}_{kk'})$, we can obtain that  
		\begin{equation*}
			\begin{split}
				\bm{\Sigma}_{kk'} 
				& = \operatorname{cov}(\bm{X}_{p_k}, \bm{X}_{p_{k'}})
				= \operatorname{E} \{(\bm{X}_{p_k} - \bm{\mu}_{p_k})(\bm{X}_{p_{k'}} - \bm{\mu}_{p_{k'}}) ^\top\}, \\
				\bm{\Sigma}_{kk'} / n
				& = \operatorname{cov}(\bar{\bm{X}}_{p_k}, \bar{\bm{X}}_{p_{k'}})
				= \operatorname{E} \{(\bar{\bm{X}}_{p_k} - \bm{\mu}_{p_k})(\bar{\bm{X}}_{p_{k'}} - \bm{\mu}_{p_{k'}}) ^\top\}.
			\end{split}
		\end{equation*}
		for every $k$ and $k'$.
		Let $\textbf{C}_{kk'} = \sum_{i = 1}^{n} (\bm{X}_{i, p_k} - \bar{\bm{X}}_{i, p_k})(\bm{X}_{i, p_{k'}} - \bar{\bm{X}}_{i, p_{k'}}) ^\top$ for every $k$ and $k'$, which can be simplified to
		\begin{equation*}
			\begin{split}
				\textbf{C}_{kk'} 
				= \sum_{i = 1}^{n} ( \bm{X}_{i, p_k} - \bm{\mu}_{p_k}) (\bm{X}_{i, p_{k'}} - \bm{\mu}_{p_{k'}}) ^\top
				- n (\bar{\bm{X}}_{p_k} - \bm{\mu}_{p_k}) (\bar{\bm{X}}_{p_{k'}} - \bm{\mu}_{p_{k'}}) ^\top.
			\end{split}
		\end{equation*}
		Taking expectation on the both sides, we have $\operatorname{E} \left(\textbf{C}_{kk'}\right) = (n - 1) \bm{\Sigma}_{kk'}$ for every $k$ and $k'$.
		Given $k$ and $k'$, by definition, $\bm{\Sigma}_{kk} = a_{kk} \textbf{I}_{p_k} + b_{kk} \textbf{J}_{p_k}$ for $k = k'$ and $\bm{\Sigma}_{kk'} = b_{kk'} \textbf{1}_{p_k \times p_{k'}}$.  
		Define 
		\begin{equation*}
			\begin{split}
				\alpha_{kk} = a_{kk} + b_{kk} =  \operatorname{tr}\left(\bm{\Sigma}_{kk'}\right) / p_k, \quad
				\beta_{kk'}  
				= \begin{cases}
					a_{kk} / p_k + b_{kk} = \operatorname{sum}\left(\bm{\Sigma}_{kk}\right) / p_k ^ 2, & k = k' \\
					b_{kk'} = \operatorname{sum}(\bm{\Sigma}_{kk'}) / (p_k p_{k'}), & k \neq k'
				\end{cases}.
			\end{split}	
		\end{equation*}
		By the definition of unbiased sample covariance matrix $(n - 1)\textbf{S}_{kk'} = \textbf{C}_{kk'}$, we have the following results: 
		\begin{equation*}
			\begin{split}
				\widetilde{\alpha}_{kk}  = p_k ^ {-1} \operatorname{tr}(\textbf{S}_{kk}), \quad
				\widetilde{\beta}_{kk} = p_k ^ {- 2} \operatorname{sum}(\textbf{S}_{kk}), \quad 
				\widetilde{\beta}_{kk'} = (p_k p_{k'}) ^ {-1} \operatorname{sum}(\textbf{S}_{kk'})
			\end{split}
		\end{equation*}
		are unbiased estimators for $\alpha_{kk}$, $\beta_{kk}$ for every $k$ and for $\beta_{kk'}$ for every $k \neq k'$.
		Furthermore, 
		\begin{equation*}
			\begin{split}
				\widetilde{a}_{kk} = \widetilde{\alpha}_{kk} - \frac{p_k \widetilde{\beta}_{kk} - \widetilde{\alpha}_{kk} }{p_k - 1}, \quad
				\widetilde{b}_{kk'} =
				\begin{cases}
					\dfrac{p_k \widetilde{\beta}_{kk} - \widetilde{\alpha}_{kk}}{p_k - 1}, & k = k' \\
					\widetilde{\beta}_{kk'}, & k \neq k'
				\end{cases},
			\end{split}
		\end{equation*}
		are unbiased estimators for $a_{kk}$ and $b_{kk'}$ every $k$ and $k'$.		
		
		\emph{Optimal property.}
		We proceed to establish the optimal property of $\widetilde{\bm{\theta}}$.
		Let $\bm{Y} = (\bm{X}_1 ^\top, \ldots, \bm{X}_n ^\top) ^\top \in \mathbb{R} ^ {pn}$ denote a vector consisting of all observations. 
		Under the normality assumption, 
		$\bm{Y} \sim N (\bm{\mu}_y, \textbf{V})$,
		where $\bm{\mu}_y = \left(\textbf{1}_{n} \otimes \textbf{I}_p \right) \bm{\mu} \in \mathbb{R} ^ {pn}$ represents the mean and $\textbf{V} = \textbf{I}_n \otimes \bm{\Sigma}(\textbf{A}, \textbf{B}, \bm{p}) \in \mathbb{R} ^ {pn \times pn}$ represents the covariance matrix.
		
		We begin by demonstrating that $\widetilde{\bm{\mu}} = \bar{\bm{X}}$ is the best linear unbiased estimator (BLUE).
		To establish this, 
		we verify the conditions outlined in Theorem 1 and Corollary 2 by \citet{Zmyslony1976}.
		Since $\bm{\Sigma}(\textbf{I}_K, \textbf{0}_{K \times K}, \bm{p}) = \textbf{I}_p$, 
		the identity matrix $\textbf{I}_n \otimes \textbf{I}_p \in \operatorname{span}(\textbf{V})$. 
		Let $M ^ {-}$ denote the generalized inverse for some matrix $M$, 
		and let $\textbf{P}_{0} = (\textbf{1}_{n} \otimes \textbf{I}_p) \{(\textbf{1}_{n} \otimes \textbf{I}_p) ^\top \left(\textbf{1}_{n} \otimes \textbf{I}_p \right)\} ^ {-}(\textbf{1}_{n} \otimes \textbf{I}_p) ^\top$, which simplifies to $n ^ {-1} \textbf{J}_n \otimes \textbf{I}_n$ using $\textbf{1}_n ^ {-} = \textbf{1}_n ^\top / n$.
		Hence, 
		$\textbf{P}_{0}$ is an orthogonal projector because 
		\begin{equation*}
			\begin{split}
				\textbf{P}_{0} ^ 2 & = (n ^ {-1} \textbf{J}_n \otimes \textbf{I}_p) 
				(n ^ {-1} \textbf{J}_n \otimes \textbf{I}_p) 
				= n ^ {-2} (\textbf{J}_n \textbf{J}_n) \otimes \textbf{I}_p = n ^ {-1} \textbf{J}_n \otimes \textbf{I}_p = \textbf{P}_{0}, \\
				\textbf{P}_{0} ^\top & = n ^ {-1} \textbf{J}_n ^\top \otimes \textbf{I}_p ^\top = \textbf{P}_{0}.
			\end{split}
		\end{equation*}
		Additionally, $\textbf{V}$ and $\textbf{P}_{0}$ are commutative because
		\begin{equation*}
			\begin{split}
				\textbf{P}_{0} \textbf{V} & = (n ^ {-1} \textbf{J}_n \otimes \textbf{I}_p) 
				\{\textbf{I}_n \otimes \bm{\Sigma}(\textbf{A}, \textbf{B}, \bm{p})\}
				= n ^ {-1} ( \textbf{J}_n \textbf{I}_n ) \otimes \{\textbf{I}_p \bm{\Sigma}(\textbf{A}, \textbf{B}, \bm{p}) \}
				= n ^ {-1} \textbf{J}_n \otimes \bm{\Sigma}(\textbf{A}, \textbf{B}, \bm{p}), \\
				\textbf{V} \textbf{P}_{0} & = \{\textbf{I}_n \otimes \bm{\Sigma}(\textbf{A}, \textbf{B}, \bm{p})\}
				(n ^ {-1} \textbf{J}_n \otimes \textbf{I}_p )
				= n ^ {-1} ( \textbf{I}_n \textbf{J}_n ) \otimes  \{\bm{\Sigma}(\textbf{A}, \textbf{B}, \bm{p}) \textbf{I}_p \}
				= n ^ {-1} \textbf{J}_n \otimes \bm{\Sigma}(\textbf{A}, \textbf{B}, \bm{p}).
			\end{split}
		\end{equation*}
		Hence, all estimable functions have their best linear unbiased estimators,
		which can be expressed in terms of the solution to the following normal equation 
		\begin{equation*}
			(\textbf{1}_{n} \otimes \textbf{I}_p) ^\top (\textbf{1}_{n} \otimes \textbf{I}_p) \bm{\mu}
			= (\textbf{1}_{n} \otimes \textbf{I}_p) ^\top \bm{Y}.
		\end{equation*}
		In other words, $\widetilde{\bm{\mu}} = \bar{\bm{X}}$ is BLUE.
		This coincides with the arguments of the proofs in \citet{RoyZmyslonyFonseca2016} and \citet{KoziolRoyZmyslony2017}.
		
		Subsequently, we can prove that $\widetilde{\alpha}_{kk}$ and $\widetilde{\beta}_{kk'}$ are the best quadratic unbiased estimators (BQUE) of $\alpha_{kk}$ and $\beta_{kk'}$ for every $k$ and $k'$.
		We need to check the conditions of Theorem 2 outlined by \citet{Zmyslony1976},
		or the analogous arguments in \citet{RoyZmyslonyFonseca2016} and \citet{KoziolRoyZmyslony2017}, 
		i.e.,
		given $\textbf{P}_{0} \textbf{V} = \textbf{V} \textbf{P}_{0}$ and $\textbf{R}_0 = \textbf{I}_{pn} - \textbf{P}_{0}$,  
		there exist BQUEs for the parameters of quadratic covariance if and only if $\operatorname{span}\left(\textbf{P}_0 \textbf{V} \textbf{P}_0 \right)$, i.e., the smallest linear space containing $\textbf{P}_0 \textbf{V} \textbf{P}_0$,
		is a quadratic space.
		Since $\bm{\Sigma}(\textbf{A}, \textbf{B}, \bm{p})$ is a uniform-block matrix, 
		$\{\bm{\Sigma}(\textbf{A}, \textbf{B}, \bm{p})\} ^ 2$ is a uniform-block matrix, expressed by $\left(\bm{\Sigma} ^ 2\right)(\textbf{A} ^ 2, \textbf{A} \textbf{B} + \textbf{B} \textbf{A} + \textbf{B} \textbf{P} \textbf{B}, \bm{p})$, 
		thus, $\operatorname{span}\{\bm{\Sigma}(\textbf{A}, \textbf{B}, \bm{p})\}$ is a quadratic subspace, and the identity matrix $\textbf{I}(\bm{p})$ belongs to it.
		Therefore, $\operatorname{span}\left(\textbf{V}\right)$ is a quadratic subspace.
		In addition, 
		it is clear that $\textbf{P}_0$ is idempotent because $\textbf{P}_0 ^ 2 = (\textbf{I}_{pn} - \textbf{P}_0) ^ 2 = \textbf{I}_{pn} ^ 2 - \textbf{I}_{pn} \textbf{P}_0 - \textbf{P}_0 \textbf{I}_{pn} + \textbf{P}_0 ^ 2 = \textbf{I}_{pn} - 2 \textbf{P}_0 + \textbf{P}_0 = \textbf{I}_{pn} - \textbf{P}_0 = \textbf{P}_0$.
		By the result (2.e) from \citet{Seely1971}, 
		$\operatorname{span}(\textbf{P}_0 \textbf{V} \textbf{P}_0) = \{\textbf{P}_0 \textbf{V} \textbf{P}_0: \textbf{V} \in \operatorname{span}(\textbf{V}) \}$ is a quadratic subspace.
		Since both $\operatorname{span}\{\bm{\Sigma}(\textbf{A}, \textbf{B}, \bm{p})\}$ and $\operatorname{span}(\textbf{V})$ are quadratic subspaces, 
		we can find bases for them respectively.
		By definition of the block Hadamard product, 
		$\bm{\Sigma}(\textbf{A}, \textbf{B}, \bm{p})$ has a base as follows:
		\begin{equation*}
			\begin{split}
				\textbf{E}_{kk} \circ \textbf{I}(\bm{p}), \quad
				\textbf{E}_{kk} \circ \textbf{J}(\bm{p}), \quad \text{for every $k$}, \quad
				(\textbf{E}_{kk'} + \textbf{E}_{k'k}) \circ \textbf{J}(\bm{p}), \quad \text{for every $k \neq k'$},
			\end{split}
		\end{equation*}
		where $\textbf{E}_{kk'}$ denotes a $K$ by $K$ matrix in which $(k, k')$ entry is $1$ and the other entries are $0$'s for every $k$ and $k'$.
		Thus, the base for $\operatorname{span}\left(\textbf{V}\right)$ is 
		\begin{equation*}
			\begin{split}
				\textbf{I}_n \otimes \left\{\textbf{E}_{kk} \circ \textbf{I}(\bm{p}) \right\}, 
				\textbf{I}_n \otimes \left\{\textbf{E}_{kk} \circ \textbf{J}(\bm{p}) \right\},  \text{for every $k$}, 
				\textbf{I}_n \otimes \left\{(\textbf{E}_{kk'} + \textbf{E}_{k'k}) \circ \textbf{J}(\bm{p}) \right\},  \text{for every $k \neq k'$}.
			\end{split}
		\end{equation*}
		By Result 2 from \citet{RoyZmyslonyFonseca2016} and \citet{KoziolRoyZmyslony2017}, $(\textbf{1}_{n} ^\top \otimes \textbf{I}_p) \bm{Y}$ is the complete and minimal sufficient statistic for $\bm{\mu}$, and 
		\begin{equation*}
			\begin{split}
				& \bm{Y} ^\top \textbf{P}_0 [\textbf{I}_n \otimes \{\textbf{E}_{kk} \circ \textbf{I}(\bm{p})\}]  \textbf{P}_0  \bm{Y}, \quad 
				\bm{Y} ^\top  \textbf{P}_0  [\textbf{I}_n \otimes \{\textbf{E}_{kk} \circ \textbf{J}(\bm{p})\}]  \textbf{P}_0  \bm{Y}, \quad \text{for every $k$}, \\
				& \bm{Y} ^\top  \textbf{P}_0  [\textbf{I}_n \otimes \{(\textbf{E}_{kk'} + \textbf{E}_{k'k}) \circ \textbf{J}(\bm{p})\}]  \textbf{P}_0  \bm{Y}, \quad \text{for every $k \neq k'$},
			\end{split}
		\end{equation*} 
		are the complete and minimal sufficient statistics for $\bm{\Sigma}$.
		Given the above base for $\operatorname{span}(\textbf{V})$,
		we follow the arguments about the coordinate-free approach \citep{RoyZmyslonyFonseca2016, KoziolRoyZmyslony2017, Wichura2006}, 
		the BQUEs for $\alpha_{kk}$ and $\beta_{kk'}$ are the least square estimators satisfying the normal equations as below:
		\begin{equation*}
			\begin{split}
				& \operatorname{vec}\left(\textbf{P}_0  \left[\textbf{I}_n \otimes \left\{\textbf{E}_{kk} \circ \textbf{I}(\bm{p})\right\}\right] \right) ^\top 
				\operatorname{vec}\left(\textbf{P}_0  \left[\textbf{I}_n \otimes \left\{\textbf{E}_{kk} \circ \textbf{I}(\bm{p})\right\}\right] \right)  \alpha_{kk} \\
				& = \operatorname{vec}\left(\textbf{P}_0  \left[\textbf{I}_n \otimes \left\{\textbf{E}_{kk} \circ \textbf{I}(\bm{p})\right\}\right] \right) ^\top 
				\operatorname{vec}\left\{(\textbf{P}_0  \bm{Y})  (\textbf{P}_0 \bm{Y}) ^\top \right\}, \\
				%%%
				& \operatorname{vec}\left(\textbf{P}_0  \left[\textbf{I}_n \otimes \left\{\textbf{E}_{kk} \circ \textbf{J}(\bm{p})\right\}\right] \right) ^\top 
				\operatorname{vec}\left(\textbf{P}_0  \left[\textbf{I}_n \otimes \left\{\textbf{E}_{kk} \circ \textbf{J}(\bm{p})\right\}\right] \right)  \beta_{kk} \\
				& = \operatorname{vec}\left(\textbf{P}_0  \left[\textbf{I}_n \otimes \left\{\textbf{E}_{kk} \circ \textbf{J}(\bm{p})\right\}\right] \right) ^\top 
				\operatorname{vec}\left\{(\textbf{P}_0  \bm{Y})  (\textbf{P}_0  \bm{Y}) ^\top \right\}, \\
				%%%
				& \operatorname{vec}\left(\textbf{P}_0  \left[\textbf{I}_n \otimes \left\{(\textbf{E}_{kk'} + \textbf{E}_{k'k}) \circ \textbf{J}(\bm{p})\right\}\right] \right) ^\top 
				\operatorname{vec}\left(\textbf{P}_0 \left[\textbf{I}_n \otimes \left\{(\textbf{E}_{kk'} + \textbf{E}_{k'k}) \circ \textbf{J}(\bm{p})\right\}\right] \right)   \beta_{kk'} \\
				& = \operatorname{vec}\left(\textbf{P}_0  \left[\textbf{I}_n \otimes \left\{(\textbf{E}_{kk'} + \textbf{E}_{k'k}) \circ \textbf{J}(\bm{p})\right\}\right] \right) ^\top 
				\operatorname{vec}\left\{(\textbf{P}_0  \bm{Y})  (\textbf{P}_0  \bm{Y}) ^\top \right\},
			\end{split}
		\end{equation*}
		for every $k$, every $k$, and every $k \neq k'$, respectively, 
		where $\operatorname{vec}(M)$ denotes a single vector by stacking the columns of $M$ \citep{HendersonSearle1979}.  	
		In addition to the fact that $\operatorname{vec}(A) ^\top \operatorname{vec}(B) = \operatorname{tr}(A^\top B)$, 
		the idempotent matrix $\textbf{P}_0$ commutes with the base of $\operatorname{span}(\textbf{V})$, 
		the above equations can be simplified as follows:
		\begin{equation*}
			\begin{split}
				& \operatorname{tr}\left(\textbf{P}_0 \left[\textbf{I}_n \otimes \left\{\textbf{E}_{kk} \circ \textbf{I}(\bm{p})\right\}\right] ^ 2 \right)
				\alpha_{kk}
				= (\textbf{P}_0 \bm{Y}) ^\top \left[\textbf{I}_n \otimes \left\{\textbf{E}_{kk} \circ \textbf{I}(\bm{p})\right\}\right] (\textbf{P}_0 \bm{Y}) 
				, \quad \text{for every $k$}, \\
				%%%
				& \operatorname{tr}\left(\textbf{P}_0 \left[\textbf{I}_n \otimes \left\{\textbf{E}_{kk} \circ \textbf{J}(\bm{p})\right\}\right] ^ 2 \right)
				\beta_{kk}
				= (\textbf{P}_0 \bm{Y}) ^\top \left[\textbf{I}_n \otimes \left\{\textbf{E}_{kk} \circ \textbf{J}(\bm{p})\right\}\right] (\textbf{P}_0 \bm{Y}) 
				, \quad \text{for every $k$}, \\
				%%%
				& \operatorname{tr}\left( \textbf{P}_0 \left[\textbf{I}_n \otimes \left\{(\textbf{E}_{kk'} + \textbf{E}_{k'k}) \circ \textbf{J}(\bm{p})\right\}\right] ^ 2 \right)
				\beta_{kk'}
				= (\textbf{P}_0 \bm{Y}) ^\top \left[\textbf{I}_n \otimes \left\{(\textbf{E}_{kk'} + \textbf{E}_{k'k}) \circ \textbf{J}(\bm{p})\right\}\right] (\textbf{P}_0 \bm{Y}) 
				, \quad \text{for $k \neq k'$}.
			\end{split}
		\end{equation*}
		Define the residual vector $\bm{r}_0 = \textbf{P}_0 \bm{Y} \in \mathbb{R} ^ {pn}$, the solutions to the simplified normal equations are  
		\begin{equation*}
			\begin{split}
				\widetilde{\alpha}_{kk} & = \frac{\bm{r}_0 ^\top \left[\textbf{I}_n \otimes \left\{\textbf{E}_{kk} \circ \textbf{I}(\bm{p})\right\}\right] \bm{r}_0}
				{\operatorname{tr}\left(\textbf{P}_0 \left[\textbf{I}_n \otimes \left\{\textbf{E}_{kk} \circ \textbf{I}(\bm{p})\right\}\right] ^ 2 \right)}, \quad
				%%%
				\widetilde{\beta}_{kk} = \frac{\bm{r}_0 ^\top \left[\textbf{I}_n \otimes \left\{\textbf{E}_{kk} \circ \textbf{J}(\bm{p})\right\}\right] \bm{r}_0}
				{\operatorname{tr}\left(\textbf{P}_0 \left[\textbf{I}_n \otimes \left\{\textbf{E}_{kk} \circ \textbf{J}(\bm{p})\right\}\right] ^ 2 \right)}
				\quad \text{for every $k$}, \\
				%%%%
				\widetilde{\beta}_{kk'} & = \frac{\bm{r}_0 ^\top \left[\textbf{I}_n \otimes \left\{(\textbf{E}_{kk'} + \textbf{E}_{k'k}) \circ \textbf{J}(\bm{p})\right\}\right] \bm{r}_0}
				{\operatorname{tr}\left( \textbf{P}_0 \left[\textbf{I}_n \otimes \left\{(\textbf{E}_{kk'} + \textbf{E}_{k'k}) \circ \textbf{J}(\bm{p})\right\}\right] ^ 2 \right)} \quad \text{for every $k \neq k'$},
			\end{split}
		\end{equation*}
		where the denominators can be further simplified respectively (using $\textbf{P} = \operatorname{diag}\left(p_1, \ldots, p_K\right)$):
		\begin{equation*}
			\begin{split}
				\operatorname{tr}\left(\textbf{P}_0 \left[\textbf{I}_n \otimes \left\{\textbf{E}_{kk} \circ \textbf{I}(\bm{p})\right\}\right] ^ 2 \right)
				& = (n - 1) \operatorname{tr}\{\textbf{E}_{kk} \circ \textbf{I}(\bm{p})\} 
				= (n - 1) p_k \quad \text{for every $k$}, \\
				%%%
				\operatorname{tr}\left(\textbf{P}_0 \left[\textbf{I}_n \otimes \left\{\textbf{E}_{kk} \circ \textbf{J}(\bm{p})\right\}\right] ^ 2 \right)
				& = (n - 1) \operatorname{tr}\{(\textbf{E}_{kk} \textbf{P} \textbf{E}_{kk}) \circ \textbf{J}(\bm{p})\} 
				= (n - 1) p_k ^ 2 \quad \text{for every $k$}, \\
				%%%
				\operatorname{tr}\left( \textbf{P}_0 \left[\textbf{I}_n \otimes \left\{(\textbf{E}_{kk'} + \textbf{E}_{k'k}) \circ \textbf{J}(\bm{p})\right\}\right] ^ 2 \right)
				& = (n - 1) (2 p_k p_{k'} )  \quad \text{for every $k \neq k'$.}
			\end{split}
		\end{equation*}
		Let $\bm{r}_i = (\bm{r}_{i, p_1} ^\top, \bm{r}_{i, p_2} ^\top, \ldots, \bm{r}_{i, p_K} ^\top) ^\top \in \mathbb{R} ^ p$ denote the $i$ th subvector of $\bm{r}_0$ satisfying that $\bm{r}_0 = (\bm{r}_1 ^\top, \bm{r}_2 ^\top, \ldots, \bm{r}_n ^\top) ^\top$.
		For every $k$ and $k'$, 
		\begin{equation*}
			\begin{split}
				\widetilde{\alpha}_{kk} & = \sum_{i = 1}^{n} \bm{r}_{i, p_k} ^\top \bm{r}_{i, p_k} / \{(n - 1) p_k\}, \\
				\widetilde{\beta}_{kk} & = \sum_{i = 1}^{n} \operatorname{sum}(\bm{r}_{i, p_k}) \operatorname{sum} (\bm{r}_{i, p_k}) / \{(n - 1) p_k ^ 2\}
				= \sum_{i = 1}^{n} \operatorname{sum} ^ 2 (\bm{r}_{i, p_k}) / \{(n - 1) p_k ^ 2\} , \\
				%%%%
				\widetilde{\beta}_{kk'} 
				& = \sum_{i = 1}^{n} 2 (\operatorname{sum}(\bm{r}_{i, p_k})\operatorname{sum}(\bm{r}_{i, p_{k'}})) / \{2 (n - 1) p_k p_{k'}\} 
				= \sum_{i = 1}^{n} \{\operatorname{sum}(\bm{r}_{i, p_k})\operatorname{sum}(\bm{r}_{i, p_{k'}})\} / \{(n - 1) p_k p_{k'}\}.
			\end{split}
		\end{equation*}
		By the facts that $\bm{r}_{i, p_k} = \bm{X}_{i, p_k} - \bar{\bm{X}}_{p_k}$ and $\textbf{C}_{kk'} = \sum_{i = 1}^{n}(\bm{X}_{i, p_k} - \bar{\bm{X}}_{p_k})(\bm{X}_{i, p_{k'}} - \bar{\bm{X}}_{p_{k'}}) ^\top = \sum_{i = 1}^{n} \bm{r}_{i, p_k} \bm{r}_{i, p_{k'}} ^\top$
		for every $k$ and $k'$, then, 
		\begin{equation*}
			\begin{split}
				k \neq k': & \operatorname{sum} (\textbf{C}_{kk'}) 
				= \operatorname{sum} \left( \sum_{\i = 1}^{n} \bm{r}_{i, p_k} \bm{r}_{i, p_{k'}} ^\top \right)
				= \sum_{i = 1}^{n} \operatorname{sum} \left (\bm{r}_{i, p_k} \bm{r}_{i, p_{k'}} ^\top \right)
				= \sum_{i = 1}^{n} \operatorname{sum}\left(\bm{r}_{i, p_k}\right) \operatorname{sum}\left(\bm{r}_{i, p_{k'}}\right), \\
				%%%%
				k = k': & \operatorname{sum} (\textbf{C}_{kk'}) 
				= \operatorname{sum} \left( \sum_{\i = 1}^{n} \bm{r}_{i, p_k} \bm{r}_{i, p_{k}} ^\top \right)
				= \sum_{i = 1}^{n} \operatorname{sum}\left(\bm{r}_{i, p_k} \bm{r}_{i, p_k} ^\top \right)
				= \sum_{i = 1}^{n} \operatorname{sum} ^ 2 (\bm{r}_{i, p_k}), \\
				%%%
				k = k': & \operatorname{tr}(\textbf{C}_{kk'})
				= \operatorname{tr} \left( \sum_{i = 1}^{n} \bm{r}_{i, p_k} \bm{r}_{i, p_k} ^\top \right)
				= \sum_{i = 1}^{n} \operatorname{tr}\left( \bm{r}_{i, p_k} \bm{r}_{i, p_k} ^\top \right)
				= \sum_{i = 1}^{n} \bm{r}_{i, p_k} ^\top \bm{r}_{i, p_k}.
			\end{split}
		\end{equation*}
		Finally, the estimators of $\bm{\mu}$ and $\bm{\Sigma}$ are respectively BLUE and BQUE,
		but also they are functions of complete statistics. 
		Therefore, we have the following best unbiased estimators (UMVUEs)
		\begin{equation*}
			\begin{split}
				\widetilde{\bm{\mu}} = \frac{1}{n} \sum_{i = 1}^{n} \bm{X}_{i}, \quad
				\widetilde{\alpha}_{kk} = \frac{1}{n - 1} \frac{\operatorname{tr}(\textbf{C}_{kk})}{p_k}, \quad 
				\widetilde{\beta}_{kk'} = 
				\begin{cases}
					\dfrac{1}{n - 1} \dfrac{\operatorname{sum}(\textbf{C}_{kk})}{p_k ^ 2}, & k = k' \\
					\dfrac{1}{n - 1} \dfrac{\operatorname{sum} (\textbf{C}_{kk'})}{p_k p_{k'}}, & k \neq k'
				\end{cases},
			\end{split}
		\end{equation*}
		for $\bm{\mu}$, $\alpha_{kk}$, and $\beta_{kk'}$ for every $k$ and $k'$.
		Using the linear transformation, 
		$\widetilde{a}_{kk}$ and $\widetilde{b}_{kk'}$ are the UMVUEs of $a_{kk}$ and $b_{kk'}$ for every $k$ and $k'$.	
	\end{proof}
	
	\subsection{Proof Corollary~\ref{Cor:vars}}
	
	\begin{proof}[Proof of Corollary~\ref{Cor:vars}]
		\citet{RoyZmyslonyFonseca2016} and \citet{KoziolRoyZmyslony2017} provided their formula (4.17) for calculating the variance of a quadratic form $\bm{Y} A \bm{Y} ^\top$:
		\begin{equation*}
			\operatorname{var}\left(\bm{Y} ^\top A \bm{Y} \right) = 2 \operatorname{tr}\left(\textbf{P}_0 A \textbf{V} A \textbf{V} \right), 
		\end{equation*}
		where matrix $A$ satisfies that $A = \textbf{P}_0 A \textbf{P}_0$, $\bm{Y} \sim N\left(\bm{\mu}_y, \textbf{V}\right)$, and $\textbf{P}_0 = \textbf{I}_{pn} - \textbf{P}_0$ is defined in the proof of Theorem 1.
		Use the result from the proof of Theorem 1, 
		\begin{equation*}
			\begin{split}
				\widetilde{\alpha}_{kk} & = \frac{\bm{r}_0 ^\top \left[\textbf{I}_n \otimes \left\{\textbf{E}_{kk} \circ \textbf{I}(\bm{p})\right\}\right] \bm{r}_0}
				{(n - 1) p_k}, \quad
				%%%
				\widetilde{\beta}_{kk} = \frac{\bm{r}_0 ^\top \left[\textbf{I}_n \otimes \left\{\textbf{E}_{kk} \circ \textbf{J}(\bm{p})\right\}\right] \bm{r}_0}
				{(n - 1) p_k ^ 2}, \quad \text{for every $k$} \\
				%%%%
				\widetilde{\beta}_{kk'} & = \frac{\bm{r}_0 ^\top \left[\textbf{I}_n \otimes \left\{(\textbf{E}_{kk'} + \textbf{E}_{k'k}) \circ \textbf{J}(\bm{p})\right\}\right] \bm{r}_0}
				{(n - 1) (2 p_k p_{k'})}, \quad \text{for every $k \neq k'$}.
			\end{split}
		\end{equation*}
		Let $x = x(k,k)$, $y = y(k, k)$, and $z = z(k, k')$ denote $\textbf{E}_{kk} \circ \textbf{I}(\bm{p})$, $\textbf{E}_{kk} \circ \textbf{J}(\bm{p})$, and $(\textbf{E}_{kk'} + \textbf{E}_{k'k}) \circ \textbf{J}(\bm{p})$, respectively.
		Additionally, let $\omega$ denote any of $x$, $y$, and $z$ and let $\textbf{Q} = \textbf{Q}(\omega) = \textbf{P}_0 (\textbf{I}_n \otimes \omega) \textbf{P}_0$. 
		So, $\textbf{P}_0  \textbf{Q}  \textbf{P}_0 = \textbf{P}_0 ^ 2  (\textbf{I}_n \otimes \omega)  \textbf{P}_0 ^ 2 = \textbf{Q}$ due to the fact that $\textbf{P}_0 ^ 2 = \textbf{P}_0$.
		In addition, recall that $\textbf{P}_0 \textbf{V} = \textbf{V} \textbf{P}_0$.
		Thus, $\operatorname{var}(\bm{Y} ^\top \textbf{Q} \bm{Y}) = 2  \operatorname{tr}(\textbf{P}_0  \textbf{Q}  \textbf{V} \textbf{Q}  \textbf{V})$, which can be simplified as below: 
		\begin{equation*}
			\begin{split}
				2  \operatorname{tr}\left(\textbf{P}_0 \textbf{Q} \textbf{V} \textbf{Q} \textbf{V} \right)
				& = 2  \operatorname{tr} \left[ \textbf{P}_0 \{\textbf{P}_0 (\textbf{I}_n \otimes \omega) \textbf{P}_0\} \textbf{V} \{\textbf{P}_0 (\textbf{I}_n \otimes \omega) \textbf{P}_0\} \textbf{V} \right] \\
				%& = 2 \times \operatorname{tr} \left\{ \textbf{P}_0 (\textbf{I}_n \otimes \omega) \textbf{P}_0 V \textbf{P}_0 (\textbf{I}_n \otimes \omega) \textbf{P}_0 V \right\} \\
				%& = 2 \times \operatorname{tr} \left\{ (\textbf{I}_n \otimes \omega) \textbf{P}_0 \textbf{P}_0 V \textbf{P}_0 \textbf{P}_0 (\textbf{I}_n \otimes \omega) V \right\} \\
				%& = 2 \times \operatorname{tr} \left\{ (\textbf{I}_n \otimes \omega) \textbf{P}_0 V \textbf{P}_0 (\textbf{I}_n \otimes \omega) V \right\} \\
				%& = 2 \times \operatorname{tr} \left\{ (\textbf{I}_n \otimes \omega) V \textbf{P}_0 \textbf{P}_0 (\textbf{I}_n \otimes \omega) V \right\} \\
				%& = 2 \times \operatorname{tr} \left\{ (\textbf{I}_n \otimes \omega) V \textbf{P}_0 (\textbf{I}_n \otimes \omega) V \right\} \\
				%& = 2 \times \operatorname{tr} \left\{ (\textbf{I}_n \otimes \omega) \textbf{P}_0 V (\textbf{I}_n \otimes \omega) V \right\} \\
				& = 2 \operatorname{tr} \left\{ \textbf{P}_0 (\textbf{I}_n \otimes \omega) \textbf{V} (\textbf{I}_n \otimes \omega) \textbf{V} \right\},
			\end{split}
		\end{equation*} 
		where we use the fact that $\textbf{P}_0$ commutes with $\textbf{I}_n \otimes \omega$ and $\textbf{V}$, respectively.
		Therefore, 
		\begin{equation*}
			\begin{split}
				\operatorname{var}\left\{\bm{r}_0 ^\top (\textbf{I}_n \otimes \omega) \bm{r}_0 \right\}
				& =
				\operatorname{var}\left[\bm{Y} ^\top \{\textbf{P}_0 (\textbf{I}_n \otimes \omega) \textbf{P}_0\} \bm{Y} \right] \\
				& = 2 \operatorname{tr} \{\textbf{P}_0 (\textbf{I}_n \otimes \omega) \textbf{V} (\textbf{I}_n \otimes \omega) \textbf{V} \} \\
				& = 2 (n - 1) \operatorname{tr}\{(\omega\bm{\Sigma}) (\omega\bm{\Sigma})\} \\
				& = 2 (n - 1) s_\omega,
			\end{split}
		\end{equation*}
		where $s_{\omega}$ denotes $\operatorname{tr}\{(\omega\bm{\Sigma}) (\omega\bm{\Sigma})\}$.
		Hence, using these notations, 
		\begin{equation*}
			\begin{split}
				\operatorname{var}\left(\widetilde{\alpha}_{kk} \right)
				& = \frac{1}{(n - 1) ^ 2 p_k ^ 2} \operatorname{var}\left\{\bm{r}_0 ^\top (\textbf{I}_n \otimes x) \bm{r}_0 \right\}
				= \frac{2 (n - 1) s_x}{(n - 1) ^ 2 p_k ^ 2}
				= \frac{2 s_x}{(n - 1) p_k ^ 2} \quad \text{for every $k$}, \\
				%%%
				\operatorname{var}\left(\widetilde{\beta}_{kk} \right)
				& = \frac{1}{(n - 1) ^ 2 p_k ^ 4} \operatorname{var}\left\{\bm{r}_0 ^\top (\textbf{I}_n \otimes y) \bm{r}_0 \right\}
				= \frac{2 (n - 1) s_y}{(n - 1) ^ 2 p_k ^ 4}
				= \frac{2 s_y}{(n - 1) p_k ^ 4} \quad \text{for every $k$}, \\
				%%%
				\operatorname{var}\left(\widetilde{\beta}_{kk'} \right)
				& = \frac{1}{4(n - 1) ^ 2 p_k ^ 2 p_{k'} ^ 2} \operatorname{var}\left\{\bm{r}_0 ^\top (\textbf{I}_n \otimes z) \bm{r}_0 \right\}
				= \frac{2 (n - 1) s_z}{4(n - 1) ^ 2 p_k ^ 2 p_{k'} ^ 2}
				= \frac{2 s_z}{4(n - 1) p_k ^ 2 p_{k'} ^ 2} \quad \text{for every $k \neq k'$},
			\end{split}
		\end{equation*}
		where $s_x, s_y$ and $s_z$ are the traces of $(x\bm{\Sigma})^2, (y\bm{\Sigma})^2$ and $(z\bm{\Sigma})^2$, respectively. 
		The expressions of the covariances among $\widetilde{\alpha}$'s and $\widetilde{\beta}$'s are associated with terms of the products of any two of $(x \bm{\Sigma})$, $(y \bm{\Sigma})$ and $(z \bm{\Sigma})$, 
		where $x = x(k, k)$, $y = y(k, k)$, and $z = z(k, k')$ for every $k$ and every $k \neq k'$.
		We focus only on the calculations of $(x\bm{\Sigma})^2$, $(y\bm{\Sigma})^2$, and $(z\bm{\Sigma})^2$.
		
		First, we calculate the $\operatorname{tr}(x \bm{\Sigma}) ^ 2$ as below:
		\begin{equation*}
			\begin{split}
				(x \bm{\Sigma}) (x \bm{\Sigma})
				& = \left\{(\textbf{E}_{kk} \textbf{A}) \circ \textbf{I}(\bm{p}) + (\textbf{E}_{kk} \textbf{B}) \circ \textbf{J}(\bm{p}) \right\}
				\left\{(\textbf{E}_{kk} \textbf{A}) \circ \textbf{I}(\bm{p}) + (\textbf{E}_{kk} \textbf{B}) \circ \textbf{J}(\bm{p}) \right\} \\
				& = \{(\textbf{E}_{kk} \textbf{A})(\textbf{E}_{kk} \textbf{A})\} \circ \textbf{I}(\bm{p}) 
				+ \{(\textbf{E}_{kk} \textbf{A})(\textbf{E}_{kk} \textbf{B})\} \circ \textbf{J}(\bm{p}) 
				+ \{(\textbf{E}_{kk} \textbf{B})(\textbf{E}_{kk} \textbf{A})\} \circ \textbf{J}(\bm{p}) \\
				& + \{(\textbf{E}_{kk} \textbf{B}) \textbf{P} (\textbf{E}_{kk} \textbf{B})\} \circ \textbf{J}(\bm{p}) \\
				& = \left\{(\textbf{E}_{kk} \textbf{A}) ^ 2 \right\} \circ \textbf{I}(\bm{p})
				+ \left\{(\textbf{E}_{kk} \textbf{A})(\textbf{E}_{kk} \textbf{B}) + (\textbf{E}_{kk} \textbf{B})(\textbf{E}_{kk} \textbf{A}) + (\textbf{E}_{kk} \textbf{B}) \textbf{P} (\textbf{E}_{kk} \textbf{B}) \right\} \circ \textbf{J}(\bm{p}),
			\end{split}
		\end{equation*}
		therefore, for every $k$, 
		\begin{equation*}
			\begin{split}
				\operatorname{tr}\left\{(x \bm{\Sigma})(x \bm{\Sigma}) \right\}
				& = p_k (a_{kk} ^ 2 + 2 a_{kk} b_{kk} + p_k b_{kk} ^ 2).
			\end{split}
		\end{equation*}
		Next, we calculate the $\operatorname{tr}\left(y \bm{\Sigma}\right) ^ 2$ as below:
		\begin{equation*}
			\begin{split}
				(y \bm{\Sigma})(y \bm{\Sigma}) & = 
				\left\{(\textbf{E}_{kk} \textbf{A} + \textbf{E}_{kk} \textbf{P} \textbf{B}) \circ \textbf{J}(\bm{p})\right\}
				\left\{(\textbf{E}_{kk} \textbf{A} + \textbf{E}_{kk} \textbf{P} \textbf{B}) \circ \textbf{J}(\bm{p})\right\} \\
				& = \left\{
				(\textbf{E}_{kk} \textbf{A} + \textbf{E}_{kk} \textbf{P} \textbf{B}) \textbf{P} (\textbf{E}_{kk} \textbf{A} + \textbf{E}_{kk} \textbf{P} \textbf{B})
				\right\} \circ \textbf{J}(\bm{p}),
			\end{split}
		\end{equation*}
		therefore, for every $k$,
		\begin{equation*}
			\begin{split}
				\operatorname{tr}\left\{(y \bm{\Sigma}) (y \bm{\Sigma})\right\} & = p_k ^ 2 (a_{kk} + p_{k} b_{kk}) ^ 2.
			\end{split}
		\end{equation*}
		Lastly, we calculate the $\operatorname{tr}(z \bm{\Sigma}) ^ 2$ as below:
		\begin{equation*}
			\begin{split}
				(z \bm{\Sigma})(z \bm{\Sigma})
				& = \left[
				\left\{(\textbf{E}_{kk'} + \textbf{E}_{k'k}) \textbf{A} + (\textbf{E}_{kk'} + \textbf{E}_{k'k}) \textbf{P} \textbf{B} \right\}
				\textbf{P}
				\left\{(\textbf{E}_{kk'} + \textbf{E}_{k'k}) \textbf{A} + (\textbf{E}_{kk'} + \textbf{E}_{k'k}) \textbf{P} \textbf{B} \right\}
				\right] \circ \textbf{J}(\bm{p}),
			\end{split}
		\end{equation*}
		therefore, for every $k \neq k'$,
		\begin{equation*}
			\begin{split}
				\operatorname{tr}\{(z_{kk'} \bm{\Sigma})(z_{kk'} \bm{\Sigma})\}
				= b_{k'k} ^ 2 p_k ^ 2 p_{k'} ^ 2 + b_{kk'} ^ 2 p_k ^ 2 p_{k'} ^ 2
				+ 2 p_k p_{k'}(a_{kk} + p_k b_{kk})(a_{k'k'} + p_{k'k'}b_{k'k'}).
			\end{split}
		\end{equation*}
		
		The expressions of the covariances among $\widetilde{\alpha}$'s and $\widetilde{\beta}$'s are associated with terms of the products of any two of $(x \bm{\Sigma})$, $(y \bm{\Sigma})$, and $(z \bm{\Sigma})$, 
		where $x = x(k, k)$, $y = y(k, k)$, and $z = z(k, k')$ for every $k$ and $k \neq k'$.
		We can obtain the covariance estimators by calculating $\operatorname{tr}\{(x\bm{\Sigma})(y\bm{\Sigma})\}$, $\operatorname{tr}\{(x\bm{\Sigma})(z\bm{\Sigma})\}$, and $\operatorname{tr}\{(y\bm{\Sigma})(z\bm{\Sigma})\}$, respectively, in a similar manner.
	\end{proof}
	
	\subsection{Proof of Theorem 2}
	
	\begin{lemmas}(Gaussian Hanson-Wright inequality)
		\label{Lem:tailboundnew}
		Let $X \in \mathbb{R} ^ p$ be a Gaussian vector with mean zero and covariance matrix $\bm{\Sigma}$ and $A \in \mathbb{R} ^ {p \times p}$.
		Then, for every $t \geq 0$, we have 
		
		\begin{equation*}
			\begin{split}
				\operatorname{pr} \left( \left \vert X ^\top A X - E\left(X ^\top A X \right) \right \vert \geq t \right) \leq 2 \exp\left(- C \min \left\{ \frac{t ^ 2}{\left \Vert \bm{\Sigma} ^ {{1/2}} A \bm{\Sigma} ^ {{1/2}} \right \Vert_{\text{F}} ^ 2}, \frac{t}{\left \Vert \bm{\Sigma} ^ {{1/2}} A \bm{\Sigma} ^ {{1/2}} \right \Vert_{\text{S}}} \right\}  \right),
			\end{split}
		\end{equation*}
		where ``F" and ``S" denote the Frobenius and spectral norms, respectively, and the positive constant $C$ does not depend on $p$, $A$, and $t$.
	\end{lemmas}
	
	\begin{proof}[Proof of Lemma~\ref{Lem:tailboundnew}]
		It can be derived from the sub-Gaussian version. 
		See the original proof in \citet{HansonWright1971} or \citet{Wright1973} and a modern version in \citet{RudelsonVershynin2013} or \citet[Theorem 6.2.1]{Vershynin2018}.
	\end{proof}
	
	We restate Theorem 2 as below.
	
	\begin{thm}[Consistency of the modified hard-thresholding estimator]
		
		%\label{Thm:hard}
		
		Consider a positive-definite covariance matrix with the uniform-block structure $\bm{\Sigma}(\textbf{A}, \textbf{B}, \bm{p}) = (\sigma_{jj'})$.
		Suppose $K = K(n) > n \to \infty$ and 
		$\log(K) / n \to 0$ as $n \to \infty$,
		the values of $p_1, \ldots, p_K$ do not depend on $K$ and $n$,  
		and there exist constants $0 < C_{p_0}, C_{q_0} < \infty$ such that $\sum_{j = 1}^{p} \sum_{j' = 1}^{p} \left \vert \sigma_{jj'} \right \vert ^ {p_0} \leq C_{p_0}$ and $\max\limits_{j} \sum_{j' = 1}^{p} \left \vert \sigma_{jj'} \right \vert ^ {q_0} \leq C_{q_0}$ for $0 < p_0 < 2$ and $0 < q_0 < 1$,
		then 
		\begin{equation*}
			\begin{split}
				\Pr \bigg\{ 
					\Vert \widehat{\bm{\Sigma}}_{\lambda} (\widehat{\textbf{A}}_{\lambda}, \widehat{\textbf{B}}_{\lambda}, \bm{p}) - \bm{\Sigma}(\textbf{A}, \textbf{B}, \bm{p}) \Vert_{\text{F}} ^ 2 
					%
					%& \leq O_P(1) C_{p_0} \left\{\frac{\log(K)}{n}\right\} ^ {1 - \frac{p_0}{2}}, \\
					& \leq C_{p_0} (2 \lambda) ^ {2 - p_0}, \\
					\Vert \widehat{\Sigma}_{\lambda} (\widehat{\textbf{A}}_{\lambda}, \widehat{\textbf{B}}_{\lambda}, \bm{p}) - \bm{\Sigma}(\textbf{A}, \textbf{B}, \bm{p}) \Vert_{\text{S}}  
					%
					%& \leq O_P(1) C_{q_0} \left\{\frac{\log(K)}{n}\right\} ^ {\frac{1 - q_0}{2}}
					& \leq C_{q_0} (2 \lambda) ^ {1 - q_0}
				\bigg\}
				\to 1,
			\end{split}
		\end{equation*}
		as $K, n \to \infty$, where we choose $\lambda = \eta \{\log(K) / n\} ^ {1 / 2}$ for some positive constant $\eta$.
	\end{thm}
	
	\begin{proof}[Proof of Theorem 2]
		
		We aim to prove it within 4 steps. 
		Let $a \vee b = \max\{a, b\}$ and $a \wedge b = \min\{a, b\}$.
		
		Step 1, we determine the upper bounds for the absolute values of the biases for the modified hard-threshold estimators under different cases.
		
		Case $1$: for a fixed $\lambda$ and every fixed $k$ and $k'$, 
		let $\sigma = b_{kk'}$ and $\widetilde{\sigma} = \widetilde{b}_{kk'}$, hence let $\widehat{\sigma} = \widetilde{\sigma} \mathbb{I}(\vert \widetilde{\sigma} \vert > \lambda)$.
		Assuming that $\vert \widetilde{\sigma} - \sigma \vert \leq \lambda / 2$, or equivalently, $\sigma - \lambda / 2 \leq \widetilde{\sigma} \leq \sigma + \lambda / 2$, 
		we obtain: 
		if $\sigma \in [-\lambda / 2, \lambda / 2]$, 
		then $\widetilde{\sigma} \in [- \lambda, \lambda]$, or equivalently, $\vert \widetilde{\sigma}  \vert \leq \lambda$, and therefore, $\widehat{\sigma} = 0$ and $\vert \widehat{\sigma} - \sigma \vert = \vert \sigma \vert \in [0, \lambda / 2]$;
		if $\sigma \in (3 \lambda / 2, \infty)$, then $\widetilde{\sigma} \in (\lambda, \infty)$, or equivalently, $\vert \widetilde{\sigma} \vert > \lambda$, and therefore, $\widehat{\sigma} = \widetilde{\sigma}$ and $\vert \widehat{\sigma} - \sigma \vert = \vert \widetilde{\sigma} - \sigma \vert \in [0, \lambda / 2]$;
		if $\sigma \in (- \infty, - 3 \lambda / 2)$, then $\widetilde{\sigma} \in (- \infty, - \lambda)$, or equivalently, $\vert \widetilde{\sigma} \vert > \lambda$, and therefore, $\widehat{\sigma} = \widetilde{\sigma}$ and $\vert \widehat{\sigma} - \sigma \vert = \vert \widetilde{\sigma} - \sigma \vert \in [0, \lambda / 2]$;
		if $\sigma \in (\lambda / 2, 3 \lambda / 2]$, then $\widetilde{\sigma} \in (0, 2\lambda ]$ and it can be larger than $\lambda$ or be smaller than $\lambda$, therefore $\widehat{\sigma} \in \{0, \widetilde{\sigma}\}$ and $\vert \widehat{\sigma} - \sigma \vert = \begin{cases}
			\vert \widetilde{\sigma} - \sigma \vert \in [0, \lambda / 2], & \widetilde{\sigma} \in (\lambda, 2 \lambda] \\ 
			\vert \sigma \vert \in (\lambda / 2, 3 \lambda / 2], & \widetilde{\sigma} \in (0, \lambda]
		\end{cases}$, which is between $0$ and $3 \lambda / 2$;
		if $\sigma \in [- 3 \lambda / 2, - \lambda / 2)$, then $\widetilde{\sigma} \in [- 2 \lambda , 0)$ and it can be larger than $- \lambda$ or be smaller than $- \lambda$, therefore $\widehat{\sigma} \in \{0, \widetilde{\sigma}\}$ and $\vert \widehat{\sigma} - \sigma \vert = \begin{cases}
			\vert \widetilde{\sigma} - \sigma \vert \in [0, \lambda / 2], & \widetilde{\sigma} \in [- 2 \lambda, - \lambda) \\
			\vert \sigma \vert \in [\lambda / 2, 3 \lambda / 2], & \widetilde{\sigma} \in [- \lambda, 0 )
		\end{cases}$, which is also between $0$ and $3 \lambda / 2$.
		Put the above arguments together, for a fixed $\lambda$, under the assumption $\vert \widetilde{\sigma} - \sigma \vert \leq \lambda / 2$, 
		\begin{equation*}
			\begin{split}
				\left \vert \widehat{\sigma} - \sigma \right \vert = \begin{cases}
					\left \vert \widetilde{\sigma} - \sigma \right \vert \in \left[0, \lambda / 2\right], & \sigma \in \left(- \infty, - 3 \lambda / 2\right) \\
					\left\{\left \vert \widetilde{\sigma} - \sigma\right \vert, \left \vert \sigma \right \vert \right\} \leq \left \vert \sigma \right \vert \vee \left(\lambda / 2\right), & \sigma \in \left[- 3 \lambda / 2, - \lambda / 2\right) \\
					\left \vert \sigma \right \vert \in \left[0, \lambda / 2\right], & \sigma \in \left[- \lambda / 2, \lambda / 2\right] \\
					\left\{\left \vert \widetilde{\sigma} - \sigma\right \vert, \left \vert \sigma \right \vert \right\} \leq \left \vert \sigma \right \vert \vee \left(\lambda / 2\right), & \sigma \in \left(\lambda / 2, 3 \lambda / 2\right] \\
					\left \vert \widetilde{\sigma} - \sigma \right \vert \in \left[0, \lambda / 2\right], & \sigma \in \left(3 \lambda / 2, \infty \right)	
				\end{cases},
			\end{split}
		\end{equation*}
		therefore the maximum of $\vert \widehat{\sigma} - \sigma \vert$ might be either $\vert \sigma \vert$ if $\lambda / 2 < \vert \sigma \vert \leq 3 \lambda / 2$ or $\lambda / 2$ otherwise.
		In other words, we may write $\vert \widehat{\sigma} - \sigma \vert \leq \vert \sigma \vert \wedge (3 \lambda / 2)$.
		
		Case $2$: for a fixed $\lambda$ and every fixed $k$, let $\sigma = a_{kk}$ and $\widetilde{\sigma} = \widetilde{a}_{kk}$, hence let $\widehat{\sigma} = \widetilde{\sigma} \mathbb{I}(\vert \widetilde{\sigma} \vert > \lambda)$.
		We assume that $\vert \widetilde{\alpha}_{kk} - \alpha_{kk} \vert \leq \lambda / 2$ and $\vert \widetilde{b}_{kk} - b_{kk} \vert \leq \lambda / 2$ holds simultaneously, then 
		\begin{equation*}
			\begin{split}
				\left \vert \widetilde{a}_{kk} - a_{kk} \right \vert 
				= \vert \widetilde{\alpha}_{kk} - \widetilde{b}_{kk} - \left(\alpha_{kk} - b_{kk} \right)  \vert
				= \vert \left(\widetilde{\alpha}_{kk} - \alpha_{kk} \right) - (\widetilde{b}_{kk} - b_{kk}) \vert 
				\leq \vert \widetilde{\alpha}_{kk} - \alpha_{kk} \vert + 
				\vert \widetilde{b}_{kk} - b_{kk}  \vert 
				\leq \lambda.
			\end{split}
		\end{equation*}
		So, the above assumption is equivalent, implying that $\vert \widetilde{\sigma} - \sigma \vert \leq \lambda$, or equivalently, $\sigma - \lambda \leq \widetilde{\sigma} \leq \sigma + \lambda$:
		if $\sigma \in (- \infty, - 2 \lambda)$, then $\widetilde{\sigma} \in (- \infty, - \lambda)$ and therefore $\vert \widetilde{\sigma} \vert > \lambda$, 
		and $\widehat{\sigma} = \widetilde{\sigma}$, and thus $\vert \widehat{\sigma} - \sigma \vert = \vert \widetilde{\sigma} - \sigma \vert \in [0, \lambda]$;
		if $\sigma = 0$, then $\vert \widetilde{\sigma} \vert \leq \lambda$ and therefore $\widehat{\sigma} = 0$, and $\vert \widehat{\sigma} - \sigma \vert = \vert \sigma \vert = 0$;
		if $\sigma \in (0, 2 \lambda]$, then $\widetilde{\sigma} \in (- \lambda, 3 \lambda]$ and therefore $\vert \widetilde{\sigma} \vert$ can be larger than $\lambda$ or be smaller than $\lambda$, thus $\widehat{\sigma} \in \{0, \widetilde{\sigma}\}$, and thus $\vert \widehat{\sigma} - \sigma \vert = \begin{cases}
			\vert \widetilde{\sigma} - \sigma \vert \in [0, \lambda], & \widetilde{\sigma} \in (\lambda, 3 \lambda] \\
			\vert \sigma \vert \in (0, 2 \lambda], & \widetilde{\sigma} \in (-\lambda,\lambda]
		\end{cases}$, which is between $0$ and $2 \lambda$;  
		if $\sigma \in [- 2 \lambda, 0)$, then $\widetilde{\sigma} \in [- 3 \lambda, \lambda)$ and therefore $\vert \widetilde{\sigma} \vert$ can be larger than $\lambda$ or be smaller than $\lambda$, thus $\widehat{\sigma} \in \{\widetilde{\sigma}, 0\}$, and thus $\vert \widehat{\sigma} - \sigma \vert = \begin{cases}
			\vert \widetilde{\sigma} - \sigma \vert \in [0, \lambda], & \widetilde{\sigma} \in [-3\lambda, - \lambda) \\
			\vert \sigma \vert \in (0, 2\lambda], & \widetilde{\sigma} \in [-\lambda, \lambda )
		\end{cases}$, which is between $0$ and $2 \lambda$;
		if $\sigma \in (2 \lambda, \infty)$, then $\widetilde{\sigma} \in (\lambda, \infty)$ and therefore $\vert \widetilde{\sigma} \vert > \lambda$, and $\widehat{\sigma} = \widetilde{\sigma}$, and thus $\vert \widehat{\sigma} - \sigma \vert = \vert \widetilde{\sigma} - \sigma \vert \in [0, \lambda]$.
		Put the above arguments together, for a fixed $\lambda$, under the assumption $\vert \widetilde{\sigma} - \sigma \vert \leq \lambda$, 
		\begin{equation*}
			\begin{split}
				\left \vert \widehat{\sigma} - \sigma \right \vert 
				= \begin{cases}
					\left \vert \widetilde{\sigma} - \sigma \right \vert \in [0, \lambda], & \sigma \in \left(- \infty, - 2 \lambda \right) \\
					\left\{\left \vert \widetilde{\sigma} - \sigma \right \vert, \left \vert \sigma \right \vert \right\} \leq \vert \sigma \vert \vee \lambda, & \sigma \in \left[-2\lambda, 0 \right) \\
					0, & \sigma = 0 \\
					\left\{\left \vert \widetilde{\sigma} - \sigma \right \vert, \left \vert \sigma \right \vert \right\} \leq \vert \sigma \vert \vee \lambda, & \sigma \in \left(0, 2 \lambda\right] \\
					\left \vert \widetilde{\sigma} - \sigma \right \vert \in [0, \lambda], & \sigma \in \left(2 \lambda, \infty \right) 
				\end{cases},
			\end{split}
		\end{equation*}
		therefore the maximum of $\vert \widehat{\sigma} - \sigma \vert$ might be either $\vert \sigma \vert$ if $0 <  \vert \sigma \vert \leq 2 \lambda$ or $\lambda$ otherwise.
		In other words, we may write $\vert \widehat{\sigma} - \sigma \vert \leq \vert \sigma \vert \wedge (2 \lambda)$.
		Thus, at Step 1, we conclude that $\vert \widehat{a}_{kk} - a_{kk} \vert \leq \vert a_{kk} \vert \wedge (2 \lambda)$ and $\vert \widehat{b}_{kk'} - b_{kk'}  \vert \leq \vert b_{kk'} \vert \wedge (3 \lambda / 2)$ for a fixed $\lambda$ and every $k$ and $k'$ under the assumptions: $\max_{1 \leq k \leq K}  \vert \widetilde{\alpha}_{kk} - \alpha_{kk} \vert \leq \lambda / 2$ and $\max_{1 \leq k,k' \leq K} \vert \widetilde{b}_{kk'} - b_{kk'} \vert \leq \lambda / 2$.
		We denote the following event sets
		\begin{equation*}
			\begin{split}
				\textbf{E}_N ^ {(0)} & = \left\{\max\limits_{1 \leq k \leq K} \left \vert \widetilde{a}_{kk} - a_{kk} \right \vert \leq \lambda \right\},  \ \ \ \
				\textbf{E}_N ^ {(1)}   = \left\{\max\limits_{1 \leq k \leq K} \left \vert \widetilde{\alpha}_{kk} - \alpha_{kk} \right \vert \leq \lambda / 2\right\},  \\
				\textbf{E}_N ^ {(2)} & = \left\{\max\limits_{1 \leq k \leq K} \left \vert \widetilde{b}_{kk} - b_{kk} \right \vert \leq \lambda / 2\right\},  \ \ \ \
				\textbf{E}_N ^ {(3)}   = \left\{\max\limits_{1 \leq k \neq k' \leq K} \left \vert \widetilde{b}_{kk'} - b_{kk'} \right \vert \leq \lambda / 2\right\},
			\end{split}
		\end{equation*}
		and the event sets $\textbf{E}_N ^ {(1)}$ and $\textbf{E}_N^{(2)}$ implies that $\textbf{E}_N^{(0)}$. 
		The probability of their complement sets can be bounded as below:
		\begin{equation*}
			\begin{split}
				\Pr \left(\textbf{E}_N ^ {(1), \complement}\right) 
				& = \Pr \left( \max\limits_{1 \leq k \leq K} \left \vert \widetilde{\alpha}_{kk} - \alpha_{kk} \right \vert > \lambda / 2 \right)
				\overset{[1]}{\leq} K \Pr \left(\left \vert \widetilde{\alpha}_{kk} - \alpha_{kk} \right \vert > \lambda / 2\right), \\
				%%%
				\Pr \left(\textbf{E}_N ^ {(2), \complement}\right) 
				& = \Pr \left( \max\limits_{1 \leq k \leq K} \left \vert \widetilde{b}_{kk} - b_{kk} \right \vert > \lambda / 2 \right)
				\leq K \Pr \left(\left \vert \widetilde{b}_{kk} - b_{kk} \right \vert > \lambda / 2\right), \\
				%%%
				\Pr \left(\textbf{E}_N ^ {(3), \complement}\right) 
				& = \Pr \left( \max\limits_{1 \leq k \neq k' \leq K} \left \vert \widetilde{b}_{kk'} - b_{kk'} \right \vert > \lambda / 2 \right)
				\leq K  (K - 1)  \Pr \left(\left \vert \widetilde{b}_{kk'} - b_{kk'} \right \vert > \lambda / 2\right),
			\end{split}
		\end{equation*}
		where $[1]$ holds due to the union bound. 
		
		Step 2, we determine the upper bounds for the above three probabilities. 
		
		Let $\textbf{E}_{kk'} \in \mathbb{R} ^ {K \times K}$ denote the matrix whose $(k,k')$ entry is $1$ and the other entries are $0$ for every $k$ and $k'$.
		Then, let $\textbf{A}_{kk} = (p_{k} ^ {-1} \textbf{E}_{kk}) \circ \textbf{I}(\bm{p})$, $\textbf{B}_{kk} = \{- p_k ^ {-1}(p_k - 1)^{-1} \textbf{E}_{kk}\} \circ \textbf{I}(\bm{p}) + \{p_k^{-1}(p_k - 1)^{-1} \textbf{E}_{kk}\} \circ \textbf{J}(\bm{p})$ and $\textbf{B}_{kk'} = \{(p_k p_{k'})^{-1} \textbf{E}_{kk'}\} \circ \textbf{J}(\bm{p})$ for every $k$ and $k \neq k'$.
		Let $\bm{W} = (\bm{X}_{1} ^\top, \ldots, \bm{X}_n ^\top) ^\top \in \mathbb{R} ^ {pn}$ and decompose $\bm{X} = (\bm{X}_{p_1} ^\top, \ldots, \bm{X}_{p_K} ^\top)^\top$, 
		where $\bm{X}_{p_k} = (\bm{X}_{\bar{p}_{k-1} + 1}, \ldots, \bm{X}_{\bar{p}_k}) ^\top \in \mathbb{R} ^ {p_k}$ for every $k$,
		$\bar{p}_0 = 0$ and $\bar{p}_{k} = \sum_{k' = 1}^{k} p_{k'}$ denote the sum of the first $k$ elements of $\bm{p}$ for $k = 1, \ldots, K$.
		Note that $\bm{\mu} = \textbf{0}_{p \times 1}$ is known and $\textbf{S} = \textbf{X} ^\top \textbf{X} / n$, thus,  
		\begin{equation*}
			\begin{split}
				\bm{W} ^\top \left(\textbf{I}_n \otimes \textbf{A}_{kk} \right) \bm{W} 
				& = \sum_{i = 1}^{n} \bm{X}_{i} ^\top \textbf{A}_{kk} \bm{X}_{i} 
				%  = \sum_{i = 1}^{N} \bm{X}_{p_k} ^ {(i),\top} \left(\frac{1}{p_k}\textbf{I}_{p_k} \right) \bm{X}_{p_k} ^ {(i)}
				= \frac{1}{p_k} \sum_{i = 1}^{n} \operatorname{tr}\left( \bm{X}_{i, p_k} \bm{X}_{i, p_k} ^\top \right)
				= \frac{n}{p_k} \operatorname{tr}(\textbf{S}_{kk}) 
				= n \widetilde{\alpha}_{kk}, \\
				%%%
				\bm{W} ^\top \left(\textbf{I}_n \otimes \textbf{B}_{kk} \right) \bm{W} 
				& = \sum_{i = 1}^{n} \bm{X}_{i} ^\top \textbf{B}_{kk} \bm{X}_{i} 
				= \frac{n}{p_k(p_k - 1)} \left\{\operatorname{sum}(\textbf{S}_{kk}) - \operatorname{tr}(\textbf{S}_{kk})\right\} 
				= n \widetilde{b}_{kk}, \\
				%%%
				\bm{W} ^\top \left(\textbf{I}_n \otimes \textbf{B}_{kk'}\right) \bm{W} 
				& = \sum_{i = 1}^{n} \bm{X}_{i} ^\top \textbf{B}_{kk'} \bm{X}_{i}
				= \frac{1}{p_k p_{k'}} \sum_{i = 1}^{n} \operatorname{sum}(\bm{X}_{i, p_k} \bm{X}_{i, p_{k'}} ^\top)
				= \frac{n}{p_k p_{k'}} \operatorname{sum}(\textbf{S}_{kk'})
				= n \widetilde{b}_{kk'},
			\end{split}
		\end{equation*}
		for every $k$ and for every $k \neq k'$. 
		
		On the one hand, $\widetilde{\alpha}_{kk}$, $\widetilde{b}_{kk'}$ are the unbiased estimators of $\alpha_{kk}$ and $b_{kk'}$ for every $k$ and $k'$.
		Let
		\begin{equation*}
			\begin{split}
				Q_{kk} ^ {(\alpha)} & = \bm{W} ^\top \left(\textbf{I}_n \otimes \textbf{A}_{kk} \right) \bm{W} - \operatorname{E} \left\{\bm{W} ^\top \left(\textbf{I}_n \otimes \textbf{A}_{kk} \right) \bm{W} \right\} = n \left(\widetilde{\alpha}_{kk} - \alpha_{kk} \right), \\
				Q_{kk'} ^ {(b)} & = \bm{W} ^\top \left(\textbf{I}_n \otimes \textbf{B}_{kk'} \right) \bm{W} - \operatorname{E} \left\{\bm{W} ^\top \left(\textbf{I}_n \otimes \textbf{B}_{kk'} \right) \bm{W} \right\} = n \left(\widetilde{b}_{kk'} - b_{kk'} \right),
			\end{split}
		\end{equation*}
		for every $k$ and $k'$.
		On the other hand, we apply Lemma~\ref{Lem:tailboundnew}, given a $\delta > 0$, 
		\begin{equation*}
			\begin{split}
				& \Pr \left(\left \vert \widetilde{\alpha}_{kk} - \alpha_{kk} \right \vert > \delta  \right) \\
				& = \Pr \left( \left \vert Q_{kk} ^ {(\alpha)} \right \vert > n \delta \right) \\
				& \leq 2 \exp\left( - C \min \left\{ \frac{(n \delta) ^ 2}{\left \Vert \left(\textbf{I}_n \otimes \bm{\Sigma} \right) ^ {1/2} \left( \textbf{I}_n \otimes \textbf{A}_{kk} \right) \left(\textbf{I}_n \otimes \bm{\Sigma} \right) ^ {1/2} \right \Vert_{\text{F}} ^ 2}, \frac{n \delta}{\left \Vert \left(\textbf{I}_n \otimes \bm{\Sigma} \right) ^ {1/2} \left( \textbf{I}_n \otimes \textbf{A}_{kk} \right) \left(\textbf{I}_n \otimes \bm{\Sigma} \right) ^ {1/2} \right \Vert_{\text{S}}} \right\} \right), \\
				%%%%
				& \Pr \left(\left \vert \widetilde{b}_{kk} - b_{kk} \right \vert > \delta  \right) \\
				& \leq 2 \exp\left( - C \min \left\{ \frac{(n \delta) ^ 2}{\left \Vert \left(\textbf{I}_n \otimes \bm{\Sigma} \right) ^ {1/2} \left( \textbf{I}_n \otimes \textbf{B}_{kk} \right) \left(\textbf{I}_n \otimes \bm{\Sigma} \right) ^ {1/2} \right \Vert_{\text{F}} ^ 2}, \frac{n \delta}{\left \Vert \left(\textbf{I}_n \otimes \bm{\Sigma} \right) ^ {1/2} \left( \textbf{I}_n \otimes \textbf{B}_{kk} \right) \left(\textbf{I}_n \otimes \bm{\Sigma} \right) ^ {1/2} \right \Vert_{\text{S}}} \right\} \right), \\
				%%%%
				& \Pr \left(\left \vert \widetilde{b}_{kk'} - b_{kk'} \right \vert > \delta  \right) \\
				& \leq 2 \exp\left( - C \min \left\{ \frac{(n \delta) ^ 2}{\left \Vert \left(\textbf{I}_n \otimes \bm{\Sigma} \right) ^ {1/2} \left( \textbf{I}_n \otimes \textbf{B}_{kk'} \right) \left(\textbf{I}_n \otimes \bm{\Sigma} \right) ^ {1/2} \right \Vert_{\text{F}} ^ 2}, \frac{n \delta}{\left \Vert \left(\textbf{I}_n \otimes \bm{\Sigma} \right) ^ {1/2} \left( \textbf{I}_n \otimes \textbf{B}_{kk'} \right) \left(\textbf{I}_n \otimes \bm{\Sigma} \right) ^ {1/2} \right \Vert_{\text{S}}} \right\} \right),
			\end{split}
		\end{equation*}
		for every $k$, for every $k$, and for every $k \neq k'$, where $C$ is a positive constant.
		By the properties of Kronecker product and Frobenius norm, 
		\begin{equation*}
			\begin{split}
				\left \Vert \left(\textbf{I}_n \otimes \bm{\Sigma} \right) ^ {1/2} \left( \textbf{I}_n \otimes \textbf{A}_{kk} \right) \left(\textbf{I}_n \otimes \bm{\Sigma} \right) ^ {1/2} \right \Vert_{\text{F}} ^ 2
				& = \left \Vert \left(\textbf{I}_n \otimes \bm{\Sigma} ^ {1/2} \right)  \left( \textbf{I}_n \otimes \textbf{A}_{kk} \right) \left(\textbf{I}_n \otimes \bm{\Sigma} ^ {1/2} \right)  \right \Vert_{\text{F}} ^ 2 \\
				%& = \left \Vert \left( \textbf{I}_n \otimes \left(\bm{\Sigma} ^ {1/2} \textbf{A}_{kk} \right) \right) \left(\textbf{I}_n \otimes \bm{\Sigma} ^ {1/2} \right)  \right \Vert_{\text{F}} ^ 2 \\
				%& = \left \Vert \textbf{I}_n \otimes \left(\bm{\Sigma} ^ {1/2} \textbf{A}_{kk} \bm{\Sigma} ^ {1/2} \right) \right \Vert_{\text{F}} ^ 2 \\
				& = n \left \Vert \bm{\Sigma} ^ {1/2} \textbf{A}_{kk} \bm{\Sigma} ^ {1/2} \right \Vert_{\text{F}} ^ 2 \\
				& = n \operatorname{tr} \left\{\left( \bm{\Sigma} ^ {1/2} \textbf{A}_{kk} \bm{\Sigma} ^ {1/2} \right) \left( \bm{\Sigma} ^ {1/2} \textbf{A}_{kk} \bm{\Sigma} ^ {1/2} \right) \right\} \\
				& = n \operatorname{tr} \left(\bm{\Sigma} ^ {1/2} \textbf{A}_{kk} \bm{\Sigma}  \textbf{A}_{kk} \bm{\Sigma} ^ {1/2} \right) \\
				%& = N \operatorname{tr} \left( \textbf{A}_{kk} \bm{\Sigma}  \textbf{A}_{kk} \bm{\Sigma} ^ {1/2} \bm{\Sigma} ^ {1/2} \right) \\
				& = n \operatorname{tr} \left( \textbf{A}_{kk} \bm{\Sigma}  \textbf{A}_{kk} \bm{\Sigma} \right),
			\end{split}
		\end{equation*}
		and
		\begin{equation*}
			\begin{split}
				\left \Vert \left(\textbf{I}_n \otimes \bm{\Sigma} \right) ^ {1/2} \left( \textbf{I}_n \otimes \textbf{B}_{kk} \right) \left(\textbf{I}_n \otimes \bm{\Sigma} \right) ^ {1/2} \right \Vert_{\text{F}} ^ 2
				& =  n \operatorname{tr} \left( \textbf{B}_{kk} \bm{\Sigma}  \textbf{B}_{kk} \bm{\Sigma} \right), \\
				\left \Vert \left(\textbf{I}_n \otimes \bm{\Sigma} \right) ^ {1/2} \left( \textbf{I}_n \otimes \textbf{B}_{kk'} \right) \left(\textbf{I}_n \otimes \bm{\Sigma} \right) ^ {1/2} \right \Vert_{\text{F}} ^ 2
				& =  n \operatorname{tr} \left( \textbf{B}_{kk'} \bm{\Sigma}  \textbf{B}_{kk'} ^ \top \bm{\Sigma} \right).
			\end{split}
		\end{equation*}
		Then, we calculate the traces,
		\begin{equation*}
			\begin{split}
				n \operatorname{tr} \left( \textbf{A}_{kk} \bm{\Sigma}  \textbf{A}_{kk} \bm{\Sigma} \right) 
				& = \frac{n}{p_k} \left(a_{kk}^2 + 2 a_{kk} b_{kk} + p_k b_{kk} ^ 2 \right), \\
				%%%
				n \operatorname{tr} \left( \textbf{B}_{kk} \bm{\Sigma}  \textbf{B}_{kk} \bm{\Sigma} \right) 
				& = \frac{n}{p_k ^ 2 (p_k - 1)^2} \left\{ p_k a_{kk} ^ 2 - 2 p_k a_{kk}(a_{kk} - b_{kk} + p_k b_{kk}) + p_k \left(a_{kk} - b_{kk} + p_k ^ 2 b_{kk} \right) ^ 2 \right\}, \\
				%%%
				n \operatorname{tr} \left( \textbf{B}_{kk'} \bm{\Sigma}  \textbf{B}_{kk'} ^ \top \bm{\Sigma} \right)
				& = \frac{n}{p_k p_{k'}} (a_{kk} + p_k b_{kk})(a_{k'k'} + p_{k'} b_{k'k'}).
			\end{split}
		\end{equation*}
		By the properties of the spectral norm, let $\lambda_{\operatorname{max}}(M)$ denote the largest eigenvalue of $M$, 
		\begin{equation*}
			\begin{split}
				\left \Vert \left(\textbf{I}_n \otimes \bm{\Sigma} \right) ^ {1/2} \left( \textbf{I}_n \otimes \textbf{A}_{kk} \right) \left(\textbf{I}_n \otimes \bm{\Sigma} \right) ^ {1/2} \right \Vert_{\text{S}} ^ 2
				& = \left \Vert \textbf{I}_n \otimes \left(\bm{\Sigma} ^ {1/2} \textbf{A}_{kk} \bm{\Sigma} ^ {1/2} \right) \right \Vert_{\text{S}} ^ 2 \\
				& = \lambda_{\text{max}} \left[ \left\{ \textbf{I}_n \otimes \left(\bm{\Sigma} ^ {1/2} \textbf{A}_{kk} \bm{\Sigma} ^ {1/2} \right)\right\} \left\{ \textbf{I}_n \otimes \left(\bm{\Sigma} ^ {1/2} \textbf{A}_{kk} \bm{\Sigma} ^ {1/2} \right)\right\}  \right] \\
				& = \lambda_{\text{max}} \left[ \textbf{I}_n \otimes \left\{\left(\bm{\Sigma} ^ {1/2} \textbf{A}_{kk} \bm{\Sigma} ^ {1/2} \right) \left(\bm{\Sigma} ^ {1/2} \textbf{A}_{kk} \bm{\Sigma} ^ {1/2} \right)\right\}  \right] \\
				& = \lambda_{\text{max}} \left\{ \left(\bm{\Sigma} ^ {1/2} \textbf{A}_{kk} \bm{\Sigma} ^ {1/2} \right) \left(\bm{\Sigma} ^ {1/2} \textbf{A}_{kk} \bm{\Sigma} ^ {1/2} \right)  \right\} \\
				& = \left \Vert \left(\bm{\Sigma} ^ {1/2} \textbf{A}_{kk} \bm{\Sigma} ^ {1/2} \right) \right \Vert_{\text{S}} ^ 2 \\
				& \leq \left \Vert \left(\bm{\Sigma} ^ {1/2} \textbf{A}_{kk} \bm{\Sigma} ^ {1/2} \right) \right \Vert_{\text{F}} ^ 2 \\
				& = \frac{1}{p_k} \left(a_{kk}^2 + 2 a_{kk} b_{kk} + p_k b_{kk} ^ 2 \right),
			\end{split}
		\end{equation*}
		and 
		\begin{equation*}
			\begin{split}
				& \left \Vert \left(\textbf{I}_n \otimes \bm{\Sigma} \right) ^ {1/2} \left( \textbf{I}_n \otimes \textbf{B}_{kk} \right) \left(\textbf{I}_n \otimes \bm{\Sigma} \right) ^ {1/2} \right \Vert_{\text{S}} ^ 2 \\
				& \leq \frac{1}{p_k ^ 2 (p_k - 1)^2} \left\{ p_k a_{kk} ^ 2 - 2 p_k a_{kk}(a_{kk} - b_{kk} + p_k b_{kk}) + p_k \left(a_{kk} - b_{kk} + p_k ^ 2 b_{kk} \right) ^ 2 \right\}, \\
				& \left \Vert \left(\textbf{I}_n \otimes \bm{\Sigma} \right) ^ {1/2} \left( \textbf{I}_n \otimes \textbf{B}_{kk'} \right) \left(\textbf{I}_n \otimes \bm{\Sigma} \right) ^ {1/2} \right \Vert_{\text{S}} ^ 2 \\
				& \leq \frac{1}{p_k p_{k'}} (a_{kk} + p_k b_{kk})(a_{k'k'} + p_{k'} b_{k'k'}).
			\end{split}
		\end{equation*}
		Since constants $C_{p_0}$ and $C_{q_0}$ and values of $p_1, \ldots, p_K$ are free of $K$ and $n$, 
		there exist positive constants $C_{1, *}$ and $C_{2, *}$, depending on them only, satisfying that 	
		\begin{equation*}
			\begin{split}
				C_{1, *} ^ 2 (C_{p_0}, C_{q_0}, \bm{p}) = 
				\max\limits_{K}
				\max\limits_{1 \leq k, k' \leq K} 
				\left\{
					\left \Vert \left(\bm{\Sigma} ^ {1/2} \textbf{A}_{kk} \bm{\Sigma} ^ {1/2} \right) \right \Vert_{\text{S}} ^ 2, 
					\left \Vert \left(\bm{\Sigma} ^ {1/2} \textbf{B}_{kk} \bm{\Sigma} ^ {1/2} \right) \right \Vert_{\text{S}} ^ 2,
					\left \Vert \left(\bm{\Sigma} ^ {1/2} \textbf{B}_{kk'} \bm{\Sigma} ^ {1/2} \right) \right \Vert_{\text{S}} ^ 2
				\right\} < \infty, 
			\end{split}
		\end{equation*}
		and
		\begin{equation*}
			C_{2, *} ^ 2 (C_{p_0}, C_{q_0}, \bm{p}) = 
			\max\limits_{K}
			\max\limits_{1 \leq k, k' \leq K} \left\{
				\operatorname{tr} \left( \textbf{A}_{kk} \bm{\Sigma}  \textbf{A}_{kk} \bm{\Sigma} \right), 
				\operatorname{tr} \left( \textbf{B}_{kk} \bm{\Sigma}  \textbf{B}_{kk} \bm{\Sigma} \right),
				\operatorname{tr} \left( \textbf{B}_{kk'} \bm{\Sigma}  \textbf{B}_{kk'} \bm{\Sigma} \right)
			\right\} < \infty. 
		\end{equation*}
		Therefore,  
		\begin{equation*}
			\begin{split}
				K \Pr \left(\left \vert \widetilde{\alpha}_{kk} - \alpha_{kk} \right \vert > \delta \right)
				& \leq 2 K \exp \left( - C \min\left\{ \frac{n \delta ^ 2}{C_{2, *} ^ 2}, \frac{n \delta}{C_{1, *}}  \right\} \right), \\
				%%%
				K \Pr \left(\left \vert \widetilde{b}_{kk} - b_{kk} \right \vert > \delta \right)
				& \leq 2 K \exp \left( - C \min\left\{ \frac{n \delta ^ 2}{C_{2, *} ^ 2}, \frac{n \delta}{C_{1, *}}  \right\} \right), \\
				%%%
				K  (K - 1) \Pr \left(\left \vert \widetilde{b}_{kk'} - b_{kk'} \right \vert > \delta \right)
				& \leq 2 K (K - 1) \exp \left( - C \min\left\{ \frac{n \delta ^ 2}{C_{2, *} ^ 2}, \frac{n \delta}{C_{1, *}}  \right\}  \right).
			\end{split}
		\end{equation*}	
		
		Step 3, 
		let $U = C n \min \left\{ \lambda ^ 2 (2 C_{2, *} ^ 2) ^ {- 1}, \lambda (2 C_{1,*}) ^ {-1} \right\} = C n \min \{\lambda ^ 2, \lambda \}$, where constants $C_{1, *}$ and $C_{2, *}$ are absorbed into constant $C$.
		Set $n ^ {-1} \log(K) \to 0$ as $K > n \to \infty$.
		Let $\lambda = \eta \sqrt{\log(K) / n} \to 0 ^ +$, then $n \lambda ^ 2 = \eta ^ 2 \log (K) \to \infty$, 
		while $\lambda ^ 2 < \lambda$ since $\lambda \to 0 ^ +$.
		So, we have
		\begin{equation*}
			\begin{split}
				\lim\limits_{n \to \infty} \frac{U}{\log(K)}
				= \lim\limits_{n \to \infty} \frac{C n \lambda ^ 2}{\log(K)} 
				= \lim\limits_{n \to \infty} \frac{C \eta ^ 2 \log (K)}{\log(K)} 
				= C \eta ^ 2.
			\end{split}
		\end{equation*}	
		Therefore, the Forbenius norm is bounded as below,
		\begin{equation*}
			\begin{split}
				\left \Vert \widehat{\bm{\Sigma}} - \bm{\Sigma} \right \Vert_{\text{F}} ^ 2
				& = \sum_{k = 1}^{K} p_k \left(\widehat{\alpha}_{kk} - \alpha_{kk} \right) ^ 2
				+ \sum_{k = 1}^{K} p_k (p_k-1) \left(\widehat{b}_{kk} - b_{kk} \right) ^ 2
				+ \sum_{k \neq k'} p_k p_{k'} \left(\widehat{b}_{kk'} - b_{kk'}\right) ^ 2 \\
				& \leq \sum_{k = 1}^{K} p_k \left\{\alpha_{kk} ^ 2 \wedge \left(2\lambda\right) ^ 2 \right\}
				+ \sum_{k = 1}^{K} p_k (p_k-1) \left\{b_{kk} ^ 2 \wedge \left(3 \lambda / 2\right) ^ 2 \right\}
				+ \sum_{k \neq k'} p_k p_{k'} \left\{b_{kk'} ^ 2 \wedge \left(3 \lambda / 2\right) ^ 2 \right\},
			\end{split}
		\end{equation*}
		and the spectral norm is bounded below
		\begin{equation*}
			\begin{split}
				\left \Vert \widehat{\bm{\Sigma}} - \bm{\Sigma} \right \Vert_{\text{S}} 
				& \leq \left \Vert \widehat{\bm{\Sigma}} - \bm{\Sigma} \right \Vert_{1} \\
				& = \sup\limits_{1 \leq j' \leq p} 
				\left\{
				\left \vert \alpha_{k(j') k(j')} \right \vert \wedge \left(2\lambda\right)
				+ \left(p_{k(j')} - 1 \right) \left \vert b_{k(j') k(j')} \right \vert \wedge \left(3 \lambda / 2\right)
				+ \sum_{k \neq k(j')} p_k \left \vert b_{k k(j')} \right \vert \wedge \left(3 \lambda / 2\right)
				\right\}, 
			\end{split}
		\end{equation*}
		where $k(j') = k$ if $j' \in \left\{\bar{p}_{k-1} + 1, \bar{p}_{k-1} + 2, \ldots, \bar{p}_k \right\}$ for every $k$.
		
		Step 4, set $0 < p_0 < 2$ and therefore $x ^ {2 - p_0}$ is monotonically increasing on $(0, \infty)$.
		Let $\omega_{kk} ^ {(\alpha), 2} = \alpha_{kk} ^ 2 \wedge (2\lambda) ^ 2$,
		$\omega_{kk'} ^ {(b), 2} = b_{kk'} ^ 2 \wedge (3 \lambda / 2) ^ 2$ for every $k$ and $k'$.
		After doing some algebra, we have
		\begin{equation*}
			\begin{split}
				\sum_{k = 1}^{K} p_k \left\{\alpha_{kk} ^ 2 \wedge \left(2\lambda\right) ^ 2 \right\}
				%& = \sum_{k = 1}^{K} p_k \omega_{kk} ^ {(\alpha), 2} \\
				& = \sum_{k = 1}^{K} p_k \left \vert \omega_{kk} ^ {(\alpha)} \right \vert ^ {p_0} 
				\left \vert \omega_{kk} ^ {(\alpha)} \right \vert ^ {2 - p_0} \\
				& \leq \sum_{k = 1}^{K} p_k \left \vert \omega_{kk} ^ {(\alpha)} \right \vert ^ {p_0} 
				\left( \max\limits_{1 \leq k \leq K} \left \vert \omega_{kk} ^ {(\alpha)} \right \vert \right) ^ {2 - p_0} \\
				& = \left( \max\limits_{1 \leq k \leq K} \left \vert \omega_{kk} ^ {(\alpha)} \right \vert \right) ^ {2 - p_0} 
				\left(\sum_{k = 1}^{K} p_k \left \vert \omega_{kk} ^ {(\alpha)} \right \vert ^ {p_0} \right) \\
				& \leq \left( 2 \lambda \right) ^ {2 - p_0} \left( \sum_{k = 1}^{K} p_k \left \vert \alpha_{kk} \right \vert ^ {p_0} \right), 
			\end{split}
		\end{equation*}
		and 
		\begin{equation*}
			\begin{split}
				\sum_{k = 1}^{K} p_k (p_k-1) \left\{b_{kk} ^ 2 \wedge \left(3 \lambda / 2\right) ^ 2 \right\}
				& \leq \left( \max\limits_{1 \leq k \leq K} \left \vert \omega_{kk} ^ {(b)} \right \vert \right) ^ {2 - p_0}
				\left\{\sum_{k = 1}^{K} p_k(p_k - 1) \left \vert \omega_{kk} ^ {(b)} \right \vert ^ {p_0} \right\} \\
				& \leq \left( 3 \lambda / 2 \right) ^ {2 - p_0} 
				\left\{\sum_{k = 1}^{K} p_k (p_k - 1) \left \vert b_{kk} \right \vert ^ {p_0} \right\}, \\
				%%%
				\sum_{k \neq k'} p_k p_{k'} \left\{b_{kk'} ^ 2 \wedge \left(3 \lambda / 2\right) ^ 2 \right\}
				& \leq \left( \max\limits_{1 \leq k \neq k' \leq K} \left \vert \omega_{kk'} ^ {(b)} \right \vert \right) ^ {2 - p_0}
				\left( \sum_{k = 1}^{K} p_k p_{k'} \left \vert \omega_{kk'} ^ {(b)} \right \vert ^ {p_0} \right) \\
				& \leq \left(3 \lambda / 2\right) ^ {2 - p_0}
				\left(\sum_{k \neq k'} p_k p_{k'} \left \vert b_{kk'} \right \vert ^ {p_0} \right).
			\end{split}
		\end{equation*}
		Put them together, 
		\begin{equation*}
			\begin{split}
				& \left \Vert \widehat{\bm{\Sigma}} - \bm{\Sigma} \right \Vert_{\text{F}} ^ 2 \\
				& \leq \sum_{k = 1}^{K} p_k \left\{\alpha_{kk} ^ 2 \wedge \left(2\lambda\right) ^ 2 \right\}
				+ \sum_{k = 1}^{K} p_k (p_k-1) \left\{b_{kk} ^ 2 \wedge \left(3 \lambda / 2\right) ^ 2 \right\}
				+ \sum_{k \neq k'} p_k p_{k'} \left\{b_{kk'} ^ 2 \wedge \left(3 \lambda / 2\right) ^ 2 \right\} \\
				& \leq \left( 2 \lambda \right) ^ {2 - p_0} \left( \sum_{k = 1}^{K} p_k \left \vert \alpha_{kk} \right \vert ^ {p_0} \right)
				+ \left( 3 \lambda / 2 \right) ^ {2 - p_0} 
				\left(\sum_{k = 1}^{K} p_k (p_k - 1) \left \vert b_{kk} \right \vert ^ {p_0} \right)
				+ \left(3 \lambda / 2\right) ^ {2 - p_0}
				\left(\sum_{k \neq k'} p_k p_{k'} \left \vert b_{kk'} \right \vert ^ {p_0} \right) \\
				& \leq \left(2 \lambda \right) ^ {2 - p_0} \left(\left \Vert \bm{\Sigma} \right \Vert_{p_0} ^ {(\text{vector})} \right) ^ {p_0},
			\end{split}
		\end{equation*}
		where $\Vert \textbf{A}_{m \times n} \Vert_{p} ^ {(\text{vector})}$ denotes the entrywise norm $\left(\sum_{i = 1}^{m} \sum_{j = 1}^{n} \vert A_{ij}\vert ^ p\right) ^ {1/p}$.
		For the spectral norm, consider $0 < q_0 < 1$, 
		\begin{equation*}
			\begin{split}
				& \left \Vert \widehat{\bm{\Sigma}} - \bm{\Sigma} \right \Vert_{\text{S}} \\
				& = \sup\limits_{1 \leq j' \leq p} 
				\left\{
				\left \vert \alpha_{k(j') k(j')} \right \vert \wedge \left(2\lambda\right)
				+ \left(p_{k(j')} - 1 \right) \left \vert b_{k(j') k(j')} \right \vert \wedge \left(3 \lambda / 2\right)
				+ \sum_{k \neq k(j')} p_k \left \vert b_{k k(j')} \right \vert \wedge \left(3 \lambda / 2\right)
				\right\} \\
				& \leq \sup\limits_{1 \leq j' \leq p} 
				\left\{
				(2 \lambda) ^ {1 - q_0} \left \vert \alpha_{k(j') k(j')} \right \vert ^ {q_0}
				+ \left(3 \lambda / 2\right) ^ {1 - q_0} \left(p_{k(j')} - 1 \right) \left \vert b_{k(j') k(j')} \right \vert ^ {q_0}
				+ \left(3 \lambda / 2\right) ^ {1 - q_0} \sum_{k \neq k(j')} p_k \left \vert b_{k k(j')} \right \vert ^ {q_0}
				\right\} \\
				& \leq (2 \lambda) ^ {1 - q_0}
				\sup\limits_{1 \leq j' \leq p}
				\left\{
				\left \vert \alpha_{k(j') k(j')} \right \vert ^ {q_0}
				+ \left(p_{k(j')} - 1 \right) \left \vert b_{k(j') k(j')} \right \vert ^ {q_0}
				+ \sum_{k \neq k(j')} p_k \left \vert b_{k k(j')} \right \vert ^ {q_0}
				\right\} \\
				& \leq (2 \lambda) ^ {1 - q_0}
				\left \Vert \bm{\Sigma} ^ {q_0} \right \Vert_{1},
			\end{split}
		\end{equation*}
		where $\Vert \textbf{A}_{m \times n} ^ {q_0} \Vert_{1}$ denotes the maximum absolute column sum (with $q_0$-th power entrywisely) $\max_{1 \leq j \leq n} \sum_{i = 1}^{m} \vert A_{ij} \vert ^ {q_0}$.
		
		In summary, given $\log(K) / n \to 0$ as $K = K_n > n \to \infty$, 
		we select $C \eta ^ 2 = 5$ and $\lambda = \eta \sqrt{\log(K) / n} \to 0$, 
		then $2K(K - 1)\exp(- U) \to 0$. 
		Furthermore, for $0 < p_0 < 2$ and $0 < q_0 < 1$, $\Vert \widehat{\bm{\Sigma}} - \bm{\Sigma} \Vert_{\text{F}} ^ 2
		\leq (2 \lambda) ^ {2 - p_0} \left(\left \Vert \bm{\Sigma} \right \Vert_{p_0} ^ {(\text{vector})} \right) ^ {p_0}$ and $\Vert \widehat{\bm{\Sigma}} - \bm{\Sigma} \Vert_{\text{S}}
		\leq (2 \lambda) ^ {1 - q_0} \Vert \bm{\Sigma} ^ {q_0} \Vert_{1}$ with probability $1$.
	\end{proof}

	\bibliographystyle{biom}	
	
	\bibliography{paper_ref_UBCovEst}